\documentclass[a4paper,fleqn,usenatbib]{mnras}

\usepackage{newtxtext,newtxmath}
\usepackage{ulem}
\usepackage[T1]{fontenc}
\usepackage{ae,aecompl}
\usepackage{bm}
\usepackage{here}

\usepackage{amsmath}	
\usepackage{amssymb}	
\usepackage{graphicx}
\usepackage{wrapfig}

\usepackage{lipsum}

\title[Cosmological simulation with dust]{Cosmological simulation with dust formation and destruction}

\author[S. Aoyama et al.]{Shohei Aoyama,$^{1,2}$\thanks{E-mail: saoyama@asiaa.sinica.edu.tw (SA)}
Kuan-Chou Hou,$^{1,3}$
Hiroyuki Hirashita,$^{1}$
Kentaro Nagamine$^{2,4}$\newauthor
and Ikkoh Shimizu$^{2}$\\
$^{1}$Institute of Astronomy, and Astrophysics, Academia Sinica, PO Box 23-141, Taipei 10617, Taiwan\\
$^{2}$Theoretical Astrophysics, Department of Earth \& Space Science, Osaka University, 1-1 Machikaneyama, Toyonaka, Osaka 560-0043, Japan\\
$^{3}$Department of Physics \& Institute of Astrophysics, National Taiwan University, Taipei 10617, Taiwan \\
$^{4}$Department of Physics \& Astronomy, University of Nevada, Las Vegas, 4505 S. Maryland Pkwy, Las Vegas, NV 89154-4002, USA \\
}

\date{Accepted XXX. Received YYY; in original form ZZZ}

\pubyear{2018}

\begin{document}
\label{firstpage}
\pagerange{\pageref{firstpage}--\pageref{lastpage}}
\maketitle

\begin{abstract}
To investigate the evolution of dust in a cosmological volume,
we perform hydrodynamic simulations, in which the enrichment of
metals and dust is treated self-consistently with star formation and
stellar feedback. 
We consider dust evolution driven by
dust production in stellar ejecta, dust destruction by sputtering,
grain growth by accretion and coagulation, 
and grain disruption by shattering, and treat small and large
grains separately to trace the grain size distribution.
After confirming that our model nicely reproduces the observed relation between
dust-to-gas ratio and metallicity for nearby galaxies, we concentrate on the
dust abundance over the cosmological volume in this paper.
The comoving dust mass density
has a peak at redshift $z\sim 1$--2,
coincident with the observationally suggested dustiest epoch in the Universe.
{In the local Universe}, roughly 10 per cent of the
dust is contained in the intergalactic medium (IGM), where
only 1/3--1/4 of the dust survives against dust destruction by sputtering.
We also show that the dust mass function is roughly reproduced at
$\lesssim 10^8$ M$_\odot$, while the massive end still has a discrepancy,
which indicates {the necessity of stronger feedback in massive galaxies}.
In addition, our model broadly reproduces the observed radial profile of dust surface density
in the circum-galactic medium (CGM).
While our model satisfies the observational constraints for the
dust extinction {on cosmological scales}, it predicts that
the dust in the CGM and IGM is
dominated by large ($> 0.03~\mu$m) grains, which is in tension with 
the steep reddening curves {observed} in the CGM.

\end{abstract}

\begin{keywords}
dust, extinction -- methods: numerical -- ISM: dust -- galaxies: evolution -- galaxies: formation -- galaxies: ISM
\end{keywords}



\section{Introduction}
\label{sec:Intro}

Dust is an essential component in understanding star formation
properties of galaxies 
both observationally and theoretically.
Because dust absorbs stellar ultraviolet (UV)--optical light and reemits it in the infrared (IR) 
\citep[e.g.][]{2000ApJ...533..682C,2002A&A...383..801B,2012ApJ...755..144T}, 
a precise estimation of star formation rate (SFR) in galaxies requires correction for dust extinction 
\citep[e.g.][]{1999ApJ...519....1S,2010A&A...514A...4T,2012ARA&A..50..531K}.
{The analysis of} \cite{2005A&A...440L..17T}
revealed that a higher fraction of star formation is hidden by
dust at $z\sim 1$ than at $z\sim 0$, where $z$ is the redshift, and
that more than half of the star formation activity is {enshrouded by dust} 
at $0.5 \le z \le 1.2$ (see also \cite{2013A&A...554A..70B}
and references therein). 

There are some interesting theoretical issues in which dust plays a significant role.
Dust is an efficient catalyst of molecular hydrogen (H$_{2}$) formation in the interstellar medium (ISM) 
\citep[e.g.][]{1963ApJ...138..393G,2004ApJ...604..222C,2009A&A...496..365C}.
\cite{2002MNRAS.337..921H} 
{showed that early dust production at high redshift}
dramatically enhances the H$_2$ abundance, which probably leads to an enhancement
of star formation {activity} in galaxies.
Dust has an impact on gas dynamics in dusty clouds
through 
radiation pressure \citep[e.g.][for a recent development]{2017MNRAS.466L.123I}.
In addition, the typical mass of the final fragments in star-forming clouds is also regulated by dust cooling 
\citep[][]{1998MNRAS.299..554W,2000ApJ...534..809O,2005MNRAS.359..211L,2005ApJ...626..627O,2006MNRAS.369.1437S}; 
this effect could have a dramatic impact on the stellar
initial mass function. 
Moreover dust eventually becomes the ingredient of planets in protostellar discs.

{
The evolution of the total dust amount in a galaxy can be broadly understood in the
chemical evolution framework since dust evolution is strongly linked to metal
enrichment \citep{1998ApJ...496..145L,2014MNRAS.440.1562M}.
As shown in a variety of chemical evolution models,
the increase of the dust amount is mainly driven by dust condensation
in stellar ejecta and dust growth in the dense ISM, while the decrease occurs when the
dust is swept by supernova (SN) shocks
\citep{1998ApJ...501..643D,2008A&A...479..453Z,2008A&A...479..669C,2011EP&S...63.1027I,2011MNRAS.416.1340H,2013EP&S...65..213A,2017MNRAS.471.4615G}.
}
When we consider the property of dust grains in the ISM, 
not only the total dust abundance but also the grain size distribution is of fundamental importance \citep[e.g.][]{1977ApJ...217..425M,2013ApJ...770...27N}. 
In particular, the extinction curve (i.e.\ the wavelength dependence of absorption and scattering cross-section) 
{depends sensitively on} the grain size distribution
\citep[][]{1983Natur.306..625B,2017MNRAS.469..870H}. 
In addition, 
{the total grain surface area,
which depends on the grain size distribution, governs the rate of grain-surface H$_2$ formation}
\citep[][]{1976ApJ...207..131B,2011ApJ...735...44Y}.

Dust evolution is driven by the interactions not only {with} gas particles but also
{with} dust itself in the ISM
\citep[see][and references therein for the processes described in what follows]{2013MNRAS.432..637A}.
Dust grains are produced by SNe and asymptotic giant branch (AGB) stars, and
after being injected into the ISM, they suffer destruction in SN shocks sweeping the ISM.
Dust grains grow by accreting surrounding gas-phase metals in the dense ISM.
Dust grains interact with themselves via collisional processes such as coagulation and shattering.
The rates of the above grain processing mechanisms in the ISM (dust destruction, accretion, coagulation,
and shattering) depend not only on the local physical condition of the gas but also on
the dust abundance and metallicity. Moreover, as found by \cite{2012MNRAS.424L..34K},
the efficiency of interstellar processing could depend strongly on the grain size distribution.
Therefore, for a complete understanding of dust evolution, we must consider not only the
evolution of dust abundance but also that of grain size distribution.

\cite{2013MNRAS.432..637A} constructed a full framework for treating the evolution of grain size distribution 
consistently with the enrichment of metals and dust in a galaxy. They treated all the above
processes of dust evolution and revealed that all of these processes are necessary for a comprehensive understanding of the observed dust-to-gas mass ratios and extinction curves in nearby galaxies (see \citealt{2015MNRAS.447L..16N} for an extension of their model to high
redshift).
To focus on the dust evolution, they treated a galaxy as a one-zone object.
As a consequence of their modelling, they succeeded in providing a tool to understand not only observed gas-to-dust mass ratios but also extinction curves in galaxies 
\citep[][]{2014MNRAS.440..134A}. 

Since the dust evolution is affected by the physical condition of the ISM where it
resides, it is important to deeply understand the hydrodynamical evolution of the ISM
in a spatially resolved way.
Hydrodynamical simulations have indeed been a powerful tool to clarify galaxy formation and evolution. 
{They provide a significant advantage over simple one-zone calculations, 
which generally need to introduce some strong assumptions such as instantaneous mixing and homogeneity.}
Many cosmological hydrodynamic simulations have reproduced and predicted the observed galaxy mass and luminosity functions 
\citep[e.g.][]{2001ApJ...558..497N,2004MNRAS.350..385N,2012MNRAS.419.1280C,2012MNRAS.427..403J,2013ApJ...766...94J,2014MNRAS.440..731S,2014ApJ...780..145T,2014MNRAS.444.1518V,2015arXiv150900800S,2015MNRAS.446..521S, 2015MNRAS.454.2277S,2017MNRAS.470.3300D,2017MNRAS.465.2936M, 2018MNRAS.473.4077P}. 
There have been some attempts to include dust evolution in cosmological hydrodynamical
simulations.
{
\cite{2010MNRAS.403..620D} calculated dust formation and destruction by SNe in their cosmological simulation
and predicted the submillimetre fluxes from high-redshift Lyman break galaxies.}
\cite{2015MNRAS.451..418Y} calculated the radiation transfer of UV light
based on the spatial distribution of metals in their zoom-in simulations, 
and estimated the IR luminosities of individual high-$z$ galaxies.
They assumed a constant dust-to-metal ratio, and did not
explicitly treat the dust evolution.

{\citet{2013MNRAS.432.2298B,2013MNRAS.436.2254B,2015MNRAS.449.1625B}
treated} dust as a separate component 
from gas, dark matter and star particles and solved the interaction between dust and gas.
They calculated H$_2$ formation on dust surfaces and dust evolution consistently to investigate the spatial
distribution of dust and molecular gas in galaxies.
\cite{2016MNRAS.457.3775M} traced dust evolution along with the hydrodynamical evolution of the
gas by performing cosmological zoom-in simulations. They revealed the importance of dust growth by accretion, and pointed out the necessity of a more realistic treatment of dust destruction and feedback by SNe. 
In addition, \cite{2017MNRAS.468.1505M} compared statistical properties of dust, 
especially, the dust mass function and the comoving dust mass density, 
and found that their simulation broadly reproduced the observation in the present-day Universe, although it tended to underestimate the dust abundance in high-$z$ dusty galaxies.
{\cite{2016ApJ...831..147Z} analyzed dust evolution in 
an isolated Milky Way-like galaxy by post-processing the simulation of \cite{2013MNRAS.432..653D}. 
They put particular focus on dust growth by accretion and examined gas-temperature-dependent sticking coefficient in accretion, in order to reproduce
the relation between silicon depletion and gas density.}

All these simulations only {traced} the dust abundance, but did not treat the
grain size distribution. As mentioned above, the grain size distribution affects the
dust evolution. For the grain size distribution, in addition to the processes included in the
above simulations, shattering and coagulation are important.
Implementation of grain size distributions in hydrodynamical simulations has not been successful, mainly because of the high computational cost. Calculating the grain size distribution in a fully self-consistent manner over the cosmic age is computationally expensive even in one-zone calculation as shown by \cite{2013MNRAS.432..637A}. {Recently, \cite{2018MNRAS.tmp.1185M} implemented a full grain size distribution in their hydrodynamical simulation. However their simulation is still limited to an isolated galaxy.}
For the purpose of treating the evolution of grain size distribution within the available computational capability, 
\citet[][hereafter A17]{2017MNRAS.466..105A} and \cite{2017MNRAS.469..870H} adopt the two-size approximation formulated by \cite{2015MNRAS.447.2937H}, 
in which the entire grain size range is represented by two sizes ranges
divided at around $a \simeq 0.03 \mu$m ($a$ is the grain radius).
\cite{2015MNRAS.447.2937H} confirmed that the two-size approximation gives 
the same evolutionary behavior of grain size distribution and extinction curve as calculated by the full treatment of \cite{2013MNRAS.432..637A,2014MNRAS.440..134A}.
Because {this two-size approximation reduces the computational cost},
it provides a feasible way to compute the evolution
of grain size evolution in hydrodynamical simulations.
Consequently, we can not only compute the spatial variations in dust abundance, 
but also examine the grain size distribution as a function of time and metallicity. 

The hydrodynamical simulation in A17{, } 
\cite{2017MNRAS.469..870H} 
{and \cite{2018arXiv180406855G}}
treated the dust evolution 
using the two size approximation in a consistent manner
with the local physical states such as the
local gas density and temperature.
They succeeded in theoretically predicting spatial inhomogeneity in
the dust abundance (A17), extinction curves
\citep{2017MNRAS.469..870H}
{and the relation between dust-to-gas mass ratio and 
oxygen abundance \citep[][]{2018arXiv180406855G}.
}
However, in {A17 and \cite{2017MNRAS.469..870H}}, only a single isolated spiral galaxy was simulated{.}
{\cite{2018arXiv180406855G} performed zoom-in simulations and analyzed only four massive clusters.}
Therefore
{no statistical information on dusty galaxies was achievable.}

In order to obtain general evolutionary features of galaxies, a simulation on a
cosmological spatial scale and time-scale is desired.
Such a cosmological simulation {is capable of predicting
the evolution of  a large number of galaxies.}
\cite{2016MNRAS.457.3775M,2017MNRAS.468.1505M}
implemented dust evolution in a cosmological hydrodynamic simulation, and
succeeded in predicting statistical properties of galaxies such as the dust mass function
and the scaling relations of dust abundance with quantities characterizing galaxies.
However, they did not include the evolution of grain size distribution. As
mentioned above, the evolution of grain size distribution is of fundamental importance in
understanding the dust evolution.

Another important point of cosmological simulations {is that they are able to predict}
the dust and metal enrichment of the intergalactic medium (IGM).
\cite{2010MNRAS.405.1025M} observationally revealed that dust grains are sure to exist 
in the circumgalactic medium (CGM) and IGM. 
{
The CGM is defined as the medium located from $\gtrsim $ a few tens of kpc
to $\sim 1$ Mpc from the galaxy centre \citep{2010MNRAS.405.1025M}.}
\cite{2012ApJ...754..116M} estimated 
the abundance and radial profile of dust in the CGM using Mg \textsc{ii} absorbers
as tracers. 
On the theoretical side, \cite{2003MNRAS.341L...7I} showed that
dust grains in the IGM {affect the thermal history of the IGM} 
through photoelectric heating. They also pointed out that the efficiency of photoelectric heating depends on the
grain size. Therefore, clarifying the evolution of dust abundance and grain size
distribution in the IGM is important. 
In addition, \citet{2016SSRv..202...79N} and \citet{2011MNRAS.412.1059Z}
mention that the metal distribution in the IGM is 
sensitive to feedback (energy injection) models. Because dust grains are created by metal condensation 
{and spread to the ISM and the IGM in a way dependent on the feedback
strength \citep{2017MNRAS.469..870H}},
the distribution of dust grains could also be useful for testing feedback models.

In this paper, we perform cosmological $N$-body/SPH simulations with \textsc{gadget3-osaka}
developed in A17 {based on the original \textsc{gadget} code 
\citep[][]{2005MNRAS.364.1105S}}.
In this paper, we particularly focus on the overall dust properties
in a cosmological volume.
{We also test the statistical properties of the dust content in galaxies, and
examine} the dust enrichment in the IGM
as a result of cosmic structure formation and SN feedback.
The cosmic history of dust enrichment and the evolution of grain size distribution
on a cosmological spatial scale and time-scale are the main topics of this paper.
The detailed analysis of individual galaxies
is described in a separate paper (Hou et al., in preparation).

This paper is organized as follows. In Section \ref{model}, we explain the model of
dust evolution in the cosmological simulation.
We present the simulation results in Section \ref{sec:result}. We discuss
the parameter dependence in Section 4. We conclude in Section 5. 
Throughout this paper, we adopt $Z_{\odot} = 0.02$ for the
solar metallicity following \cite{2015MNRAS.447.2937H}. {This value, used as simple metallicity normalization, does not affect our main results.}
{We adopt the following cosmological parameters 
\citep{2016A&A...594A..13P}:}
baryon density parameter $\Omega_{\rm b} = 0.049$, 
total matter density parameter $\Omega_{\rm m} =0.32$, 
cosmological constant parameter $\Omega_\Lambda =0.68$, 
Hubble constant $H_{0} = 67$ km s$^{-1}$ Mpc$^{-1}$, 
power spectrum index $n_{\rm s}=0.9645$, and
density fluctuation normalization $\sigma_{8}=0.831$.
In this paper, we also use $h \equiv H_{0} / (100$ km s$^{-1}$ Mpc$^{-1})=0.67$ {for the non-dimensional Hubble constant}.

\begin{table}
\centering
\begin{minipage}{90mm}
\caption{Simulation setup}
\label{table:simulation}
    \begin{tabular}{cccccc}\\ \hline 
Name & Boxsize & $N$ & $\varepsilon_{\rm grav} $ & $m_{\rm dm}$ & $m_{\rm gas}^{\rm init}$ \\ 
     & [$h^{-1}$Mpc] &&[$h^{-1}$kpc]  &[$h^{-1}{\rm M}_{\odot}$]&[$h^{-1}{\rm M}_{\odot}$] \\ \hline 
L50N512 & 50 & $2\times 512^{3}$ & 3 &$6.89 \times 10^{7}$& $1.28 \times 10^{7}$\\ 
L50N256 & 50 & $2\times 256^{3}$ & 5 &$5.51 \times 10^{8}$& $1.02 \times 10^{8}$ \\ \hline
    \end{tabular}
\textit{Notes.}
$N$, $\varepsilon_{\rm grav} $, $m_{\rm dm}$ and $m_{\rm gas}^{\rm init}$ are 
the number of particles, the gravitational softening length, the mass of dark matter particle
and the initial mass of gas particle, respectively.
\end{minipage}
\end{table}

\section{Model}\label{model}

\subsection{Galaxy evolution simulation}

We basically use the same simulation code as used in A17 (\textsc{gadget3-osaka})
{but we perform a cosmological simulation.}
{We generated initial conditions at $z=99$ with \textsc{MUSIC} \citep[][]{2011MNRAS.415.2101H}.}
The basic features in the simulation other than dust implementation
are described in a separate paper (Shimizu et al., in preparation).
Below we explain our simulation focusing on the difference from A17.
We performed cosmological hydrodynamical simulations with a comoving
50$h^{-1}$ Mpc box.
The initial number of particles are $2\times 256^{3}$ and $2\times 512^{3}$
(referred to as L50N256 and L50N512, respectively).
Other parameters of simulations are shown in Table \ref{table:simulation}. 

In our simulation, stars are formed in dense and cold gas particles.
Star particles are created from gas particles whose number density is 
greater than $n_{\rm th}=0.1$ cm$^{-3}$ and temperature is less than $T_{\rm th}=10^{4}$ K
with the following SFR:
\begin{eqnarray} 
\dfrac{d\rho_{\ast}}{dt}=\varepsilon_{\ast}\dfrac{\rho_{\rm gas}}{t_{\rm ff}},
\end{eqnarray}
where $\rho_{\ast}$ and $\rho_{\rm gas}$ are the local mass density of newly formed stars and gas,
respectively,
$\varepsilon_{\ast}$ is the star formation efficiency in a free-fall time (we adopt
$\varepsilon_{\ast}=0.05$ in this paper), and
$t_{\rm ff}\equiv \sqrt{3\pi \slash \left( 32 G \rho_{\rm gas} \right)}$ is the local free-fall time.
{The temperature threshold is used to search a region
where Lyman $\alpha$ cooling has made a favourable condition for gas collapse
(or star formation)
\citep[e.g.][]{1993ApJS...88..253S}.
}
The threshold density is determined by the
extrapolation of so-called Larson's law  \citep{1981MNRAS.194..809L} to our spatial resolution
($\sim 1$ kpc):
{this density criterion serves to choose regions where bound objects like molecular clouds are potentially formed}.
We also confirmed that the resulting SFR is
roughly consistent with the Schmidt--Kennicutt law in isolated galaxies 
(Shimizu et al., in preparation). 
{We include the metal and dust production by
not only Type II SNe but also Type Ia SNe and AGB stars.}
The formation of various {heavy} elements is treated by
implementing the \textsf{CELib} package \citep[][]{2017AJ....153...85S} 
to {\sc gadget3-osaka}.
Energy input from {massive stars} 
(stellar feedback) is also considered.
\begin{figure}
\includegraphics[width=8.3cm]{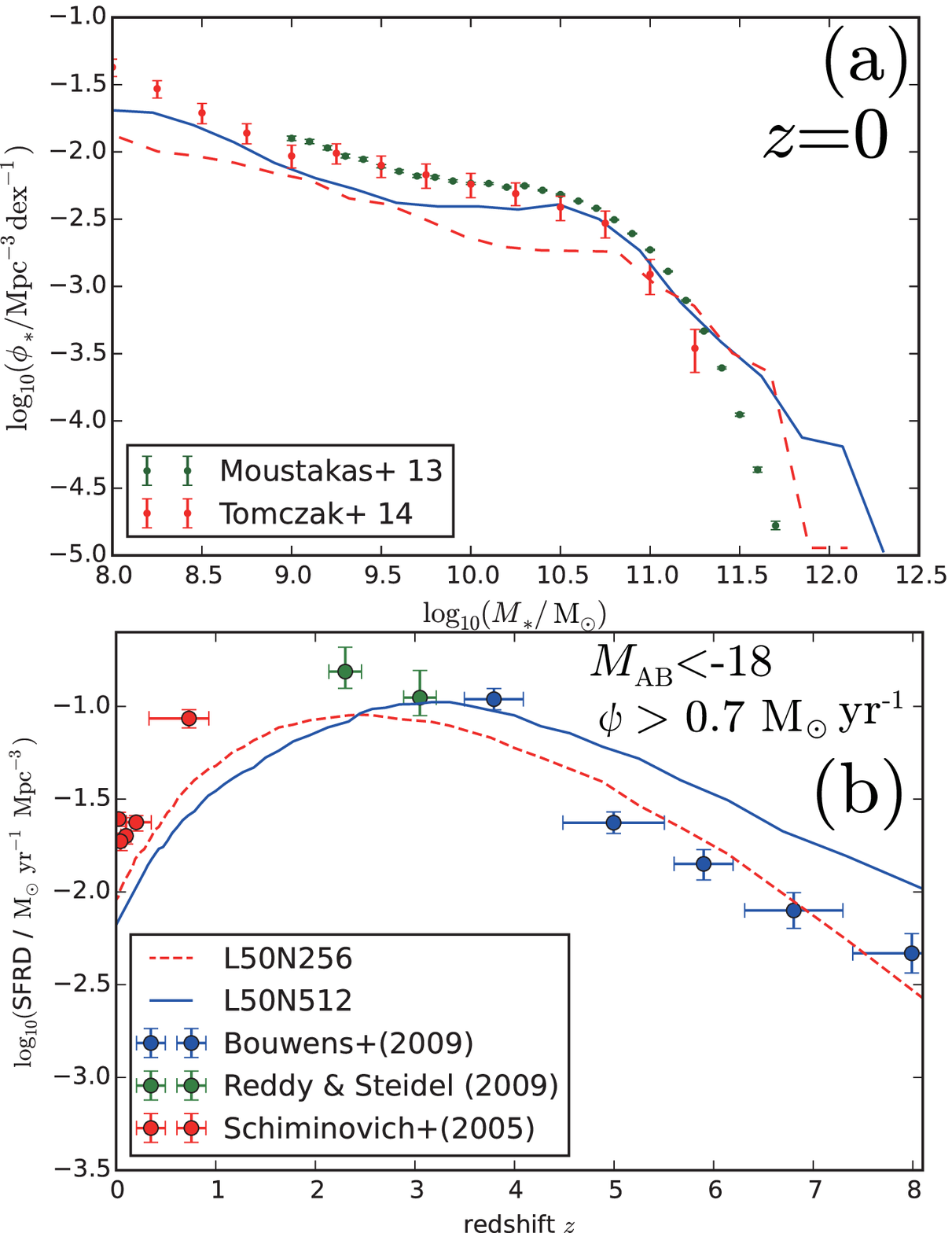}
\caption{
(a) Stellar mass function at $z=0$ for L50N512 (blue solid line)
and L50N256 (red dashed line) (see Table \ref{table:simulation}
for these two simulations).
The observational data  are taken from
\citet{2013ApJ...767...50M} and \citet{2014ApJ...783...85T} as shown in the legend. 
(b) Cosmic star formation rate density (SFRD) with
L50N512 (blue solid line) and L50N256 (red dashed line).
They are shown with observational data based on far-UV galaxy samples
\citep[][]{2005ApJ...619L..47S} and 
Lyman break galaxies \citep[][]{2009ApJ...705..936B,2009ApJ...692..778R}.
{Taking the observational detection limits into account,}
we only consider the galaxies with
star formation rate $\psi > 0.7\,{\rm M}_{\odot}\, {\rm yr}^{-1}$ 
({corresponding to the} absolute AB magnitude $M_{\rm AB}<-18$
according to the conversion formula in \citet{1998ApJ...498..106M}
) 
for a fair comparison.
}
\label{fig:SFRD}
\end{figure}
\begin{figure*}
\includegraphics[width=16.5cm]{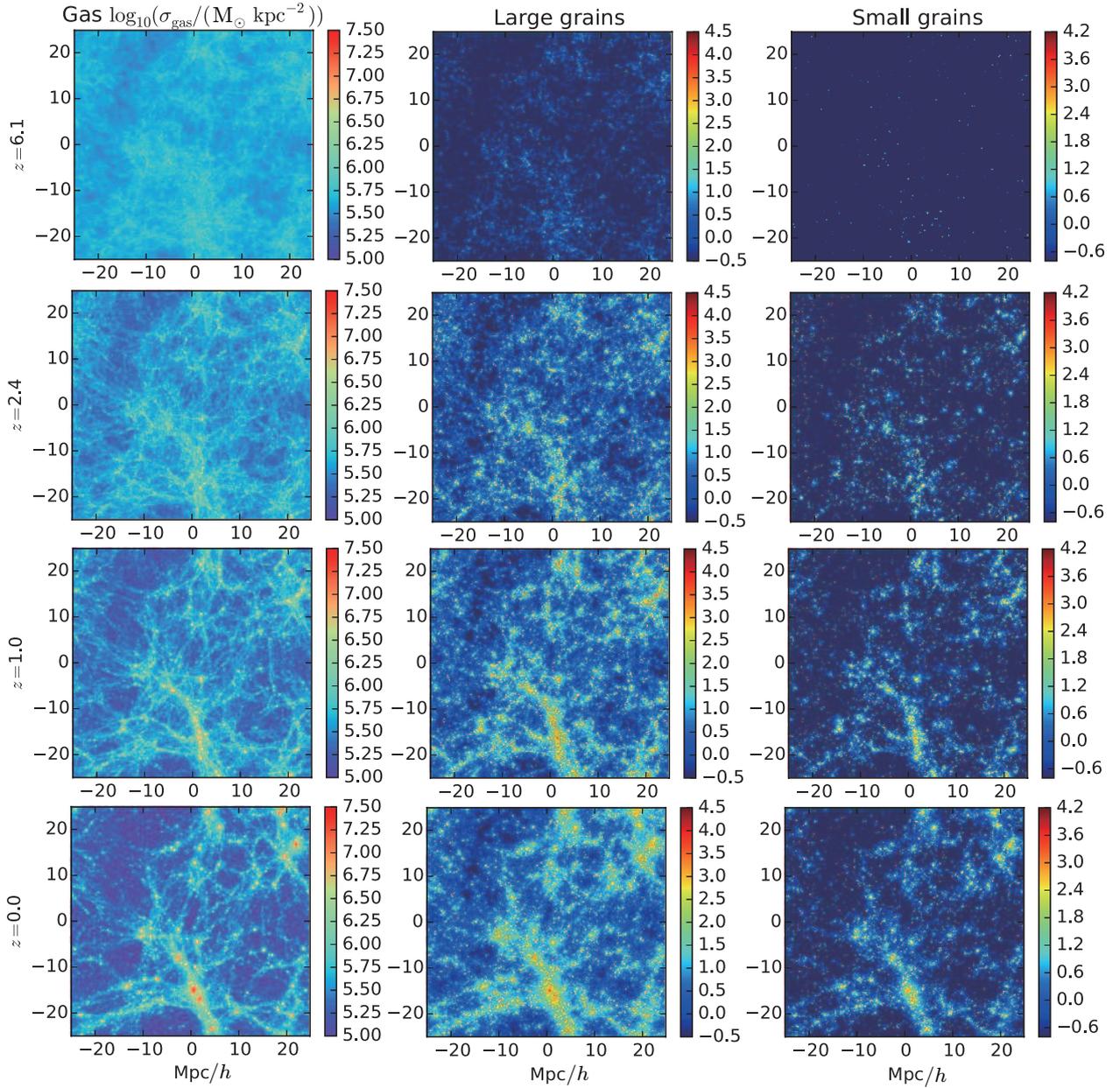}
\caption{
Time evolution of {the} projected density field 
$(\int \rho (\textbf{x}) dz)$ of gas and small and large grains
at $z=6.1,\,2.4,\,1.0,\,0.0$, where $\rho(\textbf{x})$ is the density at position $\textbf{x}$ of each component.
The depth of the integration is 50 $h^{-1}$Mpc.
The comoving box size is 50 $h^{-1}$Mpc and the colour indicates the log-scale density of each component 
in units of $M_{\odot}\, {\rm kpc}^{-2}$ (shown by the colour bar).
}
\label{fig:dustDistribution}
\end{figure*}

The model is successful in reproducing the main statistical properties of
star formation and stellar content in galaxies, especially, the cosmic star
formation rate density (SFRD) and the stellar mass function at $z=0$
\citep[][]{2005ApJ...619L..47S,2009ApJ...705..936B,2009ApJ...692..778R,2013ApJ...767...50M,2014ApJ...783...85T},
except at the very massive end of the stellar mass function, 
as shown in Fig.~\ref{fig:SFRD}. 
Comparing the results of L50N512 and L50N256, we find that 
the stellar mass function and the SFRD at low-z ($z\lesssim 3$)
do not significantly depend on the mass resolution.
{
However the SFRD at high $z$ ($z\gtrsim 3$) does depend on the mass resolution, 
because more low-mass galaxies can form in a higher resolution simulation.} 

Feedback of active galactic nuclei (AGN) is not included to avoid further complexity
not related to dust production, although it might resolve the discrepancy in the
stellar mass function at the very massive end.
AGN feedback has been implemented in many cosmological simulations and shown 
to be responsible for suppressing the formation of massive objects
\citep[e.g.][]{2013MNRAS.436.3031V, 2014MNRAS.444.1518V, 2015MNRAS.452..575S,2017MNRAS.465.3291W,2017arXiv171200023F}.
However, the treatment of AGN feedback depends on subgrid models, which involve
choice of model parameters \citep[e.g.][]{2013MNRAS.436.3031V}.
In addition, AGN feedback is also related to the formation of supermassive black holes.
Although the relation between growth of supermassive black hole and dust
enrichment is an interesting topic \citep{2011MNRAS.416.1916V},
we choose to focus on the processes directly related to dust formation in this paper. 
Therefore, we 
leave the influence of AGN feedback for the future work.

\subsection{Basic treatment of dust evolution}\label{subsec:dust_ev}
In this paper, we basically adopt the dust evolution model 
{used} in our previous
isolated-galaxy simulation \citep[A17;][]{2017MNRAS.469..870H}.
We represent the whole range of grain radii by large and small grain populations
roughly separated at $a\sim 0.03~\micron$
according to \citet{2015MNRAS.447.2937H}.
We set the typical radii of the large and small grain populations as
$0.1\, \mu$m and $5\times 10^{-3}\, \mu$m, respectively. 

The abundances of the two dust populations on a gas particle are represented by the dust-to-gas mass ratios, $\mathcal{D}_{\rm L}$ and $\mathcal{D}_{\rm S}$, as
\begin{eqnarray} 
\mathcal{D}_{\rm L} = \dfrac{m_{\rm L}}{m_{{\rm g}}}\, ,\\
\mathcal{D}_{\rm S} = \dfrac{m_{\rm S}}{m_{{\rm g}}}\, ,
\end{eqnarray}
where $m_{{\rm g}}$ is the mass of the gas particle, and $m_{\rm L}$ and $m_{\rm S}$
are the total mass of large and small grains in the gas particle,
respectively. Hereafter, we refer to $\mathcal{D}_{\rm L}$ ( $\mathcal{D}_{\rm S}$ ) as the large
(small) grain abundance. The total dust-to-gas ratio $\mathcal{D}_\mathrm{tot}$ is
defined as
\begin{eqnarray}
\mathcal{D}_{\rm tot} \equiv 
\mathcal{D}_{\rm L} + \mathcal{D}_{\rm S}.
\end{eqnarray}
In our simulation, each gas particle has its own dust abundance
$\mathcal{D}_{{\rm L}(i)}$ and $\mathcal{D}_{{\rm S}(i)}$, 
where suffix $(i)$ indicates the label for the gas particle. 
Based on the two-size model, we calculate the formation and destruction of 
large and small dust grains on each gas particle using variables and outputs in the simulation as described below (see A17, especially their equations 13 and 14 for further details).

We {calculate} the time evolution of 
the large and small grain abundances in the $i$-th particle at time $t$ as
{(Appendix \ref{appendixA})}
\begin{eqnarray} 
\dfrac{\mathrm{d}\mathcal{D}_{{\rm L}(i)}(t)}{\mathrm{d}t}
&=&- \left( \dfrac{\mathcal{D}_{{\rm L}(i)}(t)}{\tau_{\rm sh}} 
- \dfrac{\mathcal{D}_{{\rm S}(i)}(t)}{\tau_{\rm co}}\right)- \dfrac{\mathcal{D}_{{\rm L}(i)}(t)}{\tau_{\rm sp}(a_{\rm L})}\notag \\ 
& & +\left[ \dfrac{\mathrm{d}\mathcal{D}_{{\rm L}(i)}(t)}{\mathrm{d}t}\right]_{{\rm Source}}
- \left[ \dfrac{\mathrm{d}\mathcal{D}_{{\rm L}(i)}(t)}{\mathrm{d}t}\right]_{{\rm SNe}} \notag \\
& &-\frac{\mathcal{D}_{{\rm L}(i)}(t)}{m_{\mathrm{g}(i)}}\dfrac{\mathrm{d} m_{{\rm g}(i)}^{\rm return}}{\mathrm{d} t}\, ,  \label{eq:timeL} \\ 
\dfrac{\mathrm{d}\mathcal{D}_{{\rm S}(i)}(t)}{\mathrm{d}t}
&=&  \left(\dfrac{\mathcal{D}_{{\rm L}(i)}(t)}{\tau_{\rm sh}}
- \dfrac{\mathcal{D}_{{\rm S}(i)}(t)}{\tau_{\rm co }}  
+ \dfrac{\mathcal{D}_{{\rm S}(i)}(t)}{\tau_{\rm acc}}\right)\, \notag \\
& &- \dfrac{\mathcal{D}_{{\rm S}(i)}(t)}{\tau_{\rm sp}(a_{\rm S})}
-\left[ \dfrac{\mathrm{d}\mathcal{D}_{{\rm S}(i)}(t)}{\mathrm{d}t}\right]_{{\rm SNe}}\notag \\ 
& &-\frac{\mathcal{D}_{{\rm S}(i)}(t)}{m_{\mathrm{g}(i)}}
\dfrac{\mathrm{d} m_{{\rm g}(i)}^{\rm return}}{\mathrm{d} t}\, ,\label{eq:timeS} 
\end{eqnarray}
where $\tau_{\rm sh}$, $\tau_{\rm co}$, and $\tau_{\rm acc}$ 
are the time-scales of shattering, coagulation, and accretion, respectively,
{and $\mathrm{d} m_{{\rm g}(i)}^{\rm return}/{\mathrm{d} t}$
is the gas ejection rate from stars (note that the ejected gas dilutes
the dust-to-gas ratio)}.\footnote{Equations (13) and (14) in A17
need to be corrected by including this dilution term. However, A17 correctly included
this term in their {code and calculations}.}. 
The parameter $\tau_\mathrm{sp}(a)$ is the sputtering time-scale as a function of grain
radius in the hot gas not associated with SNe
(see Section \ref{subsec:sput}), and the terms with `Source' and `SNe' describe the
stellar dust production and SN destruction, respectively.
In our formulation, these time-scales depend on the gas density, dust abundances, and/or metallicity
but the dependence on those quantities
are not explicitly shown here for the brevity of notation.
The formation and destruction terms are evaluated by
\begin{eqnarray} 
\left[ \dfrac{d\Delta \mathcal{\mathcal{D}}_{{\rm L}(i)} (t)}{dt}\right]_{\rm Source}\mathrm{d}t&=&
f_{\rm in}\dfrac{\Delta m_{\rm metal}}{m_{{\rm g}(i)}}
\left( 1-\delta \right),
\label{eq:dustsource}\\
\left[ \dfrac{d\Delta {\mathcal{D}}_{{\rm L, S}(i)} (t)}{dt}\right]_{\rm SNe}\mathrm{d}t&=&
\left[ 1-( 1 - \eta )^{N_{\rm SN}}\right] {\mathcal{D}_{{S/L}(i)}(t)}, \notag\\
& &\label{eq:dustdestruct-1}
\end{eqnarray}
where
$f_\mathrm{in}$ is the dust condensation efficiency of metals in the stellar ejecta
(we assume $f_\mathrm{in}=0.1$ following A17),
$\Delta m_{\rm metal}$ is the ejected metal mass from stars, 
$\delta$ is the destroyed fraction of newly formed dust,
$N_{\rm SNe}$ is the number of SNe that affect the gas particle
of interest (note that, because a star particle represents a cluster of 
$\sim 10^{6}$--$10^{7}$ stars and contains a number of massive stars, 
a number of SNe are treated as a single explosion from the star particle), 
and $\eta$ is the destroyed fraction of preexisting dust by a single SN
(see the evaluations of $\delta$, $N_\mathrm{SNe}$ and $\eta$ in
A17).\footnote{In A17,
their equation (19) corresponding equation (\ref{eq:dustdestruct-1}) used a notation
of $\mathcal{D}_{{(\rm SNe\slash L, S)}(i)}(t)$, which should be simply
$\mathcal{D}_{{\rm S/L}(i)}(t)$.}

The time-scale parameters of accretion, coagulation, and shattering are
determined in the following way (see A17 and references therein for the detailed derivation).
Since accretion and coagulation occur in the dense clouds, which cannot
be resolved in our simulations, we
adopt a subgrid model.
We assume that dense ($n_\mathrm{gas}>1\,{\rm cm}^{-3}$, where
$n_\mathrm{gas}$ is the gas number density) and cold gas particles
($T_\mathrm{gas}<10^{4}\,{\rm K}$, where $T_\mathrm{gas}$ is the gas temperature),
which are referred to as the \textit{dense gas particles},
host dense clouds with 50 K and $10^{3}\, {\rm cm}^{-3}$.
Because the cosmological simulations in this paper only resolve less dense gas than
the single-galaxy simulation in A17, we apply a looser condition for the identification
of the dense gas particles.
{We assume that the dense clouds occupy a mass fraction of
$f_\mathrm{dense}=0.1$
in the dense gas particles (see Section 2.3 of A17 for the definition of $f_\mathrm{dense}$).}
Accretion and coagulation are assumed to occur only in the dense gas particles. 
The time-scales of accretion and coagulation ($\tau_\mathrm{acc}$ and $\tau_\mathrm{co}$)
are evaluated as follows:
\begin{eqnarray} 
\tau_{\rm acc}&=&\begin{cases}
1.2\times 10^{6}{\rm ~yr}\left(\dfrac{Z}{Z_{\odot}} \right)^{-1}\left(1-\dfrac{\mathcal{D}_{\rm tot}}{Z}\right)^{-1}\slash f_{\rm dense}\\
\hspace{3cm}(\mbox{in {\it dense gas particles}})~,\\
\infty~\mbox{(otherwise)}\,,
\end{cases}
\end{eqnarray}
\begin{eqnarray} 
\tau_{\rm co}&=&
\begin{cases}
2.71 \times 10^{5}{\rm ~ yr}\left( \dfrac{\mathcal{D}_{\rm S}}{0.01} \right)^{-1} 
\left( \dfrac{v_{\rm co}}{0.1 {\rm ~km}{\rm ~s}^{-1}} \right)^{-1} \slash f_{\rm dense}\\
\hspace{3cm}(\mbox{in {\it dense gas particles}})~,\\
\infty~({\rm otherwise})~,\label{coagulation1}
\end{cases}
\end{eqnarray}
where $Z$ is the metallicity of the gas particle.
Shattering is assumed to occur only in the diffuse gas whose number density
is smaller than
1 cm$^{-3}$:
\begin{eqnarray} 
\tau_{\rm sh}&=&
\begin{cases}
5.41 \times 10^{7}~{\rm yr}
\left(\dfrac{\mathcal{D}_{\rm L}}{0.01}\right)^{-1}
\left(\dfrac{n_{\rm gas}}{1~\mathrm{cm}^{-3}}\right)^{-1}\\
\hspace{3cm}(n_{\rm gas} < 1\,{\rm cm}^{-3})~,\\
\infty~(n_{\rm gas} \ge 1\,{\rm cm}^{-3} )~.\label{shattering2}
\end{cases}
\end{eqnarray}

\subsection{Sputtering not directly associated with SNe}\label{subsec:sput}

The sputtering terms in equations (\ref{eq:timeL}) and (\ref{eq:timeS})
were not included in our previous model (A17).
Dust grains are destroyed by sputtering in high-temperature ($\gtrsim 10^6$ K) regions
such as X-ray-emitting hot gas in the CGM or IGM
\citep{1995ApJ...448...84T}.
Note that we have already counted the dust destruction in SN shocks. Thus, to avoid
double-counting the destruction, we extract the diffuse hot gas not associated with SNe
by imposing the density threshold for sputtering at
$n_{\rm th}^{\rm sp}=0.01$ cm$^{-3}$, and consider the dust destruction by sputtering
only in regions with $n_\mathrm{gas}<n_\mathrm{th}^\mathrm{sp}$ and $T>10^{6}\,{\rm K}$.
We adopt the following destruction time-scale based on \citet{1995ApJ...448...84T}
\citep[see also][]{1979ApJ...231...77D,2006ApJ...648..435N,2015MNRAS.454.1620H}:
\begin{eqnarray} 
\tau_{\rm sp}(a, n_{\rm H}) 
=
\begin{cases}
2.1\times 10^{5}\left( \dfrac{a}{1\, \mu m} \right)\left( \dfrac{n_{\rm gas}}{1\, {\rm cm}^{-3}} \right)^{-1}\\
 ~~~(\mbox{if}~n_{\rm gas}<n_{\rm th}~\mbox{and}~T_\mathrm{gas} >10^{6}\,{\rm K}),\\
\infty~\mbox{otherwise}.
\end{cases}
\label{eq:sputtering}
\end{eqnarray}
where $n_{\rm H}$ is the hydrogen number density.

\subsection{Time-step for dust treatment}

Some of the time-scales concerning dust evolution {processes} 
could be shorter than the
hydrodynamical time-step adopted in the \textsc{gadget-3} code
\citep{2001NewA....6...79S}.
In this case, we calculate the dust evolution by dividing a single hydrodynamical
time-step into multiple sub-cycles. We
set the sub-cycle time-step $\Delta t_{\rm sub}$ as follows:
\begin{eqnarray} 
\Delta t_{\rm sub} &=&\varepsilon_{\rm sub}  \left[{\rm max}\left( \tau_{\rm hydro}^{-1},~ \tau_{\rm acc}^{-1},~ \tau_{\rm sh}^{-1},~ \tau_{\rm co}^{-1} \right)\right]^{-1}~,
\end{eqnarray}
where $\varepsilon_{\rm sub} $ is a constant which controls the accuracy of the calculation. 
Because we use the fourth-order classical Runge-Kutta method for the dust evolution, 
the error of time integration should be suppressed as
$\propto \varepsilon_{\rm sub}^{4}$. In this paper, we set $\varepsilon_{\rm sub} = 0.1$.

\subsection{Galaxy identification and definition of the IGM}\label{subsec:id}

We analyze the dust associated with galaxies and that contained in the IGM separately.
{In order to} distinguish between these two regions, 
we identify galaxies, and define the intergalactic space 
as the regions not associated with the galaxies.
In order to identify galaxies in a simulation snapshot, we use \textsc{P-Star groupfinder} 
\citep{2001MNRAS.328..726S}.  
In what follows, we give a brief summary of the algorithm following \cite{2004MNRAS.350..385N}.
First, the baryonic (gas + stars) density peaks in the smoothed density field are identified. 
Second, the densities of $N_{\rm ngb}$ nearest neighbor particles around the density peak
are measured, and the peak-density particle is considered 
as a `head particle' if all the neighbor particles have lower densities. 
In this paper, we adopt $N_{\rm ngb}= 128$ and 512 for L50N256 and L50N512, respectively.
Finally, the gas and star particles near the head particle
above a density threshold described by \cite{2004MNRAS.350..385N}
are grouped and identified as a galaxy.
In our definition, we identify the objects whose stellar mass is
greater than $10^{8} M_{\odot}$ as galaxies, since the smaller
structures are affected by our finite mass resolution.
The IGM is defined as the medium not belonging to the galaxies identified above.

\section{Results}\label{sec:result}

First, we test if the dust enrichment is
successfully implemented by comparing our theoretical prediction to the
observed dust abundance at $z=0$, where the relation between dust abundance and
metallicity is well studied.
As mentioned in the Introduction, we focus on the 
dust properties in galaxies, the CGM and the IGM. 
Second, we present the evolution of the cosmic dust abundance. 
In particular, we predict the grain size distribution 
(i.e.\ small/large grain abundance) in a cosmological volume for the first time. 
We also show
the dust mass function of galaxies and
present the radial profile of dust mass around 
massive galaxies in the CGM.
Finally, we show the cosmic extinction (extinction {on cosmological scales})
and compare it with the corresponding observational results.

\subsection{Relation between dust abundance and metallicity}

Because dust production is strongly associated with metal enrichment,
the relation between dust abundance and metallicity gives a strong test
for dust evolution models \citep{1998ApJ...496..145L,1998ApJ...501..643D}.
Some cosmological modeling with dust implementation also uses this relation
for a critical test 
\citep[][]{2017MNRAS.471.3152P, 2018MNRAS.473.4538G}.
We show the galaxy distribution on the $\mathcal{D}_\mathrm{tot}$--$Z$
(total dust-to-gas ratio vs.\ metallicity) diagram
at $z=0$ in Fig.\ \ref{fig:d-z}. 
Each point indicates an individual galaxy and its colour indicates the total stellar mass
($M_\ast$).
We observe {in Fig.\ \ref{fig:d-z}} 
that the metallicity and dust abundance are 
low for $M_{\ast}\lesssim 10^{8.5}\, M_{\odot}$.
{If the metallicity is low, dust growth by accretion is not efficient, so that
the major part of dust is produced by stellar sources.
Thus, the dust-to-gas ratio follows $\mathcal{D}_{\rm tot}\simeq f_{\rm in}Z$
for low-mass galaxies.

\begin{figure}
\includegraphics[width=8.3cm]{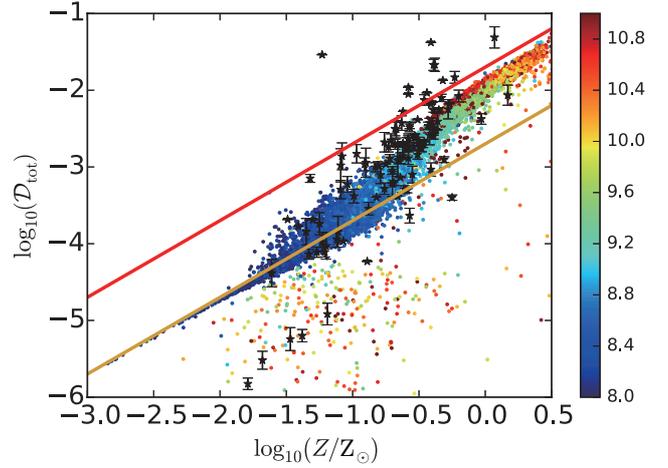}
\caption{Relation 
between dust-to-gas mass ratio $(\mathcal{D}_{\rm tot})$ and metallicity $Z$
of individual galaxies.
The color indicates the logarithmic stellar mass (M$_{\sun}$).
The yellow and red lines represent the linear relation of the stellar yield $(\mathcal{D}_{\rm tot}=f_{\rm in}Z)$ and 
the saturation limit $(\mathcal{D}_{\rm tot}=Z)$, respectively. 
The black stars denote the observational result which was reported by \citet[][]{2014A&A...563A..31R}.
}
\label{fig:d-z}
\end{figure}

In the middle mass range $10^{8.5}\,M_{\odot}\lesssim M_{\ast} \lesssim 10^{10}\,M_{\odot}$, 
$\mathcal{D}_{\rm tot}$ increases steeply at $Z\gtrsim 0.1Z_{\odot}$. 
In {these} galaxies, 
star formation occurs continuously and metal enrichment proceeds.
As a consequence, dust growth by accretion occurs and the dust-to-gas ratio
increases steeply.
{The relation between dust-to-gas ratio and metallicity in this
galaxy mass range is also consistent with}
the observational trend in a nearby
star-forming galaxy sample in \cite{2014A&A...563A..31R}.

At the massive end, $M\gtrsim 10^{10}\,M_{\odot}$, where the metallicity is
high, dust growth by accretion is saturated because of the limit $\mathcal{D}_\mathrm{tot}\leq Z$.
The $\mathcal{D}_\mathrm{tot}$--$Z$ relation of the simulated high-metallicity galaxies
lies within the dispersion of the observational data points. Some observational data
are in the area of $\mathcal{D}_\mathrm{tot}>Z$, which is unphysical. There could
still be a significant uncertainty in the observational dust mass estimate.

In our simulation, there are some outliers located far below the line of
$\mathcal{D}_\mathrm{tot}=f_\mathrm{in}Z$.
In our model, such an extremely low dust-to-metal ratio can only be produced by
SN destruction.
Interestingly, some observational data points also show such an extremely low
dust-to-metal ratio. However, we emphasize that 
those outliers 
{account for a tiny fraction} 
of the entire galaxy population in our model, 
and that most of the galaxies show a clear
correlation between dust-to-gas ratio and metallicity.

\subsection{Cosmic dust abundance}\label{subsec:Omega}

\begin{figure*}
\includegraphics[width=18cm]{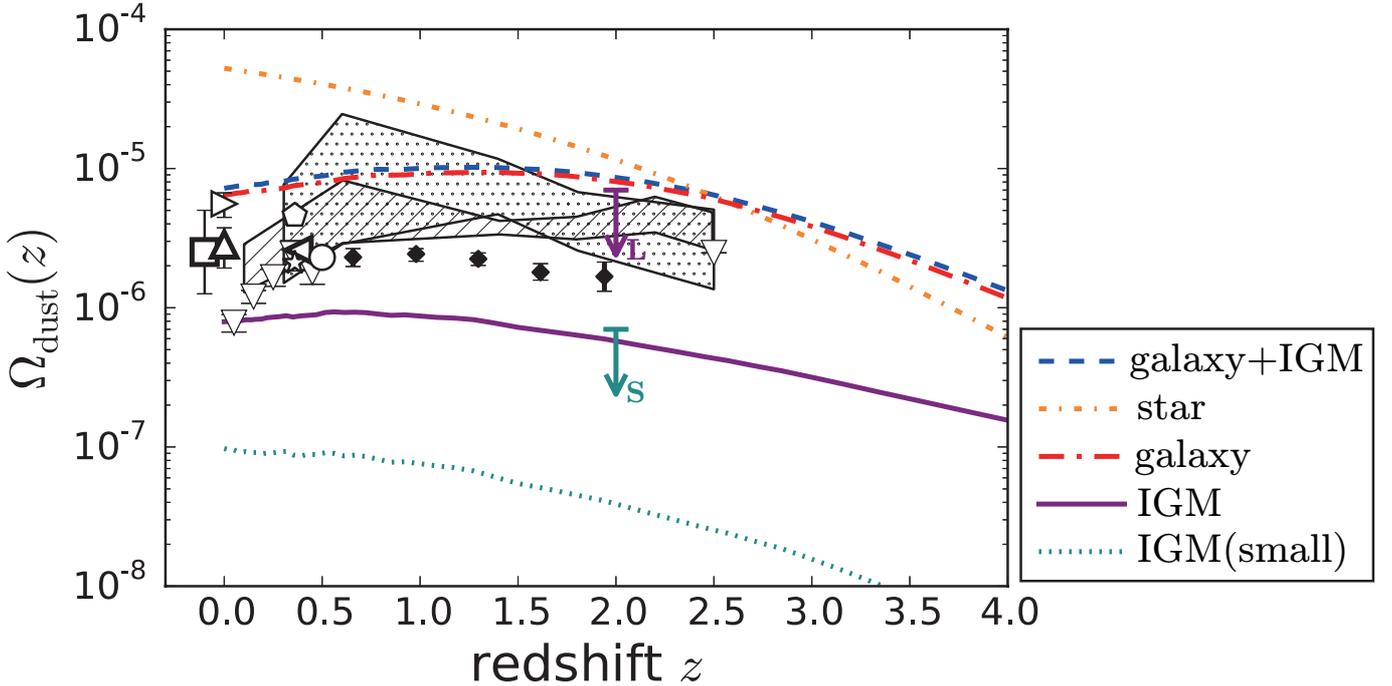}
\caption{
Redshift evolution of cosmic dust abundance $\Omega_{\rm dust}(z)$ for various components. 
The blue dashed line labelled `galaxy+IGM' shows the total amount of dust on gas particles inside simulated galaxies and the IGM. 
The red dot-long-dashed line labelled `galaxy' presents only the amount of dust on gas particles inside simulated galaxies (the ISM and a part of the CGM). 
The purple solid line labelled `IGM' is the amount of dust in the IGM. 
For the IGM component, we also show the small grain abundance with {the mosgreen}  dotted line.
The orange dot-short-dashed line labelled `star' shows the total amount of dust
{absorbed into} stars. 
We compare our results against the following observational results:  
the halo component of dust grains at $0.6\lesssim z \lesssim 2$ ($\blacklozenge$) and $z=0.5$ ($\bigcirc$) from \citet{2012ApJ...754..116M}. 
The open star 
is from \citet{2010MNRAS.405.1025M}. 
For the total abundance of dust grains in the Universe,
$\bigtriangleup, \rhd, \bigtriangledown, \lhd$, $\Box$ and pentagon are from 
\citet{2007MNRAS.379.1022D}, \citet{2013MNRAS.433..695C},
\citet{2011MNRAS.417.1510D}, \citet{2011arXiv1103.4191F},
\citet{2004ApJ...616..643F} and \citet{2010MNRAS.405.1025M}, respectively.
The hatched region and the dot-filled region are from  
\citet{2012ApJ...760...14D} and \citet{2013ApJ...768...58T}, respectively.
Upper limits of large and small grain abundance in the IGM from \citet{2003MNRAS.341L...7I} are shown {with the purple (`L') and {mosgreen} (`S') arrows}, respectively.
}
\label{fig:OmegaDust}
\end{figure*}

We show the evolution of the cosmic dust density normalized to the
critical density of the Universe, i.e.\
$\Omega_{\rm dust}(z)$ in Fig.~\ref{fig:OmegaDust}.
We also separately show the dust abundances in galaxies and the
IGM based on the galaxy identification explained in Section \ref{subsec:id};
that is, we sum up all the dust mass contained in galaxies and subtract it
from the total dust mass to obtain the IGM dust abundance.
To specify each component, we put superscript `L' and `S' for large and small grains,
respectively, no superscript for the total dust abundance,
and subscript `gal' and `IGM' for dust in the galaxies and the IGM, respectively.
For example, $\Omega^{\rm L}_{\rm dust, IGM}$ 
{and $\Omega^{\rm S}_{\rm dust, IGM}$} mean
{the comoving density of large and small grains in IGM, respectively,} and
$\Omega_\mathrm{dust,IGM}=\Omega^\mathrm{S}_\mathrm{dust,IGM}+\Omega^\mathrm{L}_\mathrm{dust,IGM}$.

Galaxies are enriched with dust through stellar dust production
and dust growth by accretion. 
Star formation starts {at} $z\sim 15$ in our simulation. 
Dust in galaxies continuously increases. 
{As galaxies grow through their star formation activity}
(Fig.~\ref{fig:SFRD}b),
they are also enriched with metals and dust.
The increase of metallicity further drives the dust enrichment through
dust growth by accretion in the dense gas.
The cosmic dust abundance continues to
increase down to $z\sim 2$ (Fig.~\ref{fig:OmegaDust}).
Since accretion dominates the abundance of small grains in the metal-rich
environment,
the small grain abundance continue to increase even at $z<2$
(Fig.~\ref{fig:OmegaDust}).
In our simulation, the comoving dust density peaks at $z=1$--2,
which coincides with the most dust enshrouded epoch in the Universe
derived from \textit{Herschel} observations \citep{2013A&A...554A..70B,2018MNRAS.475.2891D}.

{
The cosmic dust density declines slightly at $z\lesssim 1$ because of astration. 
To support this, we also show the {comoving dust density
removed by astration} in Fig.~\ref{fig:OmegaDust}.
At $z \sim 2$,}
more than half of dust grains are consumed by stars (astration).
Interestingly, more than 80 per cent of the dust grains have been
absorbed into stars by $z=0$.  

We also show the dust mass evolution in the IGM in Fig.~\ref{fig:OmegaDust}.
Both metals and dust are produced and spread into the IGM 
as a result of stellar production and feedback. 
The IGM is enriched with dust continuously; throughout all redshifts,
almost all dust grains remain in the host galaxies, but about 10 per cent of the dust
is ejected out of galaxies into the IGM. 

In Fig.~\ref{fig:OmegaDust}, we also present the observational data for
the total dust abundance in various environments, which are obtained 
from integration of galactic dust emission 
and from fluctuation analysis of the cosmic infrared background radiation (CIRB).
The total dust abundance in the simulated galaxies broadly agrees with
the observed data from CIRB
\citep{2013ApJ...768...58T, 2012ApJ...760...14D} and 
the dust abundance in disk galaxies \citep[][]{2007MNRAS.379.1022D}.
However, we tend to overestimate the galactic dust amount compared with
the observational estimates.
This is because our identification of simulated galaxies includes their circum-galactic regions. 
As we show below {in Section \ref{subsec:circum}}, a significant fraction of dust is contained in the CGM in our simulation.
Observationally, \cite{2010MNRAS.405.1025M} and \cite{2015ApJ...813....7P}
found, {based on the analysis of reddening of background QSOs,
that the sum of the dust mass contained in the CGM is comparable to
that existing in the galactic discs}.
The dust abundance estimated from Mg \textsc{ii} absorbers \citep{2012ApJ...754..116M}
also indicates that a significant amount of dust is contained in the CGM. 
They argue that Mg \textsc{ii} absorbers trace the CGM environment based on their
impact parameters.

\cite{2003MNRAS.341L...7I} constrained
the dust abundance in the IGM based on the observed thermal history of
the IGM.\footnote{They define the radii of large grains as
$10^{-2}\,\mu{\rm m}\le a \le 10^{-1}\,\mu{\rm m}$
and those of small grains as $a \sim 10^{-3}\,\mu{\rm m}$.}
They argued that, {if photoelectric heating by dust is significant, }
it heats the IGM too much to be consistent with its observed thermal history.
By calculating the gas heating rate by dust grains and comparing it with
the observed IGM temperatures at $2\lesssim z \lesssim 4$
\citep{2000MNRAS.318..817S}, 
they obtained upper limits of $\Omega^{\rm L}_{\rm dust}<7\times 10^{-6}$ and 
$\Omega^{\rm S}_{\rm dust}<7\times 10^{-7}$ at $z\gtrsim 2$.
Our simulation results are consistent with these upper limits.

A possible source of uncertainty in our galaxy/IGM dust abundance is the finite spatial resolution.
We only identify the structures with $M_\ast >10^8$ M$_\odot$ as
galaxies (Section \ref{subsec:id}). In other words, small `galaxies'
whose stellar masses are less than $10^{8} M_{\odot}$ 
are regarded not as a galaxy but as a part of the IGM.
On the other hand, observations also have a similar problem because of their finite
sensitivity and spatial resolution. Although it is extremely difficult to correct for
{the limited computational and observational} capabilities,
the above rough match between theory and observations indicates
that the dust abundance in the cosmic volume can be broadly understood by the
processes we included in the simulation (mainly stellar dust production and dust growth
by accretion).

In Fig.~\ref{fig:dustLS}, we show the time evolution of small and large grains
for galaxies and the IGM. To clarify the relative abundance, we also show the
small-to-large grain abundance ratio,
$\Omega^{\rm S}_{\rm dust}\slash \Omega^{\rm L}_{\rm dust}$. 
The dust abundance in the Universe is always dominated by large grains.
In galaxies, the redshift dependence of the small-to-large grain abundance ratio is
flat. The large grain formation is dominated by stellar dust
production and coagulation while the small grain formation is governed by
shattering and accretion. If we sum up all galaxies, the statistics is
dominated by small galaxies in which the major part of the dust is produced by
stellar sources.
The abundance of small grain is significant only in massive
($M_\ast\gtrsim 10^{10}\,{\rm M}_{\odot}$) galaxies
where shattering and accretion
{are efficient because of their high metallicity}
(Hou et al., in preparation).

In contrast, the small-to-large grain abundance ratio in the IGM
monotonically increases from high redshift down to $z\sim 0$.
The dust abundance in the IGM is fully dominated by large grains at high redshift because
dust grains are ejected into the IGM before being processed by shattering
and accretion, as discussed in \citet{2017MNRAS.469..870H}.
The increase of the small-to-large grain abundance ratio in the IGM 
{is due to the supply of small grains formed by
shattering and accretion. Massive galaxies are assembled at
low redshift {$(z \lesssim 2)$}, and the ISM in massive galaxies
contains a large amount of small grains. As a consequence, more small grains are
supplied at lower redshift.
\textit{In situ} small-grain formation in the IGM by shattering may also be possible.
However, we find that this path of small-grain formation is
negligible because the
grain--grain collision time-scale is longer than the cosmic age in the IGM.}
Indeed, we do not see any enhancement of the small grain abundance relative
to the large grain abundance in the CGM as we see in Section \ref{subsec:circum}
(Fig.\ \ref{fig:profile}).
As shown later in this section, sputtering decreases both small and large grain
abundances almost equally. Thus, the processing in the IGM is not important in determining the
grain size distribution in the IGM.

There is no observational data to compare for the grain size distribution in the
IGM, while there are some observational clues in the CGM.
Thus, the dust properties in the CGM may provide some insight into the IGM dust.
As mentioned above, the dust properties in the CGM could be traced by
Mg \textsc{ii} absorbers as argued by \citet{2012ApJ...754..116M},
who analyzed the background quasar (QSO: quasi-stellar object) data
taken by the Sloan Digital Sky Survey
\citep[SDSS;][]{2000AJ....120.1579Y}
in combination with
{\it Galaxy Evolution Explorer} \citep[\textit{GALEX};][]{2005ApJ...619L...1M} data.
The reddening curves of Mg \textsc{ii} absorbers are fitted well with the Small
Magellanic Cloud (SMC) extinction curve,
which indicates that dust grains smaller than 0.03~$\micron$ {are} abundant
\citep[][]{1992ApJ...395..130P,2001ApJ...548..296W}.
However,
our result shows that large grains dominate the IGM dust abundance (Fig.~\ref{fig:OmegaDust}).
Thus, there is a tension between our simulation results and the observed reddening curves
for Mg \textsc{ii} absorbers. We discuss this further in Sections~\ref{subsec:circum}
and \ref{subsec:ext}.

\begin{figure}
\includegraphics[width=8.3cm]{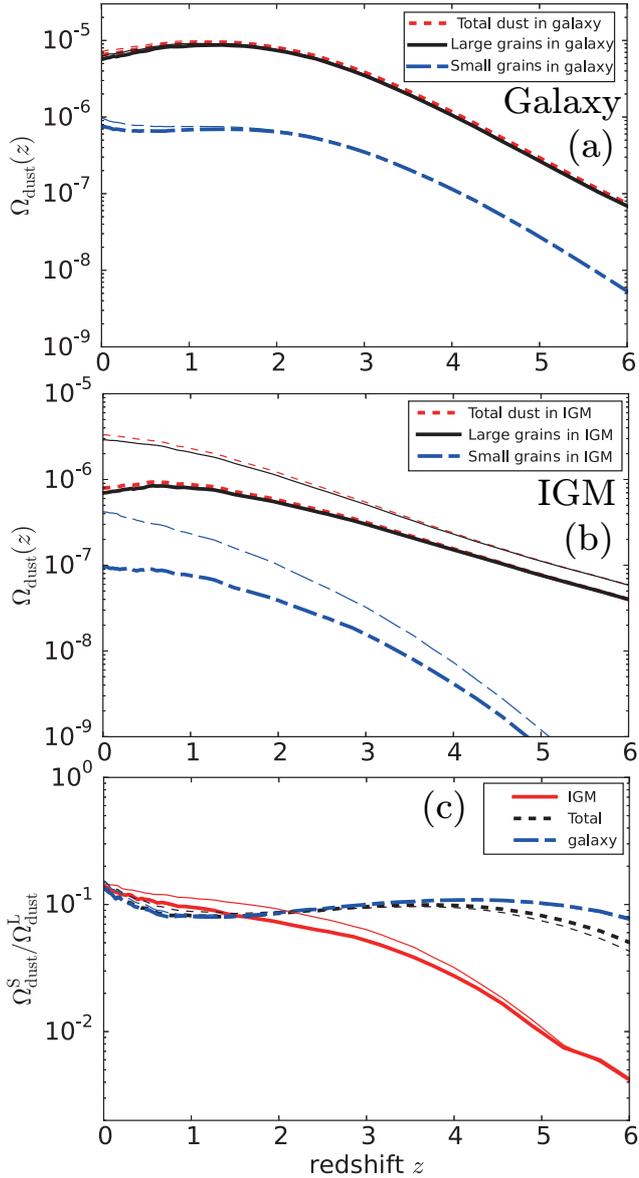}
\caption{Panels ({\it a}) and ({\it b}): 
time evolution of cosmic dust abundance of large (thick solid line) and 
small grains (thick long-dashed-short-dashed line)
in galaxies and in the IGM, respectively. 
The sum of large and small grains (`Total') is shown 
by {the} red thick dashed line in each panel. 
Panel ({\it c}): small-to-large grain abundance ratio for galaxies
(blue long-dashed-short-dashed line), the IGM (red solid line), 
and the total (black dotted line).
Thin lines in each panel represent the time evolution of each component 
without sputtering (i.e.\ $\tau_{\rm sp}(a)\to +\infty$).}
\label{fig:dustLS}
\end{figure}

We show the significance of sputtering in the hot gas (not associated with SNe)
on the time evolution of the
cosmic dust abundance in Fig.~\ref{fig:dustLS}.
{As explained in Section \ref{subsec:sput}, we
select the hot gas not associated with SNe (mainly associated with
the CGM and IGM) with {the} criterion
$n_\mathrm{gas}<0.01$ cm$^{-3}$ and $T_\mathrm{gas}>10^6$ K.}
In Fig.~\ref{fig:dustLS}, we indeed confirm that sputtering in the diffuse hot
gas affects little the dust abundance in galaxies.
In the IGM, in contrast,
70--80 per cent of the IGM dust could be destroyed by sputtering. 
Although small grains are more sensitive to sputtering than large grains,
the ratio $\Omega^\mathrm{S}_{\rm dust} / \Omega^\mathrm{L}_{\rm dust}$ is not
significantly affected by sputtering, as we observe in Fig.~\ref{fig:dustLS}c.
This is because the sputtering time-scales for both large and small grains are much shorter
than the hydrodynamical time-scale, and almost all dust grains which suffer from sputtering 
are destroyed on the spot regardless of the grain size.

\subsection{Dust mass function}

\begin{figure}
\includegraphics[width=8.3cm]{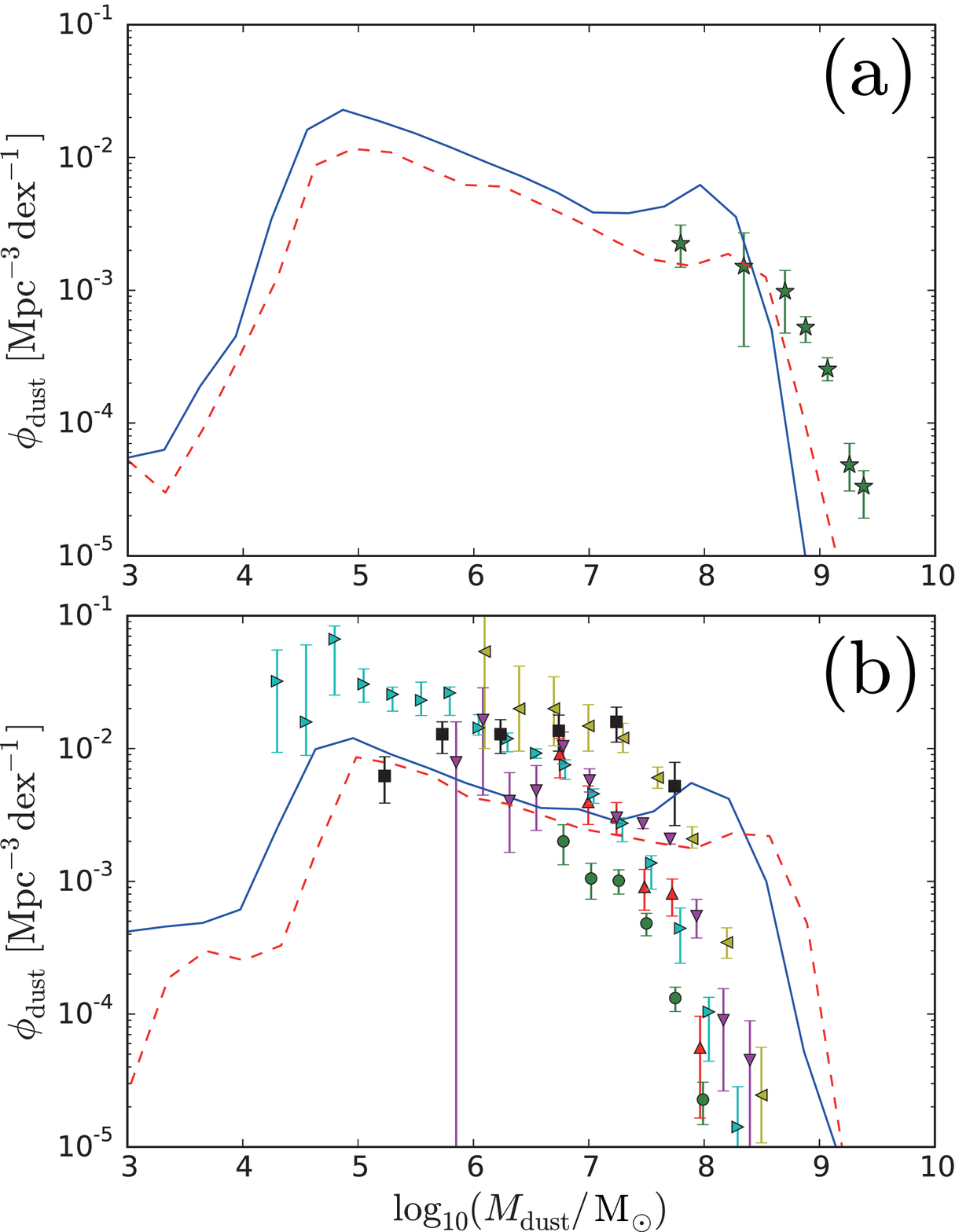}
\caption{Dust mass function of simulated galaxies at $z=2.4$ (panel ({\it a})) 
and $z=0$ (panel ({\it b})) compared with observational results.
The solid and dashed lines represent the result of L50N512 and L50N256, respectively.
The symbols $\star, \bigcirc, \bigtriangledown, \lhd, \Box$ and $\bigtriangleup$ show
the observational data taken from
\citet{2003MNRAS.341..589D,2000MNRAS.315..115D}
\citet{2011MNRAS.417.1510D,2013MNRAS.433..695C} 
\citet{2015MNRAS.452..397C} and \citet{2005MNRAS.364.1253V}, respectively, while
$\rhd$ is obtained from the {\it IRAS} PSCz-extrapolated mass function 
shown in \citet{2005MNRAS.364.1253V}.
}
\label{fig:dustMF}
\end{figure}

The dust abundance in galaxies can also be expressed in the form of the dust mass function,
which is the distribution function of dust mass in galaxies,
as shown in Fig.~\ref{fig:dustMF}.
{We compare the dust mass function of the simulated galaxies with
observed dust mass functions. We have to keep in mind that observational estimates of
dust mass depend on the adopted dust mass absorption coefficient
(dust emissivity per dust mass). 
The observationally derived dust mass is simply proportional to
$\kappa_{850}^{-1}$ ($\kappa_{850}$ is the dust mass absorption coefficient
at a wavelength of 850 $\micron$), while the dust mass in our simulation is not affected by
$\kappa_{850}$.}
We adopt a uniform dust mass absorption coefficient
$\kappa_{\rm 850}=0.77\,{\rm cm}^{2}{\rm g}^{-1}$ at 850 $\micron$ with
a wavelength dependence of $\propto\lambda^{-2}$
according to \cite{2000MNRAS.315..115D}.
This value of $\kappa_{850}$
is based on \cite{2000MNRAS.315..115D}, and is intermediate between
the values for graphite and silicates as given by \cite{1984ApJ...285...89D} and
\cite{1993MNRAS.263..607H}.
\cite{2003ARA&A..41..241D} claimed $\kappa_{850}=0.383\,{\rm cm}^{2}{\rm g}^{-1}$, 
which is approximately a half of the above value.
Thus, it is important to note that there is a factor 2--3 uncertainty in the dust mass
derived from the observed IR emission.

Our simulation produces a larger number of galaxies with high dust mass
($M_\mathrm{d}\gtrsim 10^8$\,M$_{\sun}$) at $z=0$ than at $z=2.4$, 
while the observational data indicate the opposite. 
As a result, although we reproduce the dust mass function at $z=2.4$ well
(if we consider a factor 2--3 uncertainty in the observational dust mass estimates), 
we overpredict the number of high-$M_\mathrm{d}$ galaxies at $z=0$.
Recalling that our simulation also overproduces the massive end in the galaxy stellar mass function
(Fig.~\ref{fig:SFRD}b), we argue that the overproduction of dusty galaxies is linked
to that of massive galaxies. 
As discussed above, a possible reason is that we did not include AGN feedback. 
We expect that AGN feedback suppresses the dust mass in two ways: 
(i) loss of dust by outflow, and/or (ii) suppression of star formation and subsequent chemical enrichment. 

In Fig.~\ref{fig:dustMF},
the cut-off at the low-$M_\mathrm{d}$ end ($\sim 10^4$ M$_{\odot}$) is
determined by the spatial resolution of the simulation.
Our simulation tends to underproduce the dust mass function at
$M_\mathrm{d}\sim 10^5$--$10^7$\,M$_{\sun}$, although it is marginally in the range of 
scatter of {the} observational data. 
{The slope at $M_\mathrm{d}\sim 10^5$--$10^8$ M$_{\odot} $ is similar to the
observed dust mass function.
We should recall again that, on the observational side, there is a factor 2--3
uncertainty in the dust mass.}
Therefore we just conclude that our simulation
{qualitatively accounts for the observational dust mass function.}

For comparison, we also show the dust mass function for the lower resolution run, L50N256.
Although the mass resolution of L50N256 is eight times worse than
that of L50N512, the amplitude and the cut-off of mass function at the massive end
roughly agree with each other. 
This means that the mass resolution does not affect the statistical properties of dust mass in galaxies
very much. 
As expected, the number of low-mass galaxies with $M_\mathrm{d}\lesssim 10^{5}\,M_{\odot}$ 
is higher for L50N512 than for L50N256 because of the higher mass resolution.

\subsection{Circum-galactic dust around massive galaxies}\label{subsec:circum}

\cite{2010MNRAS.405.1025M} detected a large abundance of dust
in the CGM by {analyzing the} correlation 
between the reddening of background QSOs
and a large number of galaxies in the SDSS sample whose median redshift is $z\sim 0.3$.
\cite{2015ApJ...813....7P} found a similar radial profile of CGM dust
at $z\sim 0.05$. These studies showed that the
dust mass in galaxy halos is comparable to that in galactic discs. In order to examine
if our simulation also reproduces such a large dust abundance in galaxy halos (or in the
CGM), we compare the radial profile of the surface density of dust
up to $\sim 1$ Mpc from the galaxy centre.

We select 1617 simulated galaxies in the similar stellar mass range 
as the sample of \citet{2010MNRAS.405.1025M}\footnote{\citet{2010MNRAS.405.1025M}
sampled galaxies with luminosity $L\simeq 0.45L^{\ast}$ at $z=0.3$,
where $L^{\ast}$ is the characteristic luminosity of the
luminosity function at that redshift. \cite{2014ApJ...783...85T} reported the
characteristic mass of the stellar mass function to be $\simeq 10^{11.05}$\,M$_{\odot}$.
If we assume $L\propto M_*$, a galaxy with $0.45L^*$ would have 
$M_* \simeq 5\times 10^{10}$ M$_\odot$. 
Thus, we assume that \cite{2010MNRAS.405.1025M}'s sample has
a stellar mass range of $10^{10}<M_*/{\rm M}_{\odot}<10^{11}$.},
and computed the average radial profile of dust surface densities
{in the following way}. 
We first extract the $(2\, {\rm Mpc})^{3}$ region
around each simulated galaxy at $z=0$, and 
divide each region into a $200^{3}$ cubic grid.
Thus, the effective resolution of the radial profile is $\sim 10$\,kpc.
By calculating the averaged dust mass density at each grid point for small and large grains,
we obtain the distribution of dust grains around the galaxies.
Then, the 3-dimensional dust distribution is projected onto a 2-dimensional surface to obtain
the radial profile of surface density as plotted in Fig.~\ref{fig:profile} (the projected radius is denoted as $r$).

\begin{figure}
\includegraphics[width=8.3cm]{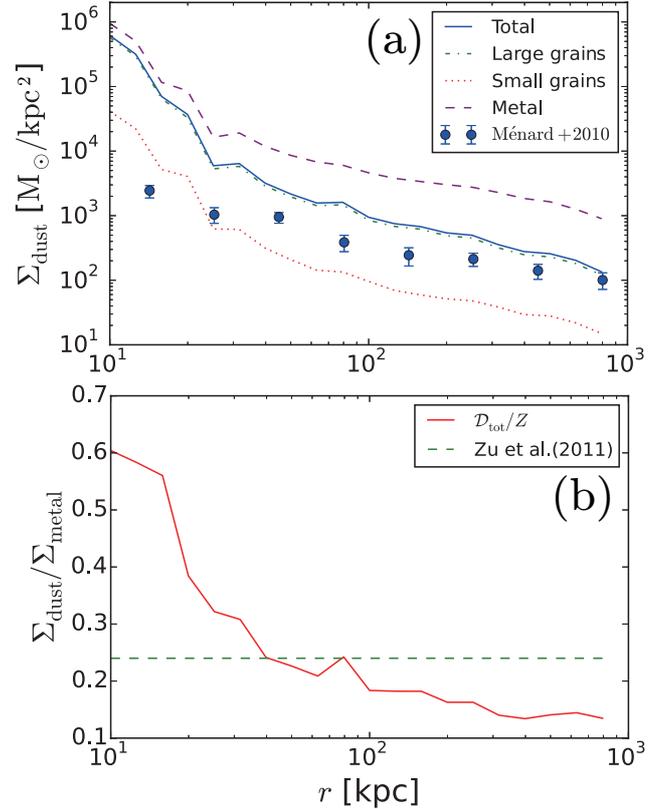}\caption{
Panel (a): Averaged surface density profile of metals and dust around massive galaxies
($r$ is the distance from the galaxy centre). 
We sampled all galaxies whose stellar mass is in the range of
$10^{10}M_{\odot}\le M_{\ast} \le 10^{11}M_{\odot}$. The number of galaxies is 1617.
The surface densities of metals and dust are presented with
the dashed (purple) and solid (blue) lines, respectively.
The blue and red dotted lines show the surface densities of large and small grains,
respectively.
Points with error bars are the observed radial profile of dust \citep{2010MNRAS.405.1025M}.
Panel (b): Radial profile of dust-to-metal surface density ratio. The dashed line shows the
constant dust-to-metal ratio adopted by \citet{2011MNRAS.412.1059Z}.
}
\label{fig:profile}
\end{figure}

In Fig.~\ref{fig:profile}a, we compare our simulation result with
{the observations}.
We find that our total dust mass (the sum of small and large grains) can account for
the observed dust abundance at $r\gtrsim 30$\,kpc.
We find that large grains dominate the total dust abundance in the entire circum-galactic region. 
The small-to-large grain mass ratio is $\sim$1/16 at $r=10$\,kpc and $\sim$1/10 at $r=1$\,Mpc. 
The dominance of large grains is also observed in the IGM as discussed in
Section~\ref{subsec:Omega}.
The slightly lower small-to-large grain abundance ratio at the small radius
is due to the continuous supply of large grains by {coagulation}.

We also plot the dust-to-metal surface density ratio, $\Sigma_{\rm dust}\slash \Sigma_{\rm metal}$, 
in Fig.~\ref{fig:profile}b. The ratio is high ($\simeq$\,0.6) in the central region
($r<20\,{\rm kpc}$), while it
dramatically drops to $\sim$0.25 at $r\simeq 20$\,kpc, and
slowly decreases at $r>20\,{\rm kpc}$
because dust grains are destroyed by sputtering in hot ($T_\mathrm{gas}>10^6$ K) gas.
At $r\gtrsim 200\, {\rm kpc}$, 
it drops to $\Sigma_{\rm dust}\slash \Sigma_{\rm metal}\simeq$\,0.14.
{A part of the dust still survives because not all the gas in the CGM
is hot.}

We compare the value of $\Sigma_{\rm dust}\slash \Sigma_{\rm metal}$
{obtained in our simulation with that derived}
by \cite{2011MNRAS.412.1059Z}.
{They} only calculated the spatial distribution 
of metals in the IGM and
assumed a fixed dust-to-metal ratio {(i.e.\ they did not
calculate dust evolution)} to obtain the dust distribution.
They showed that $\Sigma_{\rm dust}\slash \Sigma_{\rm metal} \simeq 0.24$ 
explains the dust abundance at $r\sim 1h^{-1}$\,Mpc obtained by
\cite{2010MNRAS.405.1025M}.
They only normalized the amplitude of extinction at $1 h^{-1}$Mpc, 
but their prediction agreed with the result of \cite{2010MNRAS.405.1025M} 
in a wide radius range of $0.01\,h^{-1}{\rm Mpc}\lesssim r \lesssim 8\,h^{-1}{\rm Mpc}$.
{Interestingly,
our model that incorporated dust evolution in the simulation}
also predicts $\Sigma_{\rm dust}\slash \Sigma_{\rm metal} \sim 0.25$ 
at $0.02\,h^{-1}{\rm Mpc} \lesssim r \lesssim 0.15\,h^{-1}{\rm Mpc}$
without any tuning of dust-to-metal ratio. 

\subsection{Extinction of circum-galactic and intergalactic dust}\label{subsec:ext}

The extinction over a cosmic distance (hereafter, referred to as
the cosmic extinction) also puts a constraint on the dust properties
in the cosmic volume.
We calculate the cosmic extinction up to redshift $z$, $A_{\lambda}(z)$, in
units of magnitude as 
\citep[][]{2010MNRAS.405.1025M, 2018arXiv180400848H}
\begin{eqnarray} 
A_{\lambda}(z) \!\!\!\!&=& \!\!\!\!2.5 (\log_{10}\mathrm{e})\notag\\
& \times & \!\!\!\!\displaystyle\int_{0}^{z}
\left(\kappa^{\rm S}(\lambda^{\prime })\rho_{\rm dust}^{\rm S}(z^{\prime })
    + \kappa^{\rm L}(\lambda^{\prime })\rho_{\rm dust}^{\rm L}(z^{\prime })
\right)
\dfrac{c(1+z^{\prime })^{2}}{H(z^{\prime })}dz^{\prime }~,\notag \\
\lambda &=& \lambda^{\prime }~(1+z^{\prime })\,,\notag\\
H(z)&=&H_{0}\sqrt{\Omega_{\rm m}(1+z)^{3}+\Omega_{\Lambda}\,}\, ,
\label{eq.12}
\end{eqnarray}
where $\kappa^{\rm S}(\lambda )$ and $\kappa^\mathrm{L}(\lambda )$
are the dust mass extinction coefficient for small and large grains,
respectively, as a function of wavelength,
$\rho_\mathrm{dust}^\mathrm{S}(z)$ and $\rho_\mathrm{dust}^\mathrm{L}(z)$
are the comoving mass density of small and large grains as a function of
redshift, respectively,  $c$ is the light speed,
and $H(z)$ is the Hubble parameter at $z$. 
For the expression of $H(z)$, we assumed a flat Universe.
We adopt the mass extinction coefficients
from \citet{2017MNRAS.469..870H}, who assumed
spherical and homogeneous dust grains 
with a mixture
of silicate and graphite 
\citep[][]{1984ApJ...285...89D}
and applied the Mie theory {\citep[][]{1983asls.book.....B} 
based on the same optical constants in \citet[][]{2001ApJ...548..296W}.}
In this paper, the mass fractions of silicate and
carbonaceous dust are assumed to be 0.54 and 0.46, respectively \citep[][]{2009MNRAS.394.1061H}.
Although these mass fractions give a good start (since we do not have
any knowledge on the grain composition in the IGM), they should vary
depending on the evolutionary stage of galaxy
\citep[e.g.][]{2011A&A...525A..61P, 2015ApJ...810...39B}.
The production rate of each element also changes as a function of galaxy age
\citep[e.g.][]{2014MNRAS.438.2765C}.
However, our conclusions regarding the cosmic extinction
are not affected by the detailed mixture of the
two dust species within the uncertainties in the observational constraint.
An important conclusion drawn later about a flat extinction curve shape also
holds regardless of the mixture ratio.
We consider the cosmic extinction at $z>0.3$, where some observational constraints are available.
Because the comoving distance at $z=0.3$ is 1.23 Gpc, which is much larger than 
the comoving scale of {the} large scale structure 
$\sim 100h^{-1}$\,Mpc
\citep[e.g.][]{2005ApJ...633..560E}, we are able to justify the usage of the mean density without considering the
spatial inhomogeneity. Moreover, the observational constraints we adopt are derived
from the analysis of a large number of objects in
{a large cosmic volume}; thus, only
averaged quantities are relevant for the analysis in this subsection.


We adopt the $V$-band wavelength (0.55 $\micron$) at the observer's frame for $\lambda$,
and show $A_V(z)$ in Fig.\ \ref{fig:extinction}.
For the dust mass densities, $\rho_\mathrm{dust}^\mathrm{S}$ and
$\rho_\mathrm{dust}^\mathrm{L}$, we examine two cases: in the first case,
we adopt all the dust in the simulation box, while in the second case, we count
only the dust in the IGM.
{The latter case is motivated by the fact that
the observations of background QSOs used to derive the cosmic extinction
are biased against lines of sight
passing though galaxies where the extinction is high.
That is, lines of sight without galaxies are preferentially
sampled in the measurement of the cosmic extinction
\citep[][]{2005A&A...444..461V}.}
Thus, we expect that observational results should lie between those two cases.

\begin{figure}
\includegraphics[width=8.3cm]{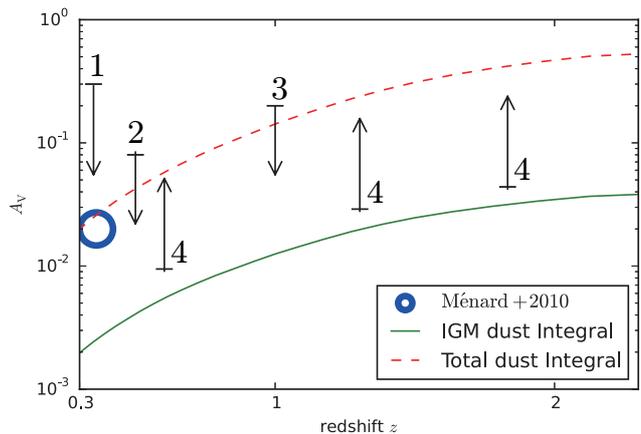}
\caption{Cosmic dust extinction in the $V$ band. 
The solid (green) and dashed (red) lines show the extinction by the IGM component
and all dust grains in the simulation box, respectively.
The numbers `1', `2', `3' and `4' are observational constraints obtained by 
\citet{2009ApJ...696.1727M}, \citet{2009JCAP...06..012A},
\citet{2003JCAP...09..009M} and \citet{2008MNRAS.385.1053M}. 
The circle indicates the estimate based on the statistics of halo extinction
by \citet{2010MNRAS.405.1025M}.
}
\label{fig:extinction}
\end{figure}
\begin{figure}
\includegraphics[width=8.3cm]{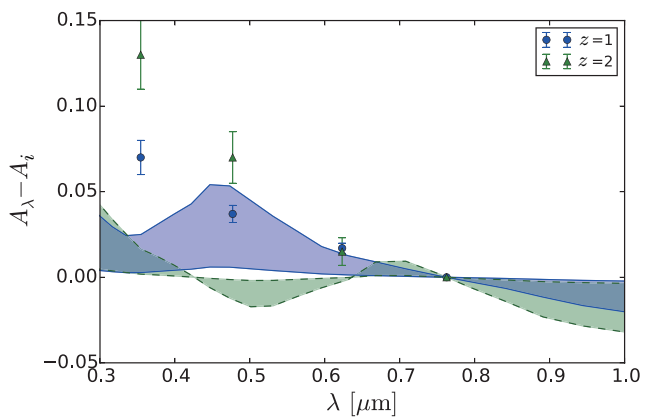}
\caption{Estimated reddening curves of Mg\,\textsc{ii} absorbers. 
The solid (blue) and dashed (green) lines represent the curves at $z=1$ and 2,
respectively. The shaded area associated with these lines show the
range of the column densities assumed in the calculation.
The filled circles and triangles with error bars
are observational data taken from \citet{2012ApJ...754..116M}.}
\label{fig:extinction8}
\end{figure}

Cosmic dust extinction has been investigated 
{in} various ways. 
\cite{2009ApJ...696.1727M} (marked by `1' in Fig.\ \ref{fig:extinction}) focused 
on the effect of dust extinction on the apparent relation between the luminosity
distance and the angular diameter distance {of} 
distant galaxies and obtained {an} 
upper limit {on} 
the cosmic dust extinction  {at} $z\sim 0.35$.
\cite{2009JCAP...06..012A} (marked by `2' in Fig.\ \ref{fig:extinction})
put an upper limit on dust extinction by using the fact that
dust grains decrease the apparent luminosity of SNe Ia and could affect the cosmological parameter
estimates.
\cite{2003JCAP...09..009M} (marked by `3' in Fig.\ \ref{fig:extinction}) made an attempt of finding 
systematic reddening for the SDSS QSO sample. 
They did not find such a systematic reddening and they set an upper limit
of $A_{\rm V}<0.20$ at $z=1$. 
\cite{2008MNRAS.385.1053M} (marked by `4' in Fig.\ \ref{fig:extinction})
observed the reddening of a statistical sample of Mg\,\textsc{ii} absorbers.
Since they only counted Mg\,\textsc{ii} absorbers, the estimated extinction is
regarded as a lower limit.
\citet{2010MNRAS.405.1025M} estimated the cosmic extinction
{at $z\sim 0.3$ on the assumption that it is
dominated by dust in galaxy halos.} 

As mentioned above, we expect that the actually
observed cosmic extinction would
lie between the cosmic extinction arising from all the dust
in the cosmic volume and that contributed from only the IGM dust
(shown by the dashed and solid lines in Fig.~\ref{fig:extinction},
respectively).
In Fig.\ \ref{fig:extinction}, we
indeed find that the observational
constraints are broadly located between those two lines.
This indicates that our dust abundance in the cosmic volume agrees 
with the observational constraints on the cosmic extinction.

The wavelength dependence of extinction could provide useful information on
the grain size \citep[e.g.][]{1977ApJ...217..425M}. Therefore, we further calculate
the extinction (or reddening) curve, which could be compared with observations.
Following \cite{2012ApJ...754..116M}, we assume that Mg \textsc{ii} absorbers
trace the medium in galaxy halos. \cite{2015ApJ...813....7P} have shown that
the reddening curves in galaxy halos are indeed similar to those in Mg \textsc{ii} absorbers.
Because the column density is important for the reddening,
using Mg \textsc{ii} absorbers is advantageous because of their known column densities.

The extinction at wavelength $\lambda$ is estimated by the following formula
\citep[][]{2018arXiv180400848H}:
\begin{eqnarray} 
A_{\lambda}=2.5(\log_{10}\mathrm{e})\mu m_{\rm H}N_{\rm H}^\mathrm{MgII}
\left(\kappa^{\rm L}(\lambda)\mathcal{D}_{\rm L}^{\rm MgII}
+\kappa^{\rm S}(\lambda)\mathcal{D}_{\rm S}^{\rm MgII}\right)\, ,
\label{eq:A_lambda}
\end{eqnarray}
where $\mu$ is the gas mass per hydrogen ($\mu =1.4$), $m_{\rm H}$
is the mass of hydrogen atom, $N_{\rm H}^\mathrm{MgII}$ is the hydrogen column density
of an Mg \textsc{ii} absorber, and
$\mathcal{D}_{\rm L}^{\rm MgII}$ and $\mathcal{D}_{\rm S}^{\rm MgII}$
are the large and small grain abundances (dust-to-gas ratios) in Mg \textsc{ii} absorbers.
We adopt $N_{\rm H}=10^{19.5}\,{\rm cm}^{-2}$, 
which is derived as a geometric mean for an Mg \textsc{ii}
absorber sample by \cite{2009MNRAS.393..808M}.
According to \cite{2009MNRAS.393..808M},
the dust-to-gas ratio is 60--80 per cent of the Milky Way value if we use
$A_V /N_\mathrm{H}^\mathrm{MgII}$ for the indicator of dust-to-gas ratio.
Assuming the typical dust-to-gas ratio of the Milky Way to be 0.01 (or slightly less)
\citep[e.g.][]{1992ApJ...395..130P}, we adopt the total dust-to-gas ratio for
Mg \textsc{ii} absorbers as $\mathcal{D}_\mathrm{tot}^\mathrm{MgII}=0.006$.
We assume that the small-to-large grain abundance ratio in Mg \textsc{ii} absorbers
is equal to that in the IGM:
\begin{eqnarray} 
\mathcal{D}_{\rm L}^{\rm MgII}&=&\dfrac{\Omega^{\rm L}_{\rm dust,IGM}}{\Omega_{\rm dust,IGM}}\mathcal{D}_{\rm tot}^{\rm MgII},\\
\mathcal{D}_{\rm S}^{\rm MgII}&=&\dfrac{\Omega^{\rm S}_{\rm dust,IGM}}{\Omega_{\rm dust,IGM}}\mathcal{D}_{\rm tot}^{\rm MgII}.
\end{eqnarray}
where $\Omega^{\rm L}_{\rm IGM\, dust}$, $\Omega^{\rm S}_{\rm IGM\, dust}$,
and $\Omega_{\rm IGM\, dust}$ are introduced in Section \ref{subsec:Omega}.
Considering the uncertainties in the column density and the dust-to-gas ratio,
we examine an order-of-magnitude range of
$N_\mathrm{H}^\mathrm{MgII}=10^{19}$--$10^{20}$ cm$^{-2}$ (while
we fix $\mathcal{D}_\mathrm{tot}^\mathrm{MgII}=0.006$ because of the degeneracy
between $N_\mathrm{H}$ and $\mathcal{D}_\mathrm{tot}^\mathrm{MgII}$).

For the reddening, we plot the difference from the extinction in the
$i$-band, $A_\lambda -A_\mathrm{i}$ (note that only such a `reddening' is measurable
in the observation) in Fig.\ \ref{fig:extinction8}. Here, we show the wavelength
in the observer's frame
[thus, apply $\lambda /(1+z)$ for $\lambda$ in equation (\ref{eq:A_lambda})].
The wavelength dependence of reddening is referred to as the reddening curve.
We plot the reddening curves of Mg \textsc{ii} absorbers at $z=1$ and 2 in Fig.\ \ref{fig:extinction}.
Our model predicts reddening curves marginally consistent with the observational data
at $z=1$ except at the shortest wavelength. There is a clear discrepancy between
the reddening curve in our simulation and the observational data at $z=2$.
The reason why the estimated extinction curve is very flat is that the dust abundance
in the IGM is dominated by
large grains in our simulation as shown in Section \ref{subsec:Omega}.
Therefore, there is a tension between the simulation and the observation
in terms of the grain size distribution at $z=2$.
On the other hand, the abundance of small grains at $z=1$ becomes greater than that at $z=2$;
accordingly, the estimated extinction curve at $z=1$ becomes relatively steep and 
able to explain a couple of observational points.
We further discuss the discrepancy between our results and the observation
data in Section \ref{subsec:issues_IGM}.

\section{Discussion}

\subsection{Comparison with other theoretical studies on CGM/IGM dust}\label{subsec:trans_halo}

The spatial distribution of dust in the CGM and IGM has been predicted by
cosmological hydrodynamical simulations
\citep[e.g.][]{2016MNRAS.457.3775M, 2017MNRAS.468.1505M} and 
semi-analytic models \citep[e.g.][]{2017MNRAS.471.3152P}. 
The radial profile of dust surface density in \cite{2017MNRAS.468.1505M} is
steeper than the observational data by
\cite{2010MNRAS.405.1025M} in the range $r<100~{\rm kpc}$ 
and flatter at $r>100~{\rm kpc}$.
\cite{2017MNRAS.468.1505M} underproduced the dust abundances 
relative to those reported by \cite{2010MNRAS.405.1025M}.
We predict a mild slope for the 
radial profile compared to that in \cite{2017MNRAS.468.1505M} except in the central region.
As a consequence, our result widely agrees with the
observed profile by \cite{2010MNRAS.405.1025M}.
\cite{2017MNRAS.471.3152P} estimated the dust abundance in hot halos 
as a function of halo mass, and predicted a comparable halo dust mass to
that observed in \cite{2015ApJ...813....7P} for galaxies with $M_\ast\sim 10^{10}$ M$_\odot$.

The spatial dust distribution in the CGM and IGM highly depends on the feedback model.
The distribution of dust grains around massive galaxies was also simulated by
\cite{2011MNRAS.412.1059Z} based on their metallicity distribution under a fixed
dust-to-metal ratio.
They adopted a momentum-driven wind model,
which was claimed originally {by} \cite{2005ApJ...621..227M} and
\citet{2005ApJ...618..569M} as
a physically reasonable feedback model. 
\citet{2011MNRAS.412.1059Z} tuned the dust-to-metal ratio and 
varied the feedback strength. 
In our work, we roughly reproduced the observed radial profile 
without fixing the dust-to-metal ratio:
in the central region ($r \lesssim 20$\,kpc), dust growth by
accretion is active and the dust-to-metal ratio is high ($\simeq 0.6$), 
whereas it drops to $\sim 0.3$ at $r\sim 20$\,kpc
and gradually decreases to
$\sim 0.14$ at $r\sim 1$\,Mpc. This decrease is due to the dust destruction
in the hot gas.
Therefore, {our model suggests that there is a significant
variation in the dust-to-metal ratio of the CGM/IGM. Importantly, there
is a systematic decrease in the dust-to-metal ratio along the distance from
the galaxy centre.
It would be interesting to reexamine \citet{2011MNRAS.412.1059Z}'s
analysis by taking the variation in the dust-to-metal ratio into account.}

\subsection{Issues for circum-galactic and intergalactic extinctions}\label{subsec:issues_IGM}

As mentioned in Section \ref{subsec:circum},
small grains are deficient in the IGM and CGM in our simulation.
This leads to a flat reddening curve, which does not fit to the actually
observed curve for Mg \textsc{ii} absorbers.
A possible reason for this discrepancy is
that our simulation failed to fully include the physical processes important
for dust in the CGM and IGM. For example, we did not include radiation pressure
that could be important in transporting dust out of the galaxies
\citep[][]{1991ApJ...381..137F,2005MNRAS.358..379B}.
However, there is no physical reason that radiation pressure preferentially
transport small grains; thus, radiation pressure will not make the
reddening curve in the IGM and CGM steeper.
Another possibility is that
{our simulation failed to treat the hydrodynamical
effects (especially, shocks and turbulence) in galaxy outflows
because of low spatial resolution.}
Some authors argue that Mg \textsc{ii} absorbers originate from outflows
driven by active star formation \citep[][]{1996ApJ...472...73N,2001ApJ...562..641B}.
In such a high-velocity environment, 
shocks or high-velocity turbulence could be induced.
Both shocks \citep[][]{1996ApJ...469..740J} and turbulence
\citep[][]{2004ApJ...616..895Y,2009MNRAS.394.1061H} can cause grain shattering.
Since our simulation is not capable of resolving
{shocks and turbulence},
{this production path of small grain by shattering
could not be successfully treated.
}
It would be interesting to investigate the possibility of shattering in outflows
using higher resolution zoom-in simulations.

The lack of small grains leads to a significantly flatter reddening curve
than the observed one by \cite{2012ApJ...754..116M} as shown in Fig.\ \ref{fig:extinction8}.
The assumption on the reddening or extinction curve is crucial in converting the
observed reddening to the dust abundance.
\cite{2010MNRAS.405.1025M} and \citet{2012ApJ...754..116M} adopted the 
SMC extinction curve
to derive the dust abundance in galaxy halos.
{A steeper extinction curve requires a smaller amount of dust
to explain a certain amount of reddening.}
Therefore, if our calculation is correct and the extinction curve
is flat, the CGM and IGM contain more dust than estimated by
\citet{2010MNRAS.405.1025M} and \citet{2012ApJ...754..116M}.
However, the observational analysis in \citet{2015ApJ...813....7P} shows that
galaxy halos at $z\sim 0.05$ show a steep extinction curve similar to the SMC curve.
Therefore, more efforts on the theoretical side would be required to investigate
the possibilities of small grain production in the CGM and IGM.

\subsection{Possible importance of AGN feedback}

In order to concentrate on the issues related to stellar processes,
we did not include AGN feedback in this paper.
As we showed above, 
the stellar mass function and the dust mass function are both overestimated at the
massive end (Figs.\ \ref{fig:SFRD} and \ref{fig:dustMF}).
Such an overestimate could be resolved by implementation of AGN feedback
as done in various models
\citep[e.g.][]{2017NatAs...1E.165H,2015MNRAS.452..575S,2017MNRAS.465.3291W,2017arXiv171200023F},
some of which took radiative feedback from AGNs into account \citep[e.g.][]{2014MNRAS.444.1518V}.
The importance for gas heating is also reported observationally \citep[e.g.][]{2012ARA&A..50..455F}.
We expect that AGN feedback suppresses the formation of dense clouds, where
dust grains grow efficiently. 
{This suppression of dust growth together with the effect of mass ejection will
decrease the number of dust-rich galaxies.}
\cite{2014MNRAS.444.1518V} claimed that 
the gas heated by AGNs could enhance dust destruction by sputtering.
This affects the dust abundance in the CGM and IGM.
The above effects overall suppress the dust abundance, especially in or around
massive galaxies.

However, the effects of AGN may not be so simple.
{\cite{2011A&A...525A..61P} already developed
a model that can treat the evolution of dust in QSOs. In their model, QSOs are
regarded as progenitors of elliptical galaxies.
Therefore, the development of AGNs could be tightly related to the host galaxy properties 
\citep[see also][]{2011MNRAS.416.1916V}.}
\cite{2017arXiv170107200H} suggest that the AGN feedback
could create a cycle of gas cooling 
and heating, which would cause 
a cyclic behaviour of dust mass increase and decrease with a period of
AGN feedback.
\cite{2002ApJ...567L.107E} suggested positive feedback to the dust
abundance by pointing out that dust could condense in
AGN-driven winds.
Therefore, clarifying the effects of AGN feedback on the
dust abundance requires inclusion of the above negative and positive
influences and {an} appropriate {modelling} of 
{the} host galaxies.


\section{Conclusion}

We investigate the dust evolution
using a cosmic hydrodynamical simulation by extending our previous
single-galaxy simulations in A17 and \cite{2017MNRAS.469..870H}.
The grain size distribution is also
treated in the bimodal form of  large and small grains, with a grain radius
boundary of $\sim 0.03~\micron$.  In our simulation,
we take into account not only dust generation by SNe and AGB stars
but also dust growth by accretion. We also include other
interstellar processing mechanisms such as dust destruction by SN shocks,
coagulation in the dense ISM, and shattering in the diffuse ISM.
For dust destruction, coagulation, and accretion,
we adopt the sub-grid models developed in A17.

We first confirm that the relation between dust-to-gas ratio and metallicity
for the simulated galaxies is consistent with the observed relation
at $z=0$. In particular the nonlinear increase of dust-to-gas ratio
as a function of metallicity at $Z\gtrsim 0.1$ Z$_\odot$ is described well by
the transition from the dust production dominated by stellar sources, 
to that dominated by accretion (dust growth).
The consistency with the observational data suggests that the implementation
of dust abundance evolution is successful in our simulation.
After confirming this, we put particular
focus on the cosmological-volume properties of dust without going into detailed
analysis of individual galaxies, which will be reported in a separate paper
(Hou et al., in preparation).

We present the comoving density of dust mass as a function of redshift. 
The comoving dust mass density in our simulation is roughly consistent with 
the one derived from the analysis of observed infrared radiation at $z<2.4$.
In our simulation, the peak of the comoving dust density lies at $z=1$--2,
which coincides with the most dust-enshrouded epoch in the Universe
derived from \textit{Herschel} observations \citep{2013A&A...554A..70B}.
We also find that the dust abundance in the IGM is always dominated by
large grains.

The statistical properties of dust in galaxies are also investigated using
the dust mass function. 
Our simulation reproduces the dust mass function at $z=2.4$ well. 
At $z=0$, it
{broadly accounts for the observational slope
of dust mass function at $M_\mathrm{d}\sim 10^5$--$10^8$ M$_{\sun}$}, but is in excess
at the massive end. The excess could be improved if we include
AGN feedback in the simulation.

We further investigate the dust properties in the IGM and CGM.
For the CGM, we examine the radial profile of dust surface density
around galaxies with $M_\ast =10^{10}$--10$^{11}$\,M$_\odot$ up to
$r=1$\,Mpc, and find that we reproduce the observed radial profile at $r>20$\,kpc.
This means that our stellar feedback model is successfully transporting
the dust formed in galaxies to the circum-galactic region.
However, we also find that the {dust abundance dominated
by large grains is not consistent with}
the steep reddening curve 
derived for Mg \textsc{ii} absorbers at $z=1$ and 2
\citep{2012ApJ...754..116M}.  This indicates that our simulation still fails
to include a mechanism
of supplying small grains to the CGM/IGM.
We predict that the dust-to-metal ratio in the halo decreases with increasing $r$, 
since dust grains in the CGM/IGM are destroyed by hot gas via sputtering.
Using the dust properties in the simulation, we predict cosmological reddening 
of $A_{V}\sim 10^{-2}$ at $z=0.3$ while $A_{V}\sim10^{-1}$ at $z=2.0$. 
Both values satisfy the observational constraints.

\section*{Acknowledgment}
We thank the anonymous referee for careful reading and useful comments.
We are grateful to Volker Springel for providing us with the original version of \textsc{gadget-3} code.
Numerical computations were carried out on Cray XC30 at
the Center for Computational Astrophysics, National Astronomical Observatory of Japan
and XL at the Theoretical Institute for Advanced Research in Astrophysics (TIARA) in
Academia Sinica.
HH thanks the Ministry of Science and Technology for financial support through MOST
105-2112-M-001-027-MY3 and MOST 107-2923-M-001-003-MY3.
This work was in part supported by JSPS KAKENHI Grant Number JP17H01111.

\bibliographystyle{mnras}
\bibliography{ken} 

\begin{thebibliography}{}
\makeatletter
\relax
\def\mn@urlcharsother{\let\do\@makeother \do\$\do\&\do\#\do\^\do\_\do\%\do\~}
\def\mn@doi{\begingroup\mn@urlcharsother \@ifnextchar [ {\mn@doi@}
  {\mn@doi@[]}}
\def\mn@doi@[#1]#2{\def\@tempa{#1}\ifx\@tempa\@empty \href
  {http://dx.doi.org/#2} {doi:#2}\else \href {http://dx.doi.org/#2} {#1}\fi
  \endgroup}
\def\mn@eprint#1#2{\mn@eprint@#1:#2::\@nil}
\def\mn@eprint@arXiv#1{\href {http://arxiv.org/abs/#1} {{\tt arXiv:#1}}}
\def\mn@eprint@dblp#1{\href {http://dblp.uni-trier.de/rec/bibtex/#1.xml}
  {dblp:#1}}
\def\mn@eprint@#1:#2:#3:#4\@nil{\def\@tempa {#1}\def\@tempb {#2}\def\@tempc
  {#3}\ifx \@tempc \@empty \let \@tempc \@tempb \let \@tempb \@tempa \fi \ifx
  \@tempb \@empty \def\@tempb {arXiv}\fi \@ifundefined
  {mn@eprint@\@tempb}{\@tempb:\@tempc}{\expandafter \expandafter \csname
  mn@eprint@\@tempb\endcsname \expandafter{\@tempc}}}

\bibitem[\protect\citeauthoryear{{Aoyama}, {Hou}, {Shimizu}, {Hirashita},
  {Todoroki}, {Choi}  \& {Nagamine}}{{Aoyama}
  et~al.}{2017}]{2017MNRAS.466..105A}
{Aoyama} S.,  {Hou} K.-C.,  {Shimizu} I.,  {Hirashita} H.,  {Todoroki} K.,
  {Choi} J.-H.,   {Nagamine} K.,  2017, \mn@doi [\mnras]
  {10.1093/mnras/stw3061}, \href
  {http://adsabs.harvard.edu/abs/2017MNRAS.466..105A} {466, 105}

\bibitem[\protect\citeauthoryear{{Asano}, {Takeuchi}, {Hirashita}  \&
  {Inoue}}{{Asano} et~al.}{2013a}]{2013EP&S...65..213A}
{Asano} R.~S.,  {Takeuchi} T.~T.,  {Hirashita} H.,   {Inoue} A.~K.,  2013a,
  \mn@doi [Earth, Planets, and Space] {10.5047/eps.2012.04.014}, \href
  {http://adsabs.harvard.edu/abs/2013EP%26S...65..213A} {65, 213}

\bibitem[\protect\citeauthoryear{{Asano}, {Takeuchi}, {Hirashita}  \&
  {Nozawa}}{{Asano} et~al.}{2013b}]{2013MNRAS.432..637A}
{Asano} R.~S.,  {Takeuchi} T.~T.,  {Hirashita} H.,   {Nozawa} T.,  2013b,
  \mn@doi [\mnras] {10.1093/mnras/stt506}, \href
  {http://adsabs.harvard.edu/abs/2013MNRAS.432..637A} {432, 637}

\bibitem[\protect\citeauthoryear{{Asano}, {Takeuchi}, {Hirashita}  \&
  {Nozawa}}{{Asano} et~al.}{2014}]{2014MNRAS.440..134A}
{Asano} R.~S.,  {Takeuchi} T.~T.,  {Hirashita} H.,   {Nozawa} T.,  2014,
  \mn@doi [\mnras] {10.1093/mnras/stu208}, \href
  {http://adsabs.harvard.edu/abs/2014MNRAS.440..134A} {440, 134}

\bibitem[\protect\citeauthoryear{{Avgoustidis}, {Verde}  \&
  {Jimenez}}{{Avgoustidis} et~al.}{2009}]{2009JCAP...06..012A}
{Avgoustidis} A.,  {Verde} L.,   {Jimenez} R.,  2009, \mn@doi [\jcap]
  {10.1088/1475-7516/2009/06/012}, \href
  {http://adsabs.harvard.edu/abs/2009JCAP...06..012A} {6, 012}

\bibitem[\protect\citeauthoryear{{Barlow} \& {Silk}}{{Barlow} \&
  {Silk}}{1976}]{1976ApJ...207..131B}
{Barlow} M.~J.,  {Silk} J.,  1976, \mn@doi [\apj] {10.1086/154477}, \href
  {http://adsabs.harvard.edu/abs/1976ApJ...207..131B} {207, 131}

\bibitem[\protect\citeauthoryear{{Bekki}}{{Bekki}}{2013a}]{2013MNRAS.432.2298B}
{Bekki} K.,  2013a, \mn@doi [\mnras] {10.1093/mnras/stt589}, \href
  {http://adsabs.harvard.edu/abs/2013MNRAS.432.2298B} {432, 2298}

\bibitem[\protect\citeauthoryear{{Bekki}}{{Bekki}}{2013b}]{2013MNRAS.436.2254B}
{Bekki} K.,  2013b, \mn@doi [\mnras] {10.1093/mnras/stt1735}, \href
  {http://adsabs.harvard.edu/abs/2013MNRAS.436.2254B} {436, 2254}

\bibitem[\protect\citeauthoryear{{Bekki}}{{Bekki}}{2015}]{2015MNRAS.449.1625B}
{Bekki} K.,  2015, \mn@doi [\mnras] {10.1093/mnras/stv165}, \href
  {http://adsabs.harvard.edu/abs/2015MNRAS.449.1625B} {449, 1625}

\bibitem[\protect\citeauthoryear{{Bekki}, {Hirashita}  \& {Tsujimoto}}{{Bekki}
  et~al.}{2015}]{2015ApJ...810...39B}
{Bekki} K.,  {Hirashita} H.,   {Tsujimoto} T.,  2015, \mn@doi [\apj]
  {10.1088/0004-637X/810/1/39}, \href
  {http://adsabs.harvard.edu/abs/2015ApJ...810...39B} {810, 39}

\bibitem[\protect\citeauthoryear{{Bianchi} \& {Ferrara}}{{Bianchi} \&
  {Ferrara}}{2005}]{2005MNRAS.358..379B}
{Bianchi} S.,  {Ferrara} A.,  2005, \mn@doi [\mnras]
  {10.1111/j.1365-2966.2005.08762.x}, \href
  {http://adsabs.harvard.edu/abs/2005MNRAS.358..379B} {358, 379}

\bibitem[\protect\citeauthoryear{{Bohren} \& {Huffman}}{{Bohren} \&
  {Huffman}}{1983}]{1983asls.book.....B}
{Bohren} C.~F.,  {Huffman} D.~R.,  1983, {Absorption and scattering of light by
  small particles}.
A Wiley-Interscience Publication

\bibitem[\protect\citeauthoryear{{Bohren}, {Huffman}  \& {Kam}}{{Bohren}
  et~al.}{1983}]{1983Natur.306..625B}
{Bohren} C.~F.,  {Huffman} D.~R.,   {Kam} Z.,  1983, \nat, \href
  {http://adsabs.harvard.edu/abs/1983Natur.306..625B} {306, 625}

\bibitem[\protect\citeauthoryear{{Bond}, {Churchill}, {Charlton}  \&
  {Vogt}}{{Bond} et~al.}{2001}]{2001ApJ...562..641B}
{Bond} N.~A.,  {Churchill} C.~W.,  {Charlton} J.~C.,   {Vogt} S.~S.,  2001,
  \mn@doi [\apj] {10.1086/323876}, \href
  {http://adsabs.harvard.edu/abs/2001ApJ...562..641B} {562, 641}

\bibitem[\protect\citeauthoryear{{Bouwens} et~al.,}{{Bouwens}
  et~al.}{2009}]{2009ApJ...705..936B}
{Bouwens} R.~J.,  et~al., 2009, \mn@doi [\apj] {10.1088/0004-637X/705/1/936},
  \href {http://adsabs.harvard.edu/abs/2009ApJ...705..936B} {705, 936}

\bibitem[\protect\citeauthoryear{{Buat}, {Boselli}, {Gavazzi}  \&
  {Bonfanti}}{{Buat} et~al.}{2002}]{2002A&A...383..801B}
{Buat} V.,  {Boselli} A.,  {Gavazzi} G.,   {Bonfanti} C.,  2002, \mn@doi [\aap]
  {10.1051/0004-6361:20011832}, \href
  {http://adsabs.harvard.edu/abs/2002A%26A...383..801B} {383, 801}

\bibitem[\protect\citeauthoryear{{Burgarella} et~al.,}{{Burgarella}
  et~al.}{2013}]{2013A&A...554A..70B}
{Burgarella} D.,  et~al., 2013, \mn@doi [\aap] {10.1051/0004-6361/201321651},
  \href {http://adsabs.harvard.edu/abs/2013A%26A...554A..70B} {554, A70}

\bibitem[\protect\citeauthoryear{{Calura}, {Pipino}  \& {Matteucci}}{{Calura}
  et~al.}{2008}]{2008A&A...479..669C}
{Calura} F.,  {Pipino} A.,   {Matteucci} F.,  2008, \mn@doi [\aap]
  {10.1051/0004-6361:20078090}, \href
  {http://adsabs.harvard.edu/abs/2008A%26A...479..669C} {479, 669}

\bibitem[\protect\citeauthoryear{{Calura}, {Gilli}, {Vignali}, {Pozzi},
  {Pipino}  \& {Matteucci}}{{Calura} et~al.}{2014}]{2014MNRAS.438.2765C}
{Calura} F.,  {Gilli} R.,  {Vignali} C.,  {Pozzi} F.,  {Pipino} A.,
  {Matteucci} F.,  2014, \mn@doi [\mnras] {10.1093/mnras/stt2329}, \href
  {http://adsabs.harvard.edu/abs/2014MNRAS.438.2765C} {438, 2765}

\bibitem[\protect\citeauthoryear{{Calzetti}, {Armus}, {Bohlin}, {Kinney},
  {Koornneef}  \& {Storchi-Bergmann}}{{Calzetti}
  et~al.}{2000}]{2000ApJ...533..682C}
{Calzetti} D.,  {Armus} L.,  {Bohlin} R.~C.,  {Kinney} A.~L.,  {Koornneef} J.,
   {Storchi-Bergmann} T.,  2000, \mn@doi [\apj] {10.1086/308692}, \href
  {http://adsabs.harvard.edu/abs/2000ApJ...533..682C} {533, 682}

\bibitem[\protect\citeauthoryear{{Cazaux} \& {Spaans}}{{Cazaux} \&
  {Spaans}}{2009}]{2009A&A...496..365C}
{Cazaux} S.,  {Spaans} M.,  2009, \mn@doi [\aap] {10.1051/0004-6361:200811302},
  \href {http://ads.nao.ac.jp/abs/2009A%26A...496..365C} {496, 365}

\bibitem[\protect\citeauthoryear{{Cazaux} \& {Tielens}}{{Cazaux} \&
  {Tielens}}{2004}]{2004ApJ...604..222C}
{Cazaux} S.,  {Tielens} A.~G.~G.~M.,  2004, \mn@doi [\apj] {10.1086/381775},
  \href {http://ads.nao.ac.jp/abs/2004ApJ...604..222C} {604, 222}

\bibitem[\protect\citeauthoryear{{Choi} \& {Nagamine}}{{Choi} \&
  {Nagamine}}{2012}]{2012MNRAS.419.1280C}
{Choi} J.-H.,  {Nagamine} K.,  2012, \mn@doi [\mnras]
  {10.1111/j.1365-2966.2011.19788.x}, \href
  {http://adsabs.harvard.edu/abs/2012MNRAS.419.1280C} {419, 1280}

\bibitem[\protect\citeauthoryear{{Clark} et~al.,}{{Clark}
  et~al.}{2015}]{2015MNRAS.452..397C}
{Clark} C.~J.~R.,  et~al., 2015, \mn@doi [\mnras] {10.1093/mnras/stv1276},
  \href {http://adsabs.harvard.edu/abs/2015MNRAS.452..397C} {452, 397}

\bibitem[\protect\citeauthoryear{{Clemens} et~al.,}{{Clemens}
  et~al.}{2013}]{2013MNRAS.433..695C}
{Clemens} M.~S.,  et~al., 2013, \mn@doi [\mnras] {10.1093/mnras/stt760}, \href
  {http://adsabs.harvard.edu/abs/2013MNRAS.433..695C} {433, 695}

\bibitem[\protect\citeauthoryear{{Dayal}, {Hirashita}  \& {Ferrara}}{{Dayal}
  et~al.}{2010}]{2010MNRAS.403..620D}
{Dayal} P.,  {Hirashita} H.,   {Ferrara} A.,  2010, \mn@doi [\mnras]
  {10.1111/j.1365-2966.2009.16164.x}, \href
  {http://adsabs.harvard.edu/abs/2010MNRAS.403..620D} {403, 620}

\bibitem[\protect\citeauthoryear{{De Bernardis} \& {Cooray}}{{De Bernardis} \&
  {Cooray}}{2012}]{2012ApJ...760...14D}
{De Bernardis} F.,  {Cooray} A.,  2012, \mn@doi [\apj]
  {10.1088/0004-637X/760/1/14}, \href
  {http://adsabs.harvard.edu/abs/2012ApJ...760...14D} {760, 14}

\bibitem[\protect\citeauthoryear{{Dobbs} \& {Pringle}}{{Dobbs} \&
  {Pringle}}{2013}]{2013MNRAS.432..653D}
{Dobbs} C.~L.,  {Pringle} J.~E.,  2013, \mn@doi [\mnras]
  {10.1093/mnras/stt508}, \href
  {http://adsabs.harvard.edu/abs/2013MNRAS.432..653D} {432, 653}

\bibitem[\protect\citeauthoryear{{Draine}}{{Draine}}{2003}]{2003ARA&A..41..241D}
{Draine} B.~T.,  2003, \mn@doi [\araa]
  {10.1146/annurev.astro.41.011802.094840}, \href
  {http://adsabs.harvard.edu/abs/2003ARA%26A..41..241D} {41, 241}

\bibitem[\protect\citeauthoryear{{Draine} \& {Lee}}{{Draine} \&
  {Lee}}{1984}]{1984ApJ...285...89D}
{Draine} B.~T.,  {Lee} H.~M.,  1984, \mn@doi [\apj] {10.1086/162480}, \href
  {http://adsabs.harvard.edu/abs/1984ApJ...285...89D} {285, 89}

\bibitem[\protect\citeauthoryear{{Draine} \& {Salpeter}}{{Draine} \&
  {Salpeter}}{1979}]{1979ApJ...231...77D}
{Draine} B.~T.,  {Salpeter} E.~E.,  1979, \mn@doi [\apj] {10.1086/157165},
  \href {http://adsabs.harvard.edu/abs/1979ApJ...231...77D} {231, 77}

\bibitem[\protect\citeauthoryear{{Driver}, {Popescu}, {Tuffs}, {Liske},
  {Graham}, {Allen}  \& {de Propris}}{{Driver}
  et~al.}{2007}]{2007MNRAS.379.1022D}
{Driver} S.~P.,  {Popescu} C.~C.,  {Tuffs} R.~J.,  {Liske} J.,  {Graham} A.~W.,
   {Allen} P.~D.,   {de Propris} R.,  2007, \mn@doi [\mnras]
  {10.1111/j.1365-2966.2007.11862.x}, \href
  {http://adsabs.harvard.edu/abs/2007MNRAS.379.1022D} {379, 1022}

\bibitem[\protect\citeauthoryear{{Driver} et~al.,}{{Driver}
  et~al.}{2018}]{2018MNRAS.475.2891D}
{Driver} S.~P.,  et~al., 2018, \mn@doi [\mnras] {10.1093/mnras/stx2728}, \href
  {http://adsabs.harvard.edu/abs/2018MNRAS.475.2891D} {475, 2891}

\bibitem[\protect\citeauthoryear{{Duffy}, {Mutch}, {Poole}, {Geil}, {Kim},
  {Mesinger}  \& {Wyithe}}{{Duffy} et~al.}{2017}]{2017MNRAS.470.3300D}
{Duffy} A.~R.,  {Mutch} S.~J.,  {Poole} G.~B.,  {Geil} P.~M.,  {Kim} H.-S.,
  {Mesinger} A.,   {Wyithe} J.~S.~B.,  2017, \mn@doi [\mnras]
  {10.1093/mnras/stx1242}, \href
  {http://adsabs.harvard.edu/abs/2017MNRAS.470.3300D} {470, 3300}

\bibitem[\protect\citeauthoryear{{Dunne}, {Eales}, {Edmunds}, {Ivison},
  {Alexander}  \& {Clements}}{{Dunne} et~al.}{2000}]{2000MNRAS.315..115D}
{Dunne} L.,  {Eales} S.,  {Edmunds} M.,  {Ivison} R.,  {Alexander} P.,
  {Clements} D.~L.,  2000, \mn@doi [\mnras] {10.1046/j.1365-8711.2000.03386.x},
  \href {http://adsabs.harvard.edu/abs/2000MNRAS.315..115D} {315, 115}

\bibitem[\protect\citeauthoryear{{Dunne}, {Eales}  \& {Edmunds}}{{Dunne}
  et~al.}{2003}]{2003MNRAS.341..589D}
{Dunne} L.,  {Eales} S.~A.,   {Edmunds} M.~G.,  2003, \mn@doi [\mnras]
  {10.1046/j.1365-8711.2003.06440.x}, \href
  {http://adsabs.harvard.edu/abs/2003MNRAS.341..589D} {341, 589}

\bibitem[\protect\citeauthoryear{{Dunne} et~al.,}{{Dunne}
  et~al.}{2011}]{2011MNRAS.417.1510D}
{Dunne} L.,  et~al., 2011, \mn@doi [\mnras] {10.1111/j.1365-2966.2011.19363.x},
  \href {http://adsabs.harvard.edu/abs/2011MNRAS.417.1510D} {417, 1510}

\bibitem[\protect\citeauthoryear{{Dwek}}{{Dwek}}{1998}]{1998ApJ...501..643D}
{Dwek} E.,  1998, \mn@doi [\apj] {10.1086/305829}, \href
  {http://adsabs.harvard.edu/abs/1998ApJ...501..643D} {501, 643}

\bibitem[\protect\citeauthoryear{{Eisenstein} et~al.,}{{Eisenstein}
  et~al.}{2005}]{2005ApJ...633..560E}
{Eisenstein} D.~J.,  et~al., 2005, \mn@doi [\apj] {10.1086/466512}, \href
  {http://adsabs.harvard.edu/abs/2005ApJ...633..560E} {633, 560}

\bibitem[\protect\citeauthoryear{{Elvis}, {Marengo}  \& {Karovska}}{{Elvis}
  et~al.}{2002}]{2002ApJ...567L.107E}
{Elvis} M.,  {Marengo} M.,   {Karovska} M.,  2002, \mn@doi [\apjl]
  {10.1086/340006}, \href {http://adsabs.harvard.edu/abs/2002ApJ...567L.107E}
  {567, L107}

\bibitem[\protect\citeauthoryear{{Fabian}}{{Fabian}}{2012}]{2012ARA&A..50..455F}
{Fabian} A.~C.,  2012, \mn@doi [\araa] {10.1146/annurev-astro-081811-125521},
  \href {http://adsabs.harvard.edu/abs/2012ARA%26A..50..455F} {50, 455}

\bibitem[\protect\citeauthoryear{{Ferrara}, {Ferrini}, {Barsella}  \&
  {Franco}}{{Ferrara} et~al.}{1991}]{1991ApJ...381..137F}
{Ferrara} A.,  {Ferrini} F.,  {Barsella} B.,   {Franco} J.,  1991, \mn@doi
  [\apj] {10.1086/170636}, \href
  {http://adsabs.harvard.edu/abs/1991ApJ...381..137F} {381, 137}

\bibitem[\protect\citeauthoryear{{Fiacconi}, {Sijacki}  \&
  {Pringle}}{{Fiacconi} et~al.}{2018}]{2017arXiv171200023F}
{Fiacconi} D.,  {Sijacki} D.,   {Pringle} J.~E.,  2018, \mn@doi [\mnras]
  {10.1093/mnras/sty893}, \href
  {http://adsabs.harvard.edu/abs/2018MNRAS.477.3807F} {477, 3807}

\bibitem[\protect\citeauthoryear{{Fukugita}}{{Fukugita}}{2011}]{2011arXiv1103.4191F}
{Fukugita} M.,  2011, preprint, \href
  {http://adsabs.harvard.edu/abs/2011arXiv1103.4191F} {} (\mn@eprint {arXiv}
  {1103.4191})

\bibitem[\protect\citeauthoryear{{Fukugita} \& {Peebles}}{{Fukugita} \&
  {Peebles}}{2004}]{2004ApJ...616..643F}
{Fukugita} M.,  {Peebles} P.~J.~E.,  2004, \mn@doi [\apj] {10.1086/425155},
  \href {http://adsabs.harvard.edu/abs/2004ApJ...616..643F} {616, 643}

\bibitem[\protect\citeauthoryear{{Ginolfi}, {Graziani}, {Schneider}, {Marassi},
  {Valiante}, {Dell'Agli}, {Ventura}  \& {Hunt}}{{Ginolfi}
  et~al.}{2018}]{2018MNRAS.473.4538G}
{Ginolfi} M.,  {Graziani} L.,  {Schneider} R.,  {Marassi} S.,  {Valiante} R.,
  {Dell'Agli} F.,  {Ventura} P.,   {Hunt} L.~K.,  2018, \mn@doi [\mnras]
  {10.1093/mnras/stx2572}, \href
  {http://adsabs.harvard.edu/abs/2018MNRAS.473.4538G} {473, 4538}

\bibitem[\protect\citeauthoryear{{Gioannini}, {Matteucci}  \&
  {Calura}}{{Gioannini} et~al.}{2017}]{2017MNRAS.471.4615G}
{Gioannini} L.,  {Matteucci} F.,   {Calura} F.,  2017, \mn@doi [\mnras]
  {10.1093/mnras/stx1914}, \href
  {http://adsabs.harvard.edu/abs/2017MNRAS.471.4615G} {471, 4615}

\bibitem[\protect\citeauthoryear{{Gjergo}, {Granato}, {Murante},
  {Ragone-Figueroa}, {Tornatore}  \& {Borgani}}{{Gjergo}
  et~al.}{2018}]{2018arXiv180406855G}
{Gjergo} E.,  {Granato} G.~L.,  {Murante} G.,  {Ragone-Figueroa} C.,
  {Tornatore} L.,   {Borgani} S.,  2018, preprint, \href
  {http://adsabs.harvard.edu/abs/2018arXiv180406855G} {} (\mn@eprint {arXiv}
  {1804.06855})

\bibitem[\protect\citeauthoryear{{Gould} \& {Salpeter}}{{Gould} \&
  {Salpeter}}{1963}]{1963ApJ...138..393G}
{Gould} R.~J.,  {Salpeter} E.~E.,  1963, \mn@doi [\apj] {10.1086/147654}, \href
  {http://adsabs.harvard.edu/abs/1963ApJ...138..393G} {138, 393}

\bibitem[\protect\citeauthoryear{{Hahn} \& {Abel}}{{Hahn} \&
  {Abel}}{2011}]{2011MNRAS.415.2101H}
{Hahn} O.,  {Abel} T.,  2011, \mn@doi [\mnras]
  {10.1111/j.1365-2966.2011.18820.x}, \href
  {http://adsabs.harvard.edu/abs/2011MNRAS.415.2101H} {415, 2101}

\bibitem[\protect\citeauthoryear{{Harrison}}{{Harrison}}{2017}]{2017NatAs...1E.165H}
{Harrison} C.~M.,  2017, \mn@doi [Nature Astronomy] {10.1038/s41550-017-0165},
  \href {http://adsabs.harvard.edu/abs/2017NatAs...1E.165H} {1, 0165}

\bibitem[\protect\citeauthoryear{{Hirashita}}{{Hirashita}}{2015}]{2015MNRAS.447.2937H}
{Hirashita} H.,  2015, \mn@doi [\mnras] {10.1093/mnras/stu2617}, \href
  {http://adsabs.harvard.edu/abs/2015MNRAS.447.2937H} {447, 2937}

\bibitem[\protect\citeauthoryear{{Hirashita} \& {Ferrara}}{{Hirashita} \&
  {Ferrara}}{2002}]{2002MNRAS.337..921H}
{Hirashita} H.,  {Ferrara} A.,  2002, \mn@doi [\mnras]
  {10.1046/j.1365-8711.2002.05968.x}, \href
  {http://adsabs.harvard.edu/abs/2002MNRAS.337..921H} {337, 921}

\bibitem[\protect\citeauthoryear{{Hirashita} \& {Kuo}}{{Hirashita} \&
  {Kuo}}{2011}]{2011MNRAS.416.1340H}
{Hirashita} H.,  {Kuo} T.-M.,  2011, \mn@doi [\mnras]
  {10.1111/j.1365-2966.2011.19131.x}, \href
  {http://adsabs.harvard.edu/abs/2011MNRAS.416.1340H} {416, 1340}

\bibitem[\protect\citeauthoryear{{Hirashita} \& {Lin}}{{Hirashita} \&
  {Lin}}{2018}]{2018arXiv180400848H}
{Hirashita} H.,  {Lin} C.-Y.,  2018, preprint, \href
  {http://adsabs.harvard.edu/abs/2018arXiv180400848H} {} (\mn@eprint {arXiv}
  {1804.00848})

\bibitem[\protect\citeauthoryear{{Hirashita} \& {Nozawa}}{{Hirashita} \&
  {Nozawa}}{2017}]{2017arXiv170107200H}
{Hirashita} H.,  {Nozawa} T.,  2017, \mn@doi [\planss]
  {10.1016/j.pss.2017.01.009}, \href
  {http://adsabs.harvard.edu/abs/2017P%26SS..149...45H} {149, 45}

\bibitem[\protect\citeauthoryear{{Hirashita} \& {Yan}}{{Hirashita} \&
  {Yan}}{2009}]{2009MNRAS.394.1061H}
{Hirashita} H.,  {Yan} H.,  2009, \mn@doi [\mnras]
  {10.1111/j.1365-2966.2009.14405.x}, \href
  {http://adsabs.harvard.edu/abs/2009MNRAS.394.1061H} {394, 1061}

\bibitem[\protect\citeauthoryear{{Hirashita}, {Nozawa}, {Villaume}  \&
  {Srinivasan}}{{Hirashita} et~al.}{2015}]{2015MNRAS.454.1620H}
{Hirashita} H.,  {Nozawa} T.,  {Villaume} A.,   {Srinivasan} S.,  2015, \mn@doi
  [\mnras] {10.1093/mnras/stv2095}, \href
  {http://adsabs.harvard.edu/abs/2015MNRAS.454.1620H} {454, 1620}

\bibitem[\protect\citeauthoryear{{Hou}, {Hirashita}, {Nagamine}, {Aoyama}  \&
  {Shimizu}}{{Hou} et~al.}{2017}]{2017MNRAS.469..870H}
{Hou} K.-C.,  {Hirashita} H.,  {Nagamine} K.,  {Aoyama} S.,   {Shimizu} I.,
  2017, \mn@doi [\mnras] {10.1093/mnras/stx877}, \href
  {http://adsabs.harvard.edu/abs/2017MNRAS.469..870H} {469, 870}

\bibitem[\protect\citeauthoryear{{Hughes}, {Robson}, {Dunlop}  \&
  {Gear}}{{Hughes} et~al.}{1993}]{1993MNRAS.263..607H}
{Hughes} D.~H.,  {Robson} E.~I.,  {Dunlop} J.~S.,   {Gear} W.~K.,  1993,
  \mn@doi [\mnras] {10.1093/mnras/263.3.607}, \href
  {http://adsabs.harvard.edu/abs/1993MNRAS.263..607H} {263, 607}

\bibitem[\protect\citeauthoryear{{Inoue}}{{Inoue}}{2011}]{2011EP&S...63.1027I}
{Inoue} A.~K.,  2011, \mn@doi [Earth, Planets, and Space]
  {10.5047/eps.2011.02.013}, \href
  {http://adsabs.harvard.edu/abs/2011EP%26S...63.1027I} {63, 1027}

\bibitem[\protect\citeauthoryear{{Inoue} \& {Kamaya}}{{Inoue} \&
  {Kamaya}}{2003}]{2003MNRAS.341L...7I}
{Inoue} A.~K.,  {Kamaya} H.,  2003, \mn@doi [\mnras]
  {10.1046/j.1365-8711.2003.06619.x}, \href
  {http://adsabs.harvard.edu/abs/2003MNRAS.341L...7I} {341, L7}

\bibitem[\protect\citeauthoryear{{Ishiki} \& {Okamoto}}{{Ishiki} \&
  {Okamoto}}{2017}]{2017MNRAS.466L.123I}
{Ishiki} S.,  {Okamoto} T.,  2017, \mn@doi [\mnras] {10.1093/mnrasl/slw253},
  \href {http://adsabs.harvard.edu/abs/2017MNRAS.466L.123I} {466, L123}

\bibitem[\protect\citeauthoryear{{Jaacks}, {Nagamine}  \& {Choi}}{{Jaacks}
  et~al.}{2012}]{2012MNRAS.427..403J}
{Jaacks} J.,  {Nagamine} K.,   {Choi} J.~H.,  2012, \mn@doi [\mnras]
  {10.1111/j.1365-2966.2012.21989.x}, \href
  {http://adsabs.harvard.edu/abs/2012MNRAS.427..403J} {427, 403}

\bibitem[\protect\citeauthoryear{{Jaacks}, {Thompson}  \& {Nagamine}}{{Jaacks}
  et~al.}{2013}]{2013ApJ...766...94J}
{Jaacks} J.,  {Thompson} R.,   {Nagamine} K.,  2013, \mn@doi [\apj]
  {10.1088/0004-637X/766/2/94}, \href
  {http://adsabs.harvard.edu/abs/2013ApJ...766...94J} {766, 94}

\bibitem[\protect\citeauthoryear{{Jones}, {Tielens}  \& {Hollenbach}}{{Jones}
  et~al.}{1996}]{1996ApJ...469..740J}
{Jones} A.~P.,  {Tielens} A.~G.~G.~M.,   {Hollenbach} D.~J.,  1996, \mn@doi
  [\apj] {10.1086/177823}, \href
  {http://adsabs.harvard.edu/abs/1996ApJ...469..740J} {469, 740}

\bibitem[\protect\citeauthoryear{{Kennicutt} \& {Evans}}{{Kennicutt} \&
  {Evans}}{2012}]{2012ARA&A..50..531K}
{Kennicutt} R.~C.,  {Evans} N.~J.,  2012, \mn@doi [\araa]
  {10.1146/annurev-astro-081811-125610}, \href
  {http://adsabs.harvard.edu/abs/2012ARA%26A..50..531K} {50, 531}

\bibitem[\protect\citeauthoryear{{Kuo} \& {Hirashita}}{{Kuo} \&
  {Hirashita}}{2012}]{2012MNRAS.424L..34K}
{Kuo} T.-M.,  {Hirashita} H.,  2012, \mn@doi [\mnras]
  {10.1111/j.1745-3933.2012.01282.x}, \href
  {http://adsabs.harvard.edu/abs/2012MNRAS.424L..34K} {424, L34}

\bibitem[\protect\citeauthoryear{{Larson}}{{Larson}}{1981}]{1981MNRAS.194..809L}
{Larson} R.~B.,  1981, \mn@doi [\mnras] {10.1093/mnras/194.4.809}, \href
  {http://adsabs.harvard.edu/abs/1981MNRAS.194..809L} {194, 809}

\bibitem[\protect\citeauthoryear{{Larson}}{{Larson}}{2005}]{2005MNRAS.359..211L}
{Larson} R.~B.,  2005, \mn@doi [\mnras] {10.1111/j.1365-2966.2005.08881.x},
  \href {http://adsabs.harvard.edu/abs/2005MNRAS.359..211L} {359, 211}

\bibitem[\protect\citeauthoryear{{Lisenfeld} \& {Ferrara}}{{Lisenfeld} \&
  {Ferrara}}{1998}]{1998ApJ...496..145L}
{Lisenfeld} U.,  {Ferrara} A.,  1998, \mn@doi [\apj] {10.1086/305354}, \href
  {http://adsabs.harvard.edu/abs/1998ApJ...496..145L} {496, 145}

\bibitem[\protect\citeauthoryear{{Madau}, {Pozzetti}  \& {Dickinson}}{{Madau}
  et~al.}{1998}]{1998ApJ...498..106M}
{Madau} P.,  {Pozzetti} L.,   {Dickinson} M.,  1998, \mn@doi [\apj]
  {10.1086/305523}, \href {http://adsabs.harvard.edu/abs/1998ApJ...498..106M}
  {498, 106}

\bibitem[\protect\citeauthoryear{{Martin}}{{Martin}}{2005}]{2005ApJ...621..227M}
{Martin} C.~L.,  2005, \mn@doi [\apj] {10.1086/427277}, \href
  {http://adsabs.harvard.edu/abs/2005ApJ...621..227M} {621, 227}

\bibitem[\protect\citeauthoryear{{Martin} et~al.,}{{Martin}
  et~al.}{2005}]{2005ApJ...619L...1M}
{Martin} D.~C.,  et~al., 2005, \mn@doi [\apjl] {10.1086/426387}, \href
  {http://adsabs.harvard.edu/abs/2005ApJ...619L...1M} {619, L1}

\bibitem[\protect\citeauthoryear{{Mathis}, {Rumpl}  \& {Nordsieck}}{{Mathis}
  et~al.}{1977}]{1977ApJ...217..425M}
{Mathis} J.~S.,  {Rumpl} W.,   {Nordsieck} K.~H.,  1977, \mn@doi [\apj]
  {10.1086/155591}, \href {http://adsabs.harvard.edu/abs/1977ApJ...217..425M}
  {217, 425}

\bibitem[\protect\citeauthoryear{{Mattsson}, {De Cia}, {Andersen}  \&
  {Zafar}}{{Mattsson} et~al.}{2014}]{2014MNRAS.440.1562M}
{Mattsson} L.,  {De Cia} A.,  {Andersen} A.~C.,   {Zafar} T.,  2014, \mn@doi
  [\mnras] {10.1093/mnras/stu370}, \href
  {http://adsabs.harvard.edu/abs/2014MNRAS.440.1562M} {440, 1562}

\bibitem[\protect\citeauthoryear{{McCarthy}, {Schaye}, {Bird}  \& {Le
  Brun}}{{McCarthy} et~al.}{2017}]{2017MNRAS.465.2936M}
{McCarthy} I.~G.,  {Schaye} J.,  {Bird} S.,   {Le Brun} A.~M.~C.,  2017,
  \mn@doi [\mnras] {10.1093/mnras/stw2792}, \href
  {http://adsabs.harvard.edu/abs/2017MNRAS.465.2936M} {465, 2936}

\bibitem[\protect\citeauthoryear{{McKinnon}, {Torrey}  \&
  {Vogelsberger}}{{McKinnon} et~al.}{2016}]{2016MNRAS.457.3775M}
{McKinnon} R.,  {Torrey} P.,   {Vogelsberger} M.,  2016, \mn@doi [\mnras]
  {10.1093/mnras/stw253}, \href
  {http://adsabs.harvard.edu/abs/2016MNRAS.457.3775M} {457, 3775}

\bibitem[\protect\citeauthoryear{{McKinnon}, {Torrey}, {Vogelsberger},
  {Hayward}  \& {Marinacci}}{{McKinnon} et~al.}{2017}]{2017MNRAS.468.1505M}
{McKinnon} R.,  {Torrey} P.,  {Vogelsberger} M.,  {Hayward} C.~C.,
  {Marinacci} F.,  2017, \mn@doi [\mnras] {10.1093/mnras/stx467}, \href
  {http://adsabs.harvard.edu/abs/2017MNRAS.468.1505M} {468, 1505}

\bibitem[\protect\citeauthoryear{{McKinnon}, {Vogelsberger}, {Torrey},
  {Marinacci}  \& {Kannan}}{{McKinnon} et~al.}{2018}]{2018MNRAS.tmp.1185M}
{McKinnon} R.,  {Vogelsberger} M.,  {Torrey} P.,  {Marinacci} F.,   {Kannan}
  R.,  2018, \mn@doi [\mnras] {10.1093/mnras/sty1248}, \href
  {http://adsabs.harvard.edu/abs/2018MNRAS.tmp.1185M} {}

\bibitem[\protect\citeauthoryear{{M{\'e}nard} \& {Chelouche}}{{M{\'e}nard} \&
  {Chelouche}}{2009}]{2009MNRAS.393..808M}
{M{\'e}nard} B.,  {Chelouche} D.,  2009, \mn@doi [\mnras]
  {10.1111/j.1365-2966.2008.14225.x}, \href
  {http://adsabs.harvard.edu/abs/2009MNRAS.393..808M} {393, 808}

\bibitem[\protect\citeauthoryear{{M{\'e}nard} \& {Fukugita}}{{M{\'e}nard} \&
  {Fukugita}}{2012}]{2012ApJ...754..116M}
{M{\'e}nard} B.,  {Fukugita} M.,  2012, \mn@doi [\apj]
  {10.1088/0004-637X/754/2/116}, \href
  {http://adsabs.harvard.edu/abs/2012ApJ...754..116M} {754, 116}

\bibitem[\protect\citeauthoryear{{M{\'e}nard}, {Nestor}, {Turnshek}, {Quider},
  {Richards}, {Chelouche}  \& {Rao}}{{M{\'e}nard}
  et~al.}{2008}]{2008MNRAS.385.1053M}
{M{\'e}nard} B.,  {Nestor} D.,  {Turnshek} D.,  {Quider} A.,  {Richards} G.,
  {Chelouche} D.,   {Rao} S.,  2008, \mn@doi [\mnras]
  {10.1111/j.1365-2966.2008.12909.x}, \href
  {http://adsabs.harvard.edu/abs/2008MNRAS.385.1053M} {385, 1053}

\bibitem[\protect\citeauthoryear{{M{\'e}nard}, {Scranton}, {Fukugita}  \&
  {Richards}}{{M{\'e}nard} et~al.}{2010}]{2010MNRAS.405.1025M}
{M{\'e}nard} B.,  {Scranton} R.,  {Fukugita} M.,   {Richards} G.,  2010,
  \mn@doi [\mnras] {10.1111/j.1365-2966.2010.16486.x}, \href
  {http://adsabs.harvard.edu/abs/2010MNRAS.405.1025M} {405, 1025}

\bibitem[\protect\citeauthoryear{{More}, {Bovy}  \& {Hogg}}{{More}
  et~al.}{2009}]{2009ApJ...696.1727M}
{More} S.,  {Bovy} J.,   {Hogg} D.~W.,  2009, \mn@doi [\apj]
  {10.1088/0004-637X/696/2/1727}, \href
  {http://adsabs.harvard.edu/abs/2009ApJ...696.1727M} {696, 1727}

\bibitem[\protect\citeauthoryear{{M{\"o}rtsell} \& {Goobar}}{{M{\"o}rtsell} \&
  {Goobar}}{2003}]{2003JCAP...09..009M}
{M{\"o}rtsell} E.,  {Goobar} A.,  2003, \mn@doi [\jcap]
  {10.1088/1475-7516/2003/09/009}, \href
  {http://adsabs.harvard.edu/abs/2003JCAP...09..009M} {9, 009}

\bibitem[\protect\citeauthoryear{{Moustakas} et~al.,}{{Moustakas}
  et~al.}{2013}]{2013ApJ...767...50M}
{Moustakas} J.,  et~al., 2013, \mn@doi [\apj] {10.1088/0004-637X/767/1/50},
  \href {http://adsabs.harvard.edu/abs/2013ApJ...767...50M} {767, 50}

\bibitem[\protect\citeauthoryear{{Murray}, {Quataert}  \& {Thompson}}{{Murray}
  et~al.}{2005}]{2005ApJ...618..569M}
{Murray} N.,  {Quataert} E.,   {Thompson} T.~A.,  2005, \mn@doi [\apj]
  {10.1086/426067}, \href {http://adsabs.harvard.edu/abs/2005ApJ...618..569M}
  {618, 569}

\bibitem[\protect\citeauthoryear{{Nagamine}, {Fukugita}, {Cen}  \&
  {Ostriker}}{{Nagamine} et~al.}{2001}]{2001ApJ...558..497N}
{Nagamine} K.,  {Fukugita} M.,  {Cen} R.,   {Ostriker} J.~P.,  2001, \mn@doi
  [\apj] {10.1086/322293}, \href
  {http://adsabs.harvard.edu/abs/2001ApJ...558..497N} {558, 497}

\bibitem[\protect\citeauthoryear{{Nagamine}, {Springel}, {Hernquist}  \&
  {Machacek}}{{Nagamine} et~al.}{2004}]{2004MNRAS.350..385N}
{Nagamine} K.,  {Springel} V.,  {Hernquist} L.,   {Machacek} M.,  2004, \mn@doi
  [\mnras] {10.1111/j.1365-2966.2004.07664.x}, \href
  {http://adsabs.harvard.edu/abs/2004MNRAS.350..385N} {350, 385}

\bibitem[\protect\citeauthoryear{{Nagamine}, {Reddy}, {Daddi}  \&
  {Sargent}}{{Nagamine} et~al.}{2016}]{2016SSRv..202...79N}
{Nagamine} K.,  {Reddy} N.,  {Daddi} E.,   {Sargent} M.~T.,  2016, \mn@doi
  [\ssr] {10.1007/s11214-016-0270-3}, \href
  {http://adsabs.harvard.edu/abs/2016SSRv..202...79N} {202, 79}

\bibitem[\protect\citeauthoryear{{Norman}, {Bowen}, {Heckman}, {Blades}  \&
  {Danly}}{{Norman} et~al.}{1996}]{1996ApJ...472...73N}
{Norman} C.~A.,  {Bowen} D.~V.,  {Heckman} T.,  {Blades} C.,   {Danly} L.,
  1996, \mn@doi [\apj] {10.1086/178042}, \href
  {http://adsabs.harvard.edu/abs/1996ApJ...472...73N} {472, 73}

\bibitem[\protect\citeauthoryear{{Nozawa} \& {Fukugita}}{{Nozawa} \&
  {Fukugita}}{2013}]{2013ApJ...770...27N}
{Nozawa} T.,  {Fukugita} M.,  2013, \mn@doi [\apj]
  {10.1088/0004-637X/770/1/27}, \href
  {http://adsabs.harvard.edu/abs/2013ApJ...770...27N} {770, 27}

\bibitem[\protect\citeauthoryear{{Nozawa}, {Kozasa}  \& {Habe}}{{Nozawa}
  et~al.}{2006}]{2006ApJ...648..435N}
{Nozawa} T.,  {Kozasa} T.,   {Habe} A.,  2006, \mn@doi [\apj] {10.1086/505639},
  \href {http://adsabs.harvard.edu/abs/2006ApJ...648..435N} {648, 435}

\bibitem[\protect\citeauthoryear{{Nozawa}, {Asano}, {Hirashita}  \&
  {Takeuchi}}{{Nozawa} et~al.}{2015}]{2015MNRAS.447L..16N}
{Nozawa} T.,  {Asano} R.~S.,  {Hirashita} H.,   {Takeuchi} T.~T.,  2015,
  \mn@doi [\mnras] {10.1093/mnrasl/slu175}, \href
  {http://adsabs.harvard.edu/abs/2015MNRAS.447L..16N} {447, L16}

\bibitem[\protect\citeauthoryear{{Omukai}}{{Omukai}}{2000}]{2000ApJ...534..809O}
{Omukai} K.,  2000, \mn@doi [\apj] {10.1086/308776}, \href
  {http://adsabs.harvard.edu/abs/2000ApJ...534..809O} {534, 809}

\bibitem[\protect\citeauthoryear{{Omukai}, {Tsuribe}, {Schneider}  \&
  {Ferrara}}{{Omukai} et~al.}{2005}]{2005ApJ...626..627O}
{Omukai} K.,  {Tsuribe} T.,  {Schneider} R.,   {Ferrara} A.,  2005, \mn@doi
  [\apj] {10.1086/429955}, \href
  {http://adsabs.harvard.edu/abs/2005ApJ...626..627O} {626, 627}

\bibitem[\protect\citeauthoryear{{Peek}, {M{\'e}nard}  \& {Corrales}}{{Peek}
  et~al.}{2015}]{2015ApJ...813....7P}
{Peek} J.~E.~G.,  {M{\'e}nard} B.,   {Corrales} L.,  2015, \mn@doi [\apj]
  {10.1088/0004-637X/813/1/7}, \href
  {http://adsabs.harvard.edu/abs/2015ApJ...813....7P} {813, 7}

\bibitem[\protect\citeauthoryear{{Pei}}{{Pei}}{1992}]{1992ApJ...395..130P}
{Pei} Y.~C.,  1992, \mn@doi [\apj] {10.1086/171637}, \href
  {http://adsabs.harvard.edu/abs/1992ApJ...395..130P} {395, 130}

\bibitem[\protect\citeauthoryear{{Pillepich} et~al.,}{{Pillepich}
  et~al.}{2018}]{2018MNRAS.473.4077P}
{Pillepich} A.,  et~al., 2018, \mn@doi [\mnras] {10.1093/mnras/stx2656}, \href
  {http://adsabs.harvard.edu/abs/2018MNRAS.473.4077P} {473, 4077}

\bibitem[\protect\citeauthoryear{{Pipino}, {Fan}, {Matteucci}, {Calura},
  {Silva}, {Granato}  \& {Maiolino}}{{Pipino}
  et~al.}{2011}]{2011A&A...525A..61P}
{Pipino} A.,  {Fan} X.~L.,  {Matteucci} F.,  {Calura} F.,  {Silva} L.,
  {Granato} G.,   {Maiolino} R.,  2011, \mn@doi [\aap]
  {10.1051/0004-6361/201014843}, \href
  {http://adsabs.harvard.edu/abs/2011A%26A...525A..61P} {525, A61}

\bibitem[\protect\citeauthoryear{{Planck Collaboration} et~al.,}{{Planck
  Collaboration} et~al.}{2016}]{2016A&A...594A..13P}
{Planck Collaboration} et~al., 2016, \mn@doi [\aap]
  {10.1051/0004-6361/201525830}, \href
  {http://adsabs.harvard.edu/abs/2016A%26A...594A..13P} {594, A13}

\bibitem[\protect\citeauthoryear{{Popping}, {Somerville}  \&
  {Galametz}}{{Popping} et~al.}{2017}]{2017MNRAS.471.3152P}
{Popping} G.,  {Somerville} R.~S.,   {Galametz} M.,  2017, \mn@doi [\mnras]
  {10.1093/mnras/stx1545}, \href
  {http://adsabs.harvard.edu/abs/2017MNRAS.471.3152P} {471, 3152}

\bibitem[\protect\citeauthoryear{{Reddy} \& {Steidel}}{{Reddy} \&
  {Steidel}}{2009}]{2009ApJ...692..778R}
{Reddy} N.~A.,  {Steidel} C.~C.,  2009, \mn@doi [\apj]
  {10.1088/0004-637X/692/1/778}, \href
  {http://adsabs.harvard.edu/abs/2009ApJ...692..778R} {692, 778}

\bibitem[\protect\citeauthoryear{{R{\'e}my-Ruyer} et~al.,}{{R{\'e}my-Ruyer}
  et~al.}{2014}]{2014A&A...563A..31R}
{R{\'e}my-Ruyer} A.,  et~al., 2014, \mn@doi [\aap]
  {10.1051/0004-6361/201322803}, \href
  {http://adsabs.harvard.edu/abs/2014A%26A...563A..31R} {563, A31}

\bibitem[\protect\citeauthoryear{{Saitoh}}{{Saitoh}}{2017}]{2017AJ....153...85S}
{Saitoh} T.~R.,  2017, \mn@doi [\aj] {10.3847/1538-3881/153/2/85}, \href
  {http://adsabs.harvard.edu/abs/2017AJ....153...85S} {153, 85}

\bibitem[\protect\citeauthoryear{{Schaller}, {Dalla Vecchia}, {Schaye},
  {Bower}, {Theuns}, {Crain}, {Furlong}  \& {McCarthy}}{{Schaller}
  et~al.}{2015}]{2015MNRAS.454.2277S}
{Schaller} M.,  {Dalla Vecchia} C.,  {Schaye} J.,  {Bower} R.~G.,  {Theuns} T.,
   {Crain} R.~A.,  {Furlong} M.,   {McCarthy} I.~G.,  2015, \mn@doi [\mnras]
  {10.1093/mnras/stv2169}, \href
  {http://adsabs.harvard.edu/abs/2015MNRAS.454.2277S} {454, 2277}

\bibitem[\protect\citeauthoryear{{Schaye}, {Theuns}, {Rauch}, {Efstathiou}  \&
  {Sargent}}{{Schaye} et~al.}{2000}]{2000MNRAS.318..817S}
{Schaye} J.,  {Theuns} T.,  {Rauch} M.,  {Efstathiou} G.,   {Sargent} W.~L.~W.,
   2000, \mn@doi [\mnras] {10.1046/j.1365-8711.2000.03815.x}, \href
  {http://adsabs.harvard.edu/abs/2000MNRAS.318..817S} {318, 817}

\bibitem[\protect\citeauthoryear{{Schaye} et~al.,}{{Schaye}
  et~al.}{2015}]{2015MNRAS.446..521S}
{Schaye} J.,  et~al., 2015, \mn@doi [\mnras] {10.1093/mnras/stu2058}, \href
  {http://adsabs.harvard.edu/abs/2015MNRAS.446..521S} {446, 521}

\bibitem[\protect\citeauthoryear{{Schiminovich} et~al.,}{{Schiminovich}
  et~al.}{2005}]{2005ApJ...619L..47S}
{Schiminovich} D.,  et~al., 2005, \mn@doi [\apjl] {10.1086/427077}, \href
  {http://adsabs.harvard.edu/abs/2005ApJ...619L..47S} {619, L47}

\bibitem[\protect\citeauthoryear{{Schneider}, {Omukai}, {Inoue}  \&
  {Ferrara}}{{Schneider} et~al.}{2006}]{2006MNRAS.369.1437S}
{Schneider} R.,  {Omukai} K.,  {Inoue} A.~K.,   {Ferrara} A.,  2006, \mn@doi
  [\mnras] {10.1111/j.1365-2966.2006.10391.x}, \href
  {http://adsabs.harvard.edu/abs/2006MNRAS.369.1437S} {369, 1437}

\bibitem[\protect\citeauthoryear{{Shimizu}, {Inoue}, {Okamoto}  \&
  {Yoshida}}{{Shimizu} et~al.}{2014}]{2014MNRAS.440..731S}
{Shimizu} I.,  {Inoue} A.~K.,  {Okamoto} T.,   {Yoshida} N.,  2014, \mn@doi
  [\mnras] {10.1093/mnras/stu265}, \href
  {http://adsabs.harvard.edu/abs/2014MNRAS.440..731S} {440, 731}

\bibitem[\protect\citeauthoryear{{Shimizu}, {Inoue}, {Okamoto}  \&
  {Yoshida}}{{Shimizu} et~al.}{2016}]{2015arXiv150900800S}
{Shimizu} I.,  {Inoue} A.~K.,  {Okamoto} T.,   {Yoshida} N.,  2016, \mn@doi
  [\mnras] {10.1093/mnras/stw1423}, \href
  {http://adsabs.harvard.edu/abs/2016MNRAS.461.3563S} {461, 3563}

\bibitem[\protect\citeauthoryear{{Sijacki}, {Vogelsberger}, {Genel},
  {Springel}, {Torrey}, {Snyder}, {Nelson}  \& {Hernquist}}{{Sijacki}
  et~al.}{2015}]{2015MNRAS.452..575S}
{Sijacki} D.,  {Vogelsberger} M.,  {Genel} S.,  {Springel} V.,  {Torrey} P.,
  {Snyder} G.~F.,  {Nelson} D.,   {Hernquist} L.,  2015, \mn@doi [\mnras]
  {10.1093/mnras/stv1340}, \href {http://ads.nao.ac.jp/abs/2015MNRAS.452..575S}
  {452, 575}

\bibitem[\protect\citeauthoryear{{Springel}}{{Springel}}{2005}]{2005MNRAS.364.1105S}
{Springel} V.,  2005, \mn@doi [\mnras] {10.1111/j.1365-2966.2005.09655.x},
  \href {http://adsabs.harvard.edu/abs/2005MNRAS.364.1105S} {364, 1105}

\bibitem[\protect\citeauthoryear{{Springel}, {Yoshida}  \& {White}}{{Springel}
  et~al.}{2001a}]{2001NewA....6...79S}
{Springel} V.,  {Yoshida} N.,   {White} S.~D.~M.,  2001a, \mn@doi [\na]
  {10.1016/S1384-1076(01)00042-2}, \href
  {http://adsabs.harvard.edu/abs/2001NewA....6...79S} {6, 79}

\bibitem[\protect\citeauthoryear{{Springel}, {White}, {Tormen}  \&
  {Kauffmann}}{{Springel} et~al.}{2001b}]{2001MNRAS.328..726S}
{Springel} V.,  {White} S.~D.~M.,  {Tormen} G.,   {Kauffmann} G.,  2001b,
  \mn@doi [\mnras] {10.1046/j.1365-8711.2001.04912.x}, \href
  {http://adsabs.harvard.edu/abs/2001MNRAS.328..726S} {328, 726}

\bibitem[\protect\citeauthoryear{{Steidel}, {Adelberger}, {Giavalisco},
  {Dickinson}  \& {Pettini}}{{Steidel} et~al.}{1999}]{1999ApJ...519....1S}
{Steidel} C.~C.,  {Adelberger} K.~L.,  {Giavalisco} M.,  {Dickinson} M.,
  {Pettini} M.,  1999, \mn@doi [\apj] {10.1086/307363}, \href
  {http://adsabs.harvard.edu/abs/1999ApJ...519....1S} {519, 1}

\bibitem[\protect\citeauthoryear{{Sutherland} \& {Dopita}}{{Sutherland} \&
  {Dopita}}{1993}]{1993ApJS...88..253S}
{Sutherland} R.~S.,  {Dopita} M.~A.,  1993, \mn@doi [\apjs] {10.1086/191823},
  \href {http://adsabs.harvard.edu/abs/1993ApJS...88..253S} {88, 253}

\bibitem[\protect\citeauthoryear{{Takeuchi}, {Buat}  \&
  {Burgarella}}{{Takeuchi} et~al.}{2005}]{2005A&A...440L..17T}
{Takeuchi} T.~T.,  {Buat} V.,   {Burgarella} D.,  2005, \mn@doi [\aap]
  {10.1051/0004-6361:200500158}, \href
  {http://adsabs.harvard.edu/abs/2005A%26A...440L..17T} {440, L17}

\bibitem[\protect\citeauthoryear{{Takeuchi}, {Buat}, {Heinis}, {Giovannoli},
  {Yuan}, {Iglesias-P{\'a}ramo}, {Murata}  \& {Burgarella}}{{Takeuchi}
  et~al.}{2010}]{2010A&A...514A...4T}
{Takeuchi} T.~T.,  {Buat} V.,  {Heinis} S.,  {Giovannoli} E.,  {Yuan} F.-T.,
  {Iglesias-P{\'a}ramo} J.,  {Murata} K.~L.,   {Burgarella} D.,  2010, \mn@doi
  [\aap] {10.1051/0004-6361/200913476}, \href
  {http://adsabs.harvard.edu/abs/2010A%26A...514A...4T} {514, A4}

\bibitem[\protect\citeauthoryear{{Takeuchi}, {Yuan}, {Ikeyama}, {Murata}  \&
  {Inoue}}{{Takeuchi} et~al.}{2012}]{2012ApJ...755..144T}
{Takeuchi} T.~T.,  {Yuan} F.-T.,  {Ikeyama} A.,  {Murata} K.~L.,   {Inoue}
  A.~K.,  2012, \mn@doi [\apj] {10.1088/0004-637X/755/2/144}, \href
  {http://adsabs.harvard.edu/abs/2012ApJ...755..144T} {755, 144}

\bibitem[\protect\citeauthoryear{{Thacker} et~al.,}{{Thacker}
  et~al.}{2013}]{2013ApJ...768...58T}
{Thacker} C.,  et~al., 2013, \mn@doi [\apj] {10.1088/0004-637X/768/1/58}, \href
  {http://adsabs.harvard.edu/abs/2013ApJ...768...58T} {768, 58}

\bibitem[\protect\citeauthoryear{{Thompson}, {Nagamine}, {Jaacks}  \&
  {Choi}}{{Thompson} et~al.}{2014}]{2014ApJ...780..145T}
{Thompson} R.,  {Nagamine} K.,  {Jaacks} J.,   {Choi} J.-H.,  2014, \mn@doi
  [\apj] {10.1088/0004-637X/780/2/145}, \href
  {http://adsabs.harvard.edu/abs/2014ApJ...780..145T} {780, 145}

\bibitem[\protect\citeauthoryear{{Tomczak} et~al.,}{{Tomczak}
  et~al.}{2014}]{2014ApJ...783...85T}
{Tomczak} A.~R.,  et~al., 2014, \mn@doi [\apj] {10.1088/0004-637X/783/2/85},
  \href {http://adsabs.harvard.edu/abs/2014ApJ...783...85T} {783, 85}

\bibitem[\protect\citeauthoryear{{Tsai} \& {Mathews}}{{Tsai} \&
  {Mathews}}{1995}]{1995ApJ...448...84T}
{Tsai} J.~C.,  {Mathews} W.~G.,  1995, \mn@doi [\apj] {10.1086/175943}, \href
  {http://adsabs.harvard.edu/abs/1995ApJ...448...84T} {448, 84}

\bibitem[\protect\citeauthoryear{{Valiante}, {Schneider}, {Salvadori}  \&
  {Bianchi}}{{Valiante} et~al.}{2011}]{2011MNRAS.416.1916V}
{Valiante} R.,  {Schneider} R.,  {Salvadori} S.,   {Bianchi} S.,  2011, \mn@doi
  [\mnras] {10.1111/j.1365-2966.2011.19168.x}, \href
  {http://adsabs.harvard.edu/abs/2011MNRAS.416.1916V} {416, 1916}

\bibitem[\protect\citeauthoryear{{Vladilo} \& {P{\'e}roux}}{{Vladilo} \&
  {P{\'e}roux}}{2005}]{2005A&A...444..461V}
{Vladilo} G.,  {P{\'e}roux} C.,  2005, \mn@doi [\aap]
  {10.1051/0004-6361:20041570}, \href
  {http://adsabs.harvard.edu/abs/2005A%26A...444..461V} {444, 461}

\bibitem[\protect\citeauthoryear{{Vlahakis}, {Dunne}  \& {Eales}}{{Vlahakis}
  et~al.}{2005}]{2005MNRAS.364.1253V}
{Vlahakis} C.,  {Dunne} L.,   {Eales} S.,  2005, \mn@doi [\mnras]
  {10.1111/j.1365-2966.2005.09666.x}, \href
  {http://adsabs.harvard.edu/abs/2005MNRAS.364.1253V} {364, 1253}

\bibitem[\protect\citeauthoryear{{Vogelsberger}, {Genel}, {Sijacki}, {Torrey},
  {Springel}  \& {Hernquist}}{{Vogelsberger}
  et~al.}{2013}]{2013MNRAS.436.3031V}
{Vogelsberger} M.,  {Genel} S.,  {Sijacki} D.,  {Torrey} P.,  {Springel} V.,
  {Hernquist} L.,  2013, \mn@doi [\mnras] {10.1093/mnras/stt1789}, \href
  {http://adsabs.harvard.edu/abs/2013MNRAS.436.3031V} {436, 3031}

\bibitem[\protect\citeauthoryear{{Vogelsberger} et~al.,}{{Vogelsberger}
  et~al.}{2014}]{2014MNRAS.444.1518V}
{Vogelsberger} M.,  et~al., 2014, \mn@doi [\mnras] {10.1093/mnras/stu1536},
  \href {http://adsabs.harvard.edu/abs/2014MNRAS.444.1518V} {444, 1518}

\bibitem[\protect\citeauthoryear{{Weinberger} et~al.,}{{Weinberger}
  et~al.}{2017}]{2017MNRAS.465.3291W}
{Weinberger} R.,  et~al., 2017, \mn@doi [\mnras] {10.1093/mnras/stw2944}, \href
  {http://ads.nao.ac.jp/abs/2017MNRAS.465.3291W} {465, 3291}

\bibitem[\protect\citeauthoryear{{Weingartner} \& {Draine}}{{Weingartner} \&
  {Draine}}{2001}]{2001ApJ...548..296W}
{Weingartner} J.~C.,  {Draine} B.~T.,  2001, \mn@doi [\apj] {10.1086/318651},
  \href {http://adsabs.harvard.edu/abs/2001ApJ...548..296W} {548, 296}

\bibitem[\protect\citeauthoryear{{Whitworth}, {Boffin}  \&
  {Francis}}{{Whitworth} et~al.}{1998}]{1998MNRAS.299..554W}
{Whitworth} A.~P.,  {Boffin} H.~M.~J.,   {Francis} N.,  1998, \mn@doi [\mnras]
  {10.1046/j.1365-8711.1998.01813.x}, \href
  {http://adsabs.harvard.edu/abs/1998MNRAS.299..554W} {299, 554}

\bibitem[\protect\citeauthoryear{{Yajima}, {Shlosman}, {Romano-D{\'{\i}}az}  \&
  {Nagamine}}{{Yajima} et~al.}{2015}]{2015MNRAS.451..418Y}
{Yajima} H.,  {Shlosman} I.,  {Romano-D{\'{\i}}az} E.,   {Nagamine} K.,  2015,
  \mn@doi [\mnras] {10.1093/mnras/stv974}, \href
  {http://adsabs.harvard.edu/abs/2015MNRAS.451..418Y} {451, 418}

\bibitem[\protect\citeauthoryear{{Yamasawa}, {Habe}, {Kozasa}, {Nozawa},
  {Hirashita}, {Umeda}  \& {Nomoto}}{{Yamasawa}
  et~al.}{2011}]{2011ApJ...735...44Y}
{Yamasawa} D.,  {Habe} A.,  {Kozasa} T.,  {Nozawa} T.,  {Hirashita} H.,
  {Umeda} H.,   {Nomoto} K.,  2011, \mn@doi [\apj]
  {10.1088/0004-637X/735/1/44}, \href
  {http://adsabs.harvard.edu/abs/2011ApJ...735...44Y} {735, 44}

\bibitem[\protect\citeauthoryear{{Yan}, {Lazarian}  \& {Draine}}{{Yan}
  et~al.}{2004}]{2004ApJ...616..895Y}
{Yan} H.,  {Lazarian} A.,   {Draine} B.~T.,  2004, \mn@doi [\apj]
  {10.1086/425111}, \href {http://adsabs.harvard.edu/abs/2004ApJ...616..895Y}
  {616, 895}

\bibitem[\protect\citeauthoryear{{York} et~al.,}{{York}
  et~al.}{2000}]{2000AJ....120.1579Y}
{York} D.~G.,  et~al., 2000, \mn@doi [\aj] {10.1086/301513}, \href
  {http://adsabs.harvard.edu/abs/2000AJ....120.1579Y} {120, 1579}

\bibitem[\protect\citeauthoryear{{Zhukovska}, {Gail}  \&
  {Trieloff}}{{Zhukovska} et~al.}{2008}]{2008A&A...479..453Z}
{Zhukovska} S.,  {Gail} H.-P.,   {Trieloff} M.,  2008, \mn@doi [\aap]
  {10.1051/0004-6361:20077789}, \href
  {http://adsabs.harvard.edu/abs/2008A%26A...479..453Z} {479, 453}

\bibitem[\protect\citeauthoryear{{Zhukovska}, {Dobbs}, {Jenkins}  \&
  {Klessen}}{{Zhukovska} et~al.}{2016}]{2016ApJ...831..147Z}
{Zhukovska} S.,  {Dobbs} C.,  {Jenkins} E.~B.,   {Klessen} R.~S.,  2016,
  \mn@doi [\apj] {10.3847/0004-637X/831/2/147}, \href
  {http://adsabs.harvard.edu/abs/2016ApJ...831..147Z} {831, 147}

\bibitem[\protect\citeauthoryear{{Zu}, {Weinberg}, {Dav{\'e}}, {Fardal},
  {Katz}, {Kere{\v s}}  \& {Oppenheimer}}{{Zu}
  et~al.}{2011}]{2011MNRAS.412.1059Z}
{Zu} Y.,  {Weinberg} D.~H.,  {Dav{\'e}} R.,  {Fardal} M.,  {Katz} N.,  {Kere{\v
  s}} D.,   {Oppenheimer} B.~D.,  2011, \mn@doi [\mnras]
  {10.1111/j.1365-2966.2010.17976.x}, \href
  {http://adsabs.harvard.edu/abs/2011MNRAS.412.1059Z} {412, 1059}

\makeatother
\end{thebibliography}


\begin{thebibliography}{}
\makeatletter
\relax
\def\mn@urlcharsother{\let\do\@makeother \do\$\do\&\do\#\do\^\do\_\do\%\do\~}
\def\mn@doi{\begingroup\mn@urlcharsother \@ifnextchar [ {\mn@doi@}
  {\mn@doi@[]}}
\def\mn@doi@[#1]#2{\def\@tempa{#1}\ifx\@tempa\@empty \href
  {http://dx.doi.org/#2} {doi:#2}\else \href {http://dx.doi.org/#2} {#1}\fi
  \endgroup}
\def\mn@eprint#1#2{\mn@eprint@#1:#2::\@nil}
\def\mn@eprint@arXiv#1{\href {http://arxiv.org/abs/#1} {{\tt arXiv:#1}}}
\def\mn@eprint@dblp#1{\href {http://dblp.uni-trier.de/rec/bibtex/#1.xml}
  {dblp:#1}}
\def\mn@eprint@#1:#2:#3:#4\@nil{\def\@tempa {#1}\def\@tempb {#2}\def\@tempc
  {#3}\ifx \@tempc \@empty \let \@tempc \@tempb \let \@tempb \@tempa \fi \ifx
  \@tempb \@empty \def\@tempb {arXiv}\fi \@ifundefined
  {mn@eprint@\@tempb}{\@tempb:\@tempc}{\expandafter \expandafter \csname
  mn@eprint@\@tempb\endcsname \expandafter{\@tempc}}}

\bibitem[\protect\citeauthoryear{{Agertz}, {Kravtsov}, {Leitner}  \&
  {Gnedin}}{{Agertz} et~al.}{2013}]{2013ApJ...770...25A}
{Agertz} O.,  {Kravtsov} A.~V.,  {Leitner} S.~N.,   {Gnedin} N.~Y.,  2013,
  \mn@doi [\apj] {10.1088/0004-637X/770/1/25}, 770, 25

\bibitem[\protect\citeauthoryear{{Asano}, {Takeuchi}, {Hirashita}  \&
  {Inoue}}{{Asano} et~al.}{2013a}]{2013EP&S...65..213A}
{Asano} R.~S.,  {Takeuchi} T.~T.,  {Hirashita} H.,   {Inoue} A.~K.,  2013a,
  \mn@doi [Earth, Planets, and Space] {10.5047/eps.2012.04.014}, 65, 213

\bibitem[\protect\citeauthoryear{{Asano}, {Takeuchi}, {Hirashita}  \&
  {Nozawa}}{{Asano} et~al.}{2013b}]{2013MNRAS.432..637A}
{Asano} R.~S.,  {Takeuchi} T.~T.,  {Hirashita} H.,   {Nozawa} T.,  2013b,
  \mn@doi [\mnras] {10.1093/mnras/stt506}, 432, 637

\bibitem[\protect\citeauthoryear{{Asano}, {Takeuchi}, {Hirashita}  \&
  {Nozawa}}{{Asano} et~al.}{2014}]{2014MNRAS.440..134A}
{Asano} R.~S.,  {Takeuchi} T.~T.,  {Hirashita} H.,   {Nozawa} T.,  2014,
  \mn@doi [\mnras] {10.1093/mnras/stu208}, 440, 134

\bibitem[\protect\citeauthoryear{{Barlow} \& {Silk}}{{Barlow} \&
  {Silk}}{1976}]{1976ApJ...207..131B}
{Barlow} M.~J.,  {Silk} J.,  1976, \mn@doi [\apj] {10.1086/154477}, 207, 131

\bibitem[\protect\citeauthoryear{{Bekki}}{{Bekki}}{2013}]{2013MNRAS.432.2298B}
{Bekki} K.,  2013, \mn@doi [\mnras] {10.1093/mnras/stt589}, 432, 2298

\bibitem[\protect\citeauthoryear{{Bekki}}{{Bekki}}{2015}]{2015MNRAS.449.1625B}
{Bekki} K.,  2015, \mn@doi [\mnras] {10.1093/mnras/stv165}, 449, 1625

\bibitem[\protect\citeauthoryear{{Bohren}, {Huffman}  \& {Kam}}{{Bohren}
  et~al.}{1983}]{1983Natur.306..625B}
{Bohren} C.~F.,  {Huffman} D.~R.,   {Kam} Z.,  1983, \nat, 306, 625

\bibitem[\protect\citeauthoryear{{Bryan} et~al.,}{{Bryan}
  et~al.}{2014}]{2014ApJS..211...19B}
{Bryan} G.~L.,  et~al., 2014, \mn@doi [\apjs] {10.1088/0067-0049/211/2/19},
  211, 19

\bibitem[\protect\citeauthoryear{{Buat}, {Boselli}, {Gavazzi}  \&
  {Bonfanti}}{{Buat} et~al.}{2002}]{2002A&A...383..801B}
{Buat} V.,  {Boselli} A.,  {Gavazzi} G.,   {Bonfanti} C.,  2002, \mn@doi [\aap]
  {10.1051/0004-6361:20011832}, 383, 801

\bibitem[\protect\citeauthoryear{{Calzetti}, {Armus}, {Bohlin}, {Kinney},
  {Koornneef}  \& {Storchi-Bergmann}}{{Calzetti}
  et~al.}{2000}]{2000ApJ...533..682C}
{Calzetti} D.,  {Armus} L.,  {Bohlin} R.~C.,  {Kinney} A.~L.,  {Koornneef} J.,
   {Storchi-Bergmann} T.,  2000, \mn@doi [\apj] {10.1086/308692}, 533, 682

\bibitem[\protect\citeauthoryear{{Cazaux} \& {Spaans}}{{Cazaux} \&
  {Spaans}}{2009}]{2009A&A...496..365C}
{Cazaux} S.,  {Spaans} M.,  2009, \mn@doi [\aap] {10.1051/0004-6361:200811302},
  496, 365

\bibitem[\protect\citeauthoryear{{Cazaux} \& {Tielens}}{{Cazaux} \&
  {Tielens}}{2004}]{2004ApJ...604..222C}
{Cazaux} S.,  {Tielens} A.~G.~G.~M.,  2004, \mn@doi [\apj] {10.1086/381775},
  604, 222

\bibitem[\protect\citeauthoryear{{Chabrier}}{{Chabrier}}{2003}]{2003PASP..115.%
.763C}
{Chabrier} G.,  2003, \mn@doi [\pasp] {10.1086/376392}, 115, 763

\bibitem[\protect\citeauthoryear{{Chevalier}}{{Chevalier}}{1974}]{1974ApJ...18%
8..501C}
{Chevalier} R.~A.,  1974, \mn@doi [\apj] {10.1086/152740}, 188, 501

\bibitem[\protect\citeauthoryear{{Choi} \& {Nagamine}}{{Choi} \&
  {Nagamine}}{2009}]{2009MNRAS.393.1595C}
{Choi} J.-H.,  {Nagamine} K.,  2009, \mn@doi [\mnras]
  {10.1111/j.1365-2966.2008.14297.x}, 393, 1595

\bibitem[\protect\citeauthoryear{{Choi} \& {Nagamine}}{{Choi} \&
  {Nagamine}}{2012}]{2012MNRAS.419.1280C}
{Choi} J.-H.,  {Nagamine} K.,  2012, \mn@doi [\mnras]
  {10.1111/j.1365-2966.2011.19788.x}, 419, 1280

\bibitem[\protect\citeauthoryear{{De Cia}, {Ledoux}, {Mattsson}, {Petitjean},
  {Srianand}, {Gavignaud}  \& {Jenkins}}{{De Cia}
  et~al.}{2016}]{2016arXiv160808621D}
{De Cia} A.,  {Ledoux} C.,  {Mattsson} L.,  {Petitjean} P.,  {Srianand} R.,
  {Gavignaud} I.,   {Jenkins} E.~B.,  2016, preprint (\mn@eprint {arXiv}
  {1608.08621})

\bibitem[\protect\citeauthoryear{{Draine} \& {Li}}{{Draine} \&
  {Li}}{2001}]{2001ApJ...551..807D}
{Draine} B.~T.,  {Li} A.,  2001, \mn@doi [\apj] {10.1086/320227}, 551, 807

\bibitem[\protect\citeauthoryear{{Durier} \& {Dalla Vecchia}}{{Durier} \&
  {Dalla Vecchia}}{2012}]{2012MNRAS.419..465D}
{Durier} F.,  {Dalla Vecchia} C.,  2012, \mn@doi [\mnras]
  {10.1111/j.1365-2966.2011.19712.x}, 419, 465

\bibitem[\protect\citeauthoryear{{Dutta}, {Begum}, {Bharadwaj}  \&
  {Chengalur}}{{Dutta} et~al.}{2013}]{2013NewA...19...89D}
{Dutta} P.,  {Begum} A.,  {Bharadwaj} S.,   {Chengalur} J.~N.,  2013, \mn@doi
  [\na] {10.1016/j.newast.2012.08.008}, 19, 89

\bibitem[\protect\citeauthoryear{{Dwek}}{{Dwek}}{1998}]{1998ApJ...501..643D}
{Dwek} E.,  1998, \mn@doi [\apj] {10.1086/305829}, 501, 643

\bibitem[\protect\citeauthoryear{{Elmegreen}}{{Elmegreen}}{1998}]{1998ggs..boo%
k.....E}
{Elmegreen} D.~M.,  1998, {Galaxies and galactic structure}.
{Upper Saddle River, NJ : Prentice Hall}

\bibitem[\protect\citeauthoryear{{Fall}, {Krumholz}  \& {Matzner}}{{Fall}
  et~al.}{2010}]{2010ApJ...710L.142F}
{Fall} S.~M.,  {Krumholz} M.~R.,   {Matzner} C.~D.,  2010, \mn@doi [\apjl]
  {10.1088/2041-8205/710/2/L142}, 710, L142

\bibitem[\protect\citeauthoryear{{Gould} \& {Salpeter}}{{Gould} \&
  {Salpeter}}{1963}]{1963ApJ...138..393G}
{Gould} R.~J.,  {Salpeter} E.~E.,  1963, \mn@doi [\apj] {10.1086/147654}, 138,
  393

\bibitem[\protect\citeauthoryear{{Heesen}, {Brinks}, {Leroy}, {Heald}, {Braun},
  {Bigiel}  \& {Beck}}{{Heesen} et~al.}{2014}]{2014AJ....147..103H}
{Heesen} V.,  {Brinks} E.,  {Leroy} A.~K.,  {Heald} G.,  {Braun} R.,  {Bigiel}
  F.,   {Beck} R.,  2014, \mn@doi [\aj] {10.1088/0004-6256/147/5/103}, 147, 103

\bibitem[\protect\citeauthoryear{{Hirashita}}{{Hirashita}}{2015}]{2015MNRAS.44%
7.2937H}
{Hirashita} H.,  2015, \mn@doi [\mnras] {10.1093/mnras/stu2617}, 447, 2937

\bibitem[\protect\citeauthoryear{{Hirashita} \& {Kuo}}{{Hirashita} \&
  {Kuo}}{2011}]{2011MNRAS.416.1340H}
{Hirashita} H.,  {Kuo} T.-M.,  2011, \mn@doi [\mnras]
  {10.1111/j.1365-2966.2011.19131.x}, 416, 1340

\bibitem[\protect\citeauthoryear{{Hirashita} \& {Voshchinnikov}}{{Hirashita} \&
  {Voshchinnikov}}{2014}]{2014MNRAS.437.1636H}
{Hirashita} H.,  {Voshchinnikov} N.~V.,  2014, \mn@doi [\mnras]
  {10.1093/mnras/stt1997}, 437, 1636

\bibitem[\protect\citeauthoryear{{Hirashita} \& {Yan}}{{Hirashita} \&
  {Yan}}{2009}]{2009MNRAS.394.1061H}
{Hirashita} H.,  {Yan} H.,  2009, \mn@doi [\mnras]
  {10.1111/j.1365-2966.2009.14405.x}, 394, 1061

\bibitem[\protect\citeauthoryear{{Hopkins}}{{Hopkins}}{2013}]{2013MNRAS.428.28%
40H}
{Hopkins} P.~F.,  2013, \mn@doi [\mnras] {10.1093/mnras/sts210}, 428, 2840

\bibitem[\protect\citeauthoryear{{Hopkins}, {Quataert}  \& {Murray}}{{Hopkins}
  et~al.}{2011}]{2011MNRAS.417..950H}
{Hopkins} P.~F.,  {Quataert} E.,   {Murray} N.,  2011, \mn@doi [\mnras]
  {10.1111/j.1365-2966.2011.19306.x}, 417, 950

\bibitem[\protect\citeauthoryear{{Hou}, {Hirashita}  \& {Micha{\l}owski}}{{Hou}
  et~al.}{2016}]{2016arXiv160806099H}
{Hou} K.-C.,  {Hirashita} H.,   {Micha{\l}owski} M.~J.,  2016, preprint
  (\mn@eprint {arXiv} {1608.06099})

\bibitem[\protect\citeauthoryear{{Inoue}}{{Inoue}}{2003}]{2003PASJ...55..901I}
{Inoue} A.~K.,  2003, \mn@doi [\pasj] {10.1093/pasj/55.5.901}, 55, 901

\bibitem[\protect\citeauthoryear{{Inoue}}{{Inoue}}{2011a}]{2011EP&S...63.1027I}
{Inoue} A.~K.,  2011a, \mn@doi [Earth, Planets, and Space]
  {10.5047/eps.2011.02.013}, 63, 1027

\bibitem[\protect\citeauthoryear{{Inoue}}{{Inoue}}{2011b}]{2011MNRAS.415.2920I}
{Inoue} A.~K.,  2011b, \mn@doi [\mnras] {10.1111/j.1365-2966.2011.18906.x},
  415, 2920

\bibitem[\protect\citeauthoryear{{Jaacks}, {Nagamine}  \& {Choi}}{{Jaacks}
  et~al.}{2012}]{2012MNRAS.427..403J}
{Jaacks} J.,  {Nagamine} K.,   {Choi} J.~H.,  2012, \mn@doi [\mnras]
  {10.1111/j.1365-2966.2012.21989.x}, 427, 403

\bibitem[\protect\citeauthoryear{{Jaacks}, {Thompson}  \& {Nagamine}}{{Jaacks}
  et~al.}{2013}]{2013ApJ...766...94J}
{Jaacks} J.,  {Thompson} R.,   {Nagamine} K.,  2013, \mn@doi [\apj]
  {10.1088/0004-637X/766/2/94}, 766, 94

\bibitem[\protect\citeauthoryear{{Jenkins}}{{Jenkins}}{2009}]{2009ApJ...700.12%
99J}
{Jenkins} E.~B.,  2009, \mn@doi [\apj] {10.1088/0004-637X/700/2/1299}, 700,
  1299

\bibitem[\protect\citeauthoryear{{Jones}, {Tielens}, {Hollenbach}  \&
  {McKee}}{{Jones} et~al.}{1994}]{1994ApJ...433..797J}
{Jones} A.~P.,  {Tielens} A.~G.~G.~M.,  {Hollenbach} D.~J.,   {McKee} C.~F.,
  1994, \mn@doi [\apj] {10.1086/174689}, 433, 797

\bibitem[\protect\citeauthoryear{{Jones}, {Tielens}  \& {Hollenbach}}{{Jones}
  et~al.}{1996}]{1996ApJ...469..740J}
{Jones} A.~P.,  {Tielens} A.~G.~G.~M.,   {Hollenbach} D.~J.,  1996, \mn@doi
  [\apj] {10.1086/177823}, 469, 740

\bibitem[\protect\citeauthoryear{{Kataoka}, {Okuzumi}, {Tanaka}  \&
  {Nomura}}{{Kataoka} et~al.}{2014}]{2014A&A...568A..42K}
{Kataoka} A.,  {Okuzumi} S.,  {Tanaka} H.,   {Nomura} H.,  2014, \mn@doi [\aap]
  {10.1051/0004-6361/201323199}, 568, A42

\bibitem[\protect\citeauthoryear{{Kennicutt} \& {Evans}}{{Kennicutt} \&
  {Evans}}{2012}]{2012ARA&A..50..531K}
{Kennicutt} R.~C.,  {Evans} N.~J.,  2012, \mn@doi [\araa]
  {10.1146/annurev-astro-081811-125610}, 50, 531

\bibitem[\protect\citeauthoryear{{Kim} et~al.,}{{Kim}
  et~al.}{2014}]{2014ApJS..210...14K}
{Kim} J.-h.,  et~al., 2014, \mn@doi [\apjs] {10.1088/0067-0049/210/1/14}, 210,
  14

\bibitem[\protect\citeauthoryear{{Kuo} \& {Hirashita}}{{Kuo} \&
  {Hirashita}}{2012}]{2012MNRAS.424L..34K}
{Kuo} T.-M.,  {Hirashita} H.,  2012, \mn@doi [\mnras]
  {10.1111/j.1745-3933.2012.01282.x}, 424, L34

\bibitem[\protect\citeauthoryear{{Kuo}, {Hirashita}  \& {Zafar}}{{Kuo}
  et~al.}{2013}]{2013MNRAS.436.1238K}
{Kuo} T.-M.,  {Hirashita} H.,   {Zafar} T.,  2013, \mn@doi [\mnras]
  {10.1093/mnras/stt1648}, 436, 1238

\bibitem[\protect\citeauthoryear{{Larson}}{{Larson}}{2005}]{2005MNRAS.359..211%
L}
{Larson} R.~B.,  2005, \mn@doi [\mnras] {10.1111/j.1365-2966.2005.08881.x},
  359, 211

\bibitem[\protect\citeauthoryear{{Li} \& {Draine}}{{Li} \&
  {Draine}}{2001}]{2001ApJ...554..778L}
{Li} A.,  {Draine} B.~T.,  2001, \mn@doi [\apj] {10.1086/323147}, 554, 778

\bibitem[\protect\citeauthoryear{{Mathis}}{{Mathis}}{1990}]{1990ARA&A..28...37%
M}
{Mathis} J.~S.,  1990, \mn@doi [\araa] {10.1146/annurev.aa.28.090190.000345},
  28, 37

\bibitem[\protect\citeauthoryear{{Mathis}, {Rumpl}  \& {Nordsieck}}{{Mathis}
  et~al.}{1977}]{1977ApJ...217..425M}
{Mathis} J.~S.,  {Rumpl} W.,   {Nordsieck} K.~H.,  1977, \mn@doi [\apj]
  {10.1086/155591}, 217, 425

\bibitem[\protect\citeauthoryear{{Mattsson} \& {Andersen}}{{Mattsson} \&
  {Andersen}}{2012}]{2012MNRAS.423...38M}
{Mattsson} L.,  {Andersen} A.~C.,  2012, \mn@doi [\mnras]
  {10.1111/j.1365-2966.2012.20574.x}, 423, 38

\bibitem[\protect\citeauthoryear{{McKee}}{{McKee}}{1989}]{1989IAUS..135..431M}
{McKee} C.,  1989, in {Allamandola} L.~J.,  {Tielens} A.~G.~G.~M.,  eds,  IAU
  Symposium Vol. 135, Interstellar Dust. p.~431

\bibitem[\protect\citeauthoryear{{McKee} \& {Ostriker}}{{McKee} \&
  {Ostriker}}{1977}]{1977ApJ...218..148M}
{McKee} C.~F.,  {Ostriker} J.~P.,  1977, \mn@doi [\apj] {10.1086/155667}, 218,
  148

\bibitem[\protect\citeauthoryear{{McKee}, {Hollenbach}, {Seab}  \&
  {Tielens}}{{McKee} et~al.}{1987}]{1987ApJ...318..674M}
{McKee} C.~F.,  {Hollenbach} D.~J.,  {Seab} G.~C.,   {Tielens} A.~G.~G.~M.,
  1987, \mn@doi [\apj] {10.1086/165403}, 318, 674

\bibitem[\protect\citeauthoryear{{McKinnon}, {Torrey}, {Vogelsberger},
  {Hayward}  \& {Marinacci}}{{McKinnon} et~al.}{2016a}]{2016arXiv160602714M}
{McKinnon} R.,  {Torrey} P.,  {Vogelsberger} M.,  {Hayward} C.~C.,
  {Marinacci} F.,  2016a, preprint (\mn@eprint {arXiv} {1606.02714})

\bibitem[\protect\citeauthoryear{{McKinnon}, {Torrey}  \&
  {Vogelsberger}}{{McKinnon} et~al.}{2016b}]{2016MNRAS.457.3775M}
{McKinnon} R.,  {Torrey} P.,   {Vogelsberger} M.,  2016b, \mn@doi [\mnras]
  {10.1093/mnras/stw253}, 457, 3775

\bibitem[\protect\citeauthoryear{{Morris}}{{Morris}}{1996}]{1996PASA...13...97%
M}
{Morris} J.~P.,  1996, \pasa, 13, 97

\bibitem[\protect\citeauthoryear{{Moustakas}, {Kennicutt}, {Tremonti}, {Dale},
  {Smith}  \& {Calzetti}}{{Moustakas} et~al.}{2010}]{2010ApJS..190..233M}
{Moustakas} J.,  {Kennicutt} Jr. R.~C.,  {Tremonti} C.~A.,  {Dale} D.~A.,
  {Smith} J.-D.~T.,   {Calzetti} D.,  2010, \mn@doi [\apjs]
  {10.1088/0067-0049/190/2/233}, 190, 233

\bibitem[\protect\citeauthoryear{{Nagamine}, {Fukugita}, {Cen}  \&
  {Ostriker}}{{Nagamine} et~al.}{2001}]{2001ApJ...558..497N}
{Nagamine} K.,  {Fukugita} M.,  {Cen} R.,   {Ostriker} J.~P.,  2001, \mn@doi
  [\apj] {10.1086/322293}, 558, 497

\bibitem[\protect\citeauthoryear{{Nagamine}, {Springel}, {Hernquist}  \&
  {Machacek}}{{Nagamine} et~al.}{2004}]{2004MNRAS.350..385N}
{Nagamine} K.,  {Springel} V.,  {Hernquist} L.,   {Machacek} M.,  2004, \mn@doi
  [\mnras] {10.1111/j.1365-2966.2004.07664.x}, 350, 385

\bibitem[\protect\citeauthoryear{{Nozawa} \& {Fukugita}}{{Nozawa} \&
  {Fukugita}}{2013}]{2013ApJ...770...27N}
{Nozawa} T.,  {Fukugita} M.,  2013, \mn@doi [\apj]
  {10.1088/0004-637X/770/1/27}, 770, 27

\bibitem[\protect\citeauthoryear{{Nozawa}, {Kozasa}  \& {Habe}}{{Nozawa}
  et~al.}{2006}]{2006ApJ...648..435N}
{Nozawa} T.,  {Kozasa} T.,   {Habe} A.,  2006, \mn@doi [\apj] {10.1086/505639},
  648, 435

\bibitem[\protect\citeauthoryear{{Nozawa}, {Asano}, {Hirashita}  \&
  {Takeuchi}}{{Nozawa} et~al.}{2015}]{2015MNRAS.447L..16N}
{Nozawa} T.,  {Asano} R.~S.,  {Hirashita} H.,   {Takeuchi} T.~T.,  2015,
  \mn@doi [\mnras] {10.1093/mnrasl/slu175}, 447, L16

\bibitem[\protect\citeauthoryear{{Okuzumi}, {Tanaka}  \& {Sakagami}}{{Okuzumi}
  et~al.}{2009}]{2009ApJ...707.1247O}
{Okuzumi} S.,  {Tanaka} H.,   {Sakagami} M.-a.,  2009, \mn@doi [\apj]
  {10.1088/0004-637X/707/2/1247}, 707, 1247

\bibitem[\protect\citeauthoryear{{Omukai}, {Tsuribe}, {Schneider}  \&
  {Ferrara}}{{Omukai} et~al.}{2005}]{2005ApJ...626..627O}
{Omukai} K.,  {Tsuribe} T.,  {Schneider} R.,   {Ferrara} A.,  2005, \mn@doi
  [\apj] {10.1086/429955}, 626, 627

\bibitem[\protect\citeauthoryear{{Popping}, {Somerville}  \&
  {Galametz}}{{Popping} et~al.}{2016}]{2016arXiv160908622P}
{Popping} G.,  {Somerville} R.~S.,   {Galametz} M.,  2016, preprint (\mn@eprint
  {arXiv} {1609.08622})

\bibitem[\protect\citeauthoryear{{R{\'e}my-Ruyer} et~al.,}{{R{\'e}my-Ruyer}
  et~al.}{2014}]{2014A&A...563A..31R}
{R{\'e}my-Ruyer} A.,  et~al., 2014, \mn@doi [\aap]
  {10.1051/0004-6361/201322803}, 563, A31

\bibitem[\protect\citeauthoryear{{Saitoh} \& {Makino}}{{Saitoh} \&
  {Makino}}{2013}]{2013ApJ...768...44S}
{Saitoh} T.~R.,  {Makino} J.,  2013, \mn@doi [\apj]
  {10.1088/0004-637X/768/1/44}, 768, 44

\bibitem[\protect\citeauthoryear{{Sandstrom} et~al.,}{{Sandstrom}
  et~al.}{2013}]{2013ApJ...777....5S}
{Sandstrom} K.~M.,  et~al., 2013, \mn@doi [\apj] {10.1088/0004-637X/777/1/5},
  777, 5

\bibitem[\protect\citeauthoryear{{Schaye} et~al.,}{{Schaye}
  et~al.}{2015}]{2015MNRAS.446..521S}
{Schaye} J.,  et~al., 2015, \mn@doi [\mnras] {10.1093/mnras/stu2058}, 446, 521

\bibitem[\protect\citeauthoryear{{Schneider}, {Omukai}, {Inoue}  \&
  {Ferrara}}{{Schneider} et~al.}{2006}]{2006MNRAS.369.1437S}
{Schneider} R.,  {Omukai} K.,  {Inoue} A.~K.,   {Ferrara} A.,  2006, \mn@doi
  [\mnras] {10.1111/j.1365-2966.2006.10391.x}, 369, 1437

\bibitem[\protect\citeauthoryear{{Shimizu}, {Inoue}, {Okamoto}  \&
  {Yoshida}}{{Shimizu} et~al.}{2014}]{2014MNRAS.440..731S}
{Shimizu} I.,  {Inoue} A.~K.,  {Okamoto} T.,   {Yoshida} N.,  2014, \mn@doi
  [\mnras] {10.1093/mnras/stu265}, 440, 731

\bibitem[\protect\citeauthoryear{{Shimizu}, {Inoue}, {Yoshida}  \&
  {Okamoto}}{{Shimizu} et~al.}{2015}]{2015arXiv150900800S}
{Shimizu} I.,  {Inoue} A.~K.,  {Yoshida} N.,   {Okamoto} T.,  2015, preprint
  (\mn@eprint {arXiv} {1509.00800})

\bibitem[\protect\citeauthoryear{{Smith} et~al.,}{{Smith}
  et~al.}{2016}]{2016MNRAS.462..331S}
{Smith} M.~W.~L.,  et~al., 2016, \mn@doi [\mnras] {10.1093/mnras/stw1611}, 462,
  331

\bibitem[\protect\citeauthoryear{{Springel}}{{Springel}}{2005}]{2005MNRAS.364.%
1105S}
{Springel} V.,  2005, \mn@doi [\mnras] {10.1111/j.1365-2966.2005.09655.x}, 364,
  1105

\bibitem[\protect\citeauthoryear{{Springel} \& {Hernquist}}{{Springel} \&
  {Hernquist}}{2002}]{2002MNRAS.333..649S}
{Springel} V.,  {Hernquist} L.,  2002, \mn@doi [\mnras]
  {10.1046/j.1365-8711.2002.05445.x}, 333, 649

\bibitem[\protect\citeauthoryear{{Springel} \& {Hernquist}}{{Springel} \&
  {Hernquist}}{2003}]{2003MNRAS.339..289S}
{Springel} V.,  {Hernquist} L.,  2003, \mn@doi [\mnras]
  {10.1046/j.1365-8711.2003.06206.x}, 339, 289

\bibitem[\protect\citeauthoryear{{Steidel}, {Adelberger}, {Giavalisco},
  {Dickinson}  \& {Pettini}}{{Steidel} et~al.}{1999}]{1999ApJ...519....1S}
{Steidel} C.~C.,  {Adelberger} K.~L.,  {Giavalisco} M.,  {Dickinson} M.,
  {Pettini} M.,  1999, \mn@doi [\apj] {10.1086/307363}, 519, 1

\bibitem[\protect\citeauthoryear{{Stinson}, {Brook}, {Macci{\`o}}, {Wadsley},
  {Quinn}  \& {Couchman}}{{Stinson} et~al.}{2013}]{2013MNRAS.428..129S}
{Stinson} G.~S.,  {Brook} C.,  {Macci{\`o}} A.~V.,  {Wadsley} J.,  {Quinn}
  T.~R.,   {Couchman} H.~M.~P.,  2013, \mn@doi [\mnras] {10.1093/mnras/sts028},
  428, 129

\bibitem[\protect\citeauthoryear{{Takeuchi}, {Buat}, {Heinis}, {Giovannoli},
  {Yuan}, {Iglesias-P{\'a}ramo}, {Murata}  \& {Burgarella}}{{Takeuchi}
  et~al.}{2010}]{2010A&A...514A...4T}
{Takeuchi} T.~T.,  {Buat} V.,  {Heinis} S.,  {Giovannoli} E.,  {Yuan} F.-T.,
  {Iglesias-P{\'a}ramo} J.,  {Murata} K.~L.,   {Burgarella} D.,  2010, \mn@doi
  [\aap] {10.1051/0004-6361/200913476}, 514, A4

\bibitem[\protect\citeauthoryear{{Takeuchi}, {Yuan}, {Ikeyama}, {Murata}  \&
  {Inoue}}{{Takeuchi} et~al.}{2012}]{2012ApJ...755..144T}
{Takeuchi} T.~T.,  {Yuan} F.-T.,  {Ikeyama} A.,  {Murata} K.~L.,   {Inoue}
  A.~K.,  2012, \mn@doi [\apj] {10.1088/0004-637X/755/2/144}, 755, 144

\bibitem[\protect\citeauthoryear{{Thompson}, {Nagamine}, {Jaacks}  \&
  {Choi}}{{Thompson} et~al.}{2014}]{2014ApJ...780..145T}
{Thompson} R.,  {Nagamine} K.,  {Jaacks} J.,   {Choi} J.-H.,  2014, \mn@doi
  [\apj] {10.1088/0004-637X/780/2/145}, 780, 145

\bibitem[\protect\citeauthoryear{{Todoroki}}{{Todoroki}}{2014}]{2014MsT.......%
...1T}
{Todoroki} K.,  2014, Master's thesis, University of Nevada, Las Vegas

\bibitem[\protect\citeauthoryear{{Valiante}, {Schneider}, {Bianchi}  \&
  {Andersen}}{{Valiante} et~al.}{2009}]{2009MNRAS.397.1661V}
{Valiante} R.,  {Schneider} R.,  {Bianchi} S.,   {Andersen} A.~C.,  2009,
  \mn@doi [\mnras] {10.1111/j.1365-2966.2009.15076.x}, 397, 1661

\bibitem[\protect\citeauthoryear{{Vogelsberger} et~al.,}{{Vogelsberger}
  et~al.}{2014}]{2014MNRAS.444.1518V}
{Vogelsberger} M.,  et~al., 2014, \mn@doi [\mnras] {10.1093/mnras/stu1536},
  444, 1518

\bibitem[\protect\citeauthoryear{{Voshchinnikov} \&
  {Hirashita}}{{Voshchinnikov} \& {Hirashita}}{2014}]{2014MNRAS.445..301V}
{Voshchinnikov} N.~V.,  {Hirashita} H.,  2014, \mn@doi [\mnras]
  {10.1093/mnras/stu1720}, 445, 301

\bibitem[\protect\citeauthoryear{{Whitworth}, {Boffin}  \&
  {Francis}}{{Whitworth} et~al.}{1998}]{1998MNRAS.299..554W}
{Whitworth} A.~P.,  {Boffin} H.~M.~J.,   {Francis} N.,  1998, \mn@doi [\mnras]
  {10.1046/j.1365-8711.1998.01813.x}, 299, 554

\bibitem[\protect\citeauthoryear{{Wiersma}, {Schaye}, {Dalla Vecchia}, {Booth},
  {Theuns}  \& {Aguirre}}{{Wiersma} et~al.}{2010}]{2010MNRAS.409..132W}
{Wiersma} R.~P.~C.,  {Schaye} J.,  {Dalla Vecchia} C.,  {Booth} C.~M.,
  {Theuns} T.,   {Aguirre} A.,  2010, \mn@doi [\mnras]
  {10.1111/j.1365-2966.2010.17299.x}, 409, 132

\bibitem[\protect\citeauthoryear{{Wise}, {Abel}, {Turk}, {Norman}  \&
  {Smith}}{{Wise} et~al.}{2012}]{2012MNRAS.427..311W}
{Wise} J.~H.,  {Abel} T.,  {Turk} M.~J.,  {Norman} M.~L.,   {Smith} B.~D.,
  2012, \mn@doi [\mnras] {10.1111/j.1365-2966.2012.21809.x}, 427, 311

\bibitem[\protect\citeauthoryear{{Wiseman}, {Schady}, {Bolmer}, {Kr{\"u}hler},
  {Yates}, {Greiner}  \& {Fynbo}}{{Wiseman} et~al.}{2016}]{2016arXiv160700288W}
{Wiseman} P.,  {Schady} P.,  {Bolmer} J.,  {Kr{\"u}hler} T.,  {Yates} R.~M.,
  {Greiner} J.,   {Fynbo} J.~P.~U.,  2016, preprint (\mn@eprint {arXiv}
  {1607.00288})

\bibitem[\protect\citeauthoryear{{Woosley} \& {Heger}}{{Woosley} \&
  {Heger}}{2007}]{2007PhR...442..269W}
{Woosley} S.~E.,  {Heger} A.,  2007, \mn@doi [\physrep]
  {10.1016/j.physrep.2007.02.009}, 442, 269

\bibitem[\protect\citeauthoryear{{Yajima}, {Nagamine}, {Thompson}  \&
  {Choi}}{{Yajima} et~al.}{2014}]{2014MNRAS.439.3073Y}
{Yajima} H.,  {Nagamine} K.,  {Thompson} R.,   {Choi} J.-H.,  2014, \mn@doi
  [\mnras] {10.1093/mnras/stu169}, 439, 3073

\bibitem[\protect\citeauthoryear{{Yajima}, {Shlosman}, {Romano-D{\'{\i}}az}  \&
  {Nagamine}}{{Yajima} et~al.}{2015}]{2015MNRAS.451..418Y}
{Yajima} H.,  {Shlosman} I.,  {Romano-D{\'{\i}}az} E.,   {Nagamine} K.,  2015,
  \mn@doi [\mnras] {10.1093/mnras/stv974}, 451, 418

\bibitem[\protect\citeauthoryear{{Yamasawa}, {Habe}, {Kozasa}, {Nozawa},
  {Hirashita}, {Umeda}  \& {Nomoto}}{{Yamasawa}
  et~al.}{2011}]{2011ApJ...735...44Y}
{Yamasawa} D.,  {Habe} A.,  {Kozasa} T.,  {Nozawa} T.,  {Hirashita} H.,
  {Umeda} H.,   {Nomoto} K.,  2011, \mn@doi [\apj]
  {10.1088/0004-637X/735/1/44}, 735, 44

\bibitem[\protect\citeauthoryear{{Yan}, {Lazarian}  \& {Draine}}{{Yan}
  et~al.}{2004}]{2004ApJ...616..895Y}
{Yan} H.,  {Lazarian} A.,   {Draine} B.~T.,  2004, \mn@doi [\apj]
  {10.1086/425111}, 616, 895

\bibitem[\protect\citeauthoryear{{Zhou}, {Cao}  \& {Wu}}{{Zhou}
  et~al.}{2015}]{2015AJ....149....1Z}
{Zhou} Z.-M.,  {Cao} C.,   {Wu} H.,  2015, \mn@doi [\aj]
  {10.1088/0004-6256/149/1/1}, 149, 1

\bibitem[\protect\citeauthoryear{{Zhukovska}}{{Zhukovska}}{2014}]{2014A&A...56%
2A..76Z}
{Zhukovska} S.,  2014, \mn@doi [\aap] {10.1051/0004-6361/201322989}, 562, A76

\bibitem[\protect\citeauthoryear{{Zhukovska}, {Gail}  \&
  {Trieloff}}{{Zhukovska} et~al.}{2008}]{2008A&A...479..453Z}
{Zhukovska} S.,  {Gail} H.-P.,   {Trieloff} M.,  2008, \mn@doi [\aap]
  {10.1051/0004-6361:20077789}, 479, 453

\bibitem[\protect\citeauthoryear{{Zhukovska}, {Dobbs}, {Jenkins}  \&
  {Klessen}}{{Zhukovska} et~al.}{2016}]{2016arXiv160804781Z}
{Zhukovska} S.,  {Dobbs} C.,  {Jenkins} E.~B.,   {Klessen} R.,  2016, preprint
  (\mn@eprint {arXiv} {1608.04781})

\bibitem[\protect\citeauthoryear{{de Blok}, {Walter}, {Brinks}, {Trachternach},
  {Oh}  \& {Kennicutt}}{{de Blok} et~al.}{2008}]{2008AJ....136.2648D}
{de Blok} W.~J.~G.,  {Walter} F.,  {Brinks} E.,  {Trachternach} C.,  {Oh}
  S.-H.,   {Kennicutt} Jr. R.~C.,  2008, \mn@doi [\aj]
  {10.1088/0004-6256/136/6/2648}, 136, 2648

\bibitem[\protect\citeauthoryear{{de Vaucouleurs} \& {Pence}}{{de Vaucouleurs}
  \& {Pence}}{1978}]{1978AJ.....83.1163D}
{de Vaucouleurs} G.,  {Pence} W.~D.,  1978, \mn@doi [\aj] {10.1086/112305}, 83,
  1163

\makeatother
\end{thebibliography}

\appendix
\section{Evolution of dust abundances}\label{appendixA}
{We show the derivation of equations (\ref{eq:timeL}) and (\ref{eq:timeS}).
The time evolution of gas mass [$m_{{\rm g}(i)}(t)$], large grain mass
[$m_{{\rm L}(i)}(t)$], and small grain mass [$m_{{\rm S}(i)}(t)$] in the $i$th gas
particle is written as
(following equations 1, 3, and 4 in \citealt{2015MNRAS.447.2937H})
\begin{eqnarray}
\dfrac{\mathrm{d}m_{{\rm g}(i)}(t)}{\mathrm{d}t}&=&-\psi_{i}+
\dfrac{\mathrm{d}m^{\rm return}_{{\rm g}(i)}}{\mathrm{d}t},\\
\dfrac{\mathrm{d}m_{{\rm L}(i)}(t)}{\mathrm{d}t}&=&-\mathcal{D}_{{\rm L}(i)}\psi_{i}\\
& &- \left( \dfrac{m_{{\rm L}(i)}(t)}{\tau_{\rm sh}} 
- \dfrac{m_{{\rm S}(i)}(t)}{\tau_{\rm co}}\right)
- \dfrac{m_{{\rm L}(i)}(t)}{\tau_{\rm sp}(a_{\rm L})}\notag \\ 
& & +\left[ \dfrac{\mathrm{d}m_{{\rm L}(i)}(t)}{\mathrm{d}t}\right]_{{\rm Source}}
- \left[ \dfrac{\mathrm{d}m_{{\rm L}(i)}(t)}{\mathrm{d}t}\right]_{{\rm SNe}} \, ,
\label{eq:timeL0a} \\ 
\dfrac{\mathrm{d}m_{{\rm S}(i)}(t)}{\mathrm{d}t}&=&-\mathcal{D}_{{\rm S}(i)}\psi_{i}\\
& &  \left(\dfrac{m_{{\rm L}(i)}(t)}{\tau_{\rm sh}}
- \dfrac{m_{{\rm S}(i)}(t)}{\tau_{\rm co }}  
+ \dfrac{m_{{\rm S}(i)}(t)}{\tau_{\rm acc}}\right)\, \notag \\
& &- \dfrac{m_{{\rm S}(i)}(t)}{\tau_{\rm sp}(a_{\rm S})}
-\left[ \dfrac{\mathrm{d}m_{{\rm S}(i)}(t)}{\mathrm{d}t}\right]_{{\rm SNe}}\, ,
\label{eq:timeS0a} 
\end{eqnarray}
where $\psi_{i}$ is the star formation rate of the $i$th gas particle.
The time derivatives of dust mass can be converted to those of dust-to-gas ratio
using the following relation between
$\mathrm{d}m_{{\rm L/S}(i)}/\mathrm{d}t$ and $\mathrm{d}\mathcal{D}_{{\rm L/S}(i)}/\mathrm{d}t$:
\begin{eqnarray} 
m_{{\rm g}(i)}\dfrac{\mathrm{d}\mathcal{D}_{{\rm L/S}(i)}}{\mathrm{d}t}&=&
m_{{\rm g}(i)}\dfrac{\mathrm{d}}{\mathrm{d}t}\left(\dfrac{m_{{\rm L/S}(i)}}{m_{{\rm g}(i)}} \right)\\
&=&\dfrac{\mathrm{d}m_{{\rm L/S}(i)}}{\mathrm{d}t}-\mathcal{D}_{{\rm L/S}(i)}
\dfrac{\mathrm{d}m_{{\rm g}(i)}}{\mathrm{d}t}\\
&=&\dfrac{\mathrm{d}m_{{\rm L/S}(i)}}{\mathrm{d}t}+\mathcal{D}_{{\rm L/S}(i)}\psi 
-\mathcal{D}_{{\rm L/S}(i)}\dfrac{\mathrm{d}m^{\rm return}_{{\rm g}(i)}}{\mathrm{d}t}
\label{eq:relation}.\notag \\
\end{eqnarray}
By using this relation and rewriting
$[\mathrm{d}m_{\mathrm{L/S}(i)}/\mathrm{d}t]_\mathrm{Source}/m_{\mathrm{g}(i)}=
[\mathrm{d}\mathcal{D}_{\mathrm{L/S}(i)}/\mathrm{d}t]_\mathrm{Source}$ and
$[\mathrm{d}m_{\mathrm{L/S}(i)}/\mathrm{d}t]_\mathrm{SNe}/m_{\mathrm{g}(i)}=
[\mathrm{d}\mathcal{D}_{\mathrm{L/S}(i)}/\mathrm{d}t]_\mathrm{SNe}$, we
obtain equations \eqref{eq:timeL} and \eqref{eq:timeS}.
The astration terms do not appear in these equations because
astration does not change the dust-to-gas ratio (i.e.\ dust and gas are
both consumed without changing the dust-to-gas ratio of the remaining gas).
}

\bsp	
\label{lastpage}
\end{document}